\documentclass[journal,10pt]{IEEEtran}

\usepackage{amsmath,amssymb}
\usepackage[euler]{textgreek}
\usepackage{esint}
\usepackage{subfigure}
\usepackage{psfrag}
\usepackage{graphicx}
\usepackage{color}
\usepackage[usenames,dvipsnames]{xcolor}
\usepackage{pifont}
\usepackage{fourier}
\usepackage{xr}
\usepackage{stmaryrd}
\usepackage{cancel}
\usepackage{upgreek}
\usepackage{enumitem}
\usepackage{upgreek}
\usepackage{cite}
\usepackage{hyperref}

\usepackage{fancyhdr}
\fancypagestyle{plain}{
\fancyhf{}
\lfoot{Page \thepage}
}
\usepackage[process=auto]{pstool}
\usepackage[normalem]{ulem}
\definecolor{redg}{rgb}{1,0,0}
\definecolor{blueg}{rgb}{0.22,0.33,0.64}
\definecolor{greeng}{rgb}{0,0.63,0.29}
\definecolor{orangeg}{rgb}{0.96,0.47,0.13}

\DeclareMathAlphabet\mathbfcal{OMS}{cmsy}{b}{n}

\newcommand{\figref}{Fig.~\ref}

\newcommand{\ve}[1]{\mathbf{#1}}

\newcommand{\ves}[1]{\boldsymbol{#1}}

\newcommand{\tx}[1]{\text{#1}}
\newcommand{\te}[1]{\overline{\overline{#1}}}

\newcommand{\hatv}[1]{\hat{\mathbf{#1}}}

\begin{document}

\title{Spacetime Metamaterials}

\author{Christophe Caloz,~\IEEEmembership{Fellow,~IEEE,}
        and~Zo\'{e}-Lise Deck-L\'{e}ger,~\IEEEmembership{Student,~IEEE}
        \\ \vspace{0.5cm}
\thanks{The authors are with Polytechnique Montr\'{e}al, Montr\'{e}al,
Canada.}
\thanks{Manuscript received April, 2019; revised Month Day, 2019.}}


\maketitle

\tableofcontents

\vspace{5mm}
\begin{abstract}
This paper presents the authors' vision of the emerging field of spacetime metamaterials in a cohesive and pedagogical perspective. For this purpose, it systematically builds up the physics, modeling and applications of these media upon the foundation of their pure-space and pure-time counterparts. First, it introduces spacetime metamaterials as a generalization of (bianisotropic) metamaterials, presented in the holistic perspective of direct and inverse spacetime scattering, where spacetime variance and dispersion offer unprecedented medium diversity despite some limitations related to the uncertainty principle. Then, it describes the fundamental physical phenomena occurring in spacetime systems, such as frequency transitions, nonreciprocity, Fizeau dragging, bianisotropy transformation, and superluminality, allowed when the medium moves perpendicularly to the direction of the wave. Next, it extends some principles and tools of relativity physics, particularly a medium-extended version of the spacetime (or Minkowski) diagrams, elaborates a general strategy to compute the fields scattered by spacetime media, and presents a gallery of possible spacetime media, including the spacetime step discontinuity, which constitutes the building brick of any spacetime metamaterial. From this point, it describes the phenomenology of such a spacetime interface, deduces from it the related electromagnetic boundary conditions, and derives the corresponding scattering (Fresnel-like) coefficients and frequency transitions, culminating with a section that generalizes time reversal to spacetime compansion (compression and expansion). Then, a section illustrates the new physics of spacetime metamaterials with the examples of spacetime mirrors and cavities, the inverse prism and chromatic birefringence, and spacetime crystals. Finally, the paper discusses various applications -- categorized as frequency multiplication and mixing, matching and filtering, nonreciprocity and absorption, cloaking, electromagnetic processing, and radiation. Ultimately, the conclusion section provides a 23-item list that concisely summarizes the key results and teachings of the overall document. 
\end{abstract}


\IEEEpeerreviewmaketitle

\section{Introduction}\label{sec:intro}

A metamaterial is an artificial structure consisting of a subwavelength lattice of scattering particles -- or metaparticles -- engineered to provide desired medium properties that are generally \emph{beyond} (Greek prefix \textmu\textepsilon\texttau$\acute{\text{\textalpha}}$) those available in other materials. Their constituent metaparticles are naturally themselves composed of molecules, atoms and subatomic particles, but it is rather the larger-scale geometry, orientation and arrangement of these metaparticles that determine the essential macroscopic properties of the metamaterial. While they most often refer to homogenizable structures, i.e., subwavelengh-lattice structures that can be described by media parameters, their subwavelength regime is always restricted in terms of frequency range, and a metamaterial is hence really a metamaterial only in a small fraction of the electromagnetic spectrum. For this reason, we envisage a somewhat broader definition of metamaterials, which encompasses their response beyond their actual medium range.

\begin{figure}[h]
\centering
\includegraphics[width=0.95\linewidth]{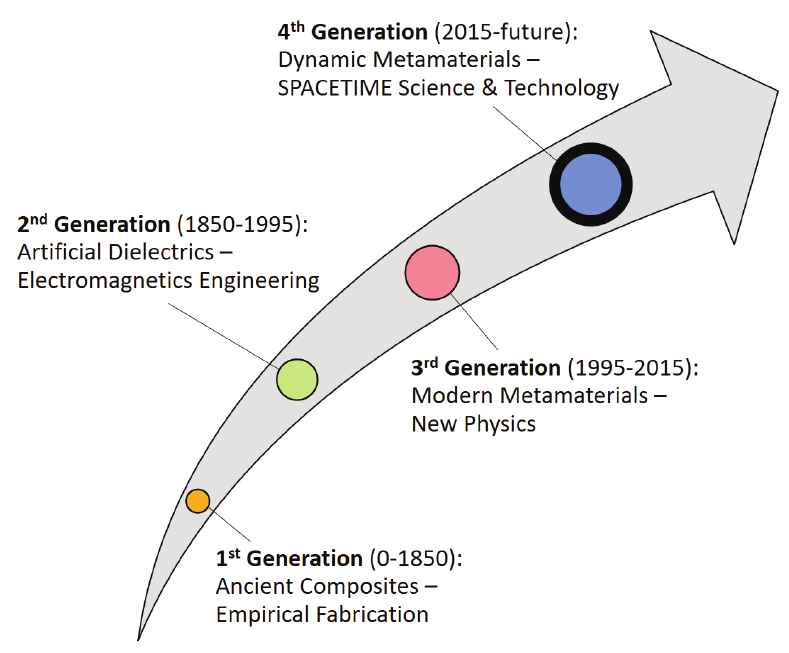}{
}
\vspace{-3mm}
\caption{Metamaterial evolution in four consecutive generations, where the recently started $4^\tx{th}$ generation is largely represented by spacetime metamaterials.}
\label{fig:generations}
\end{figure}

Although the term `metamaterial' is hardly 20 years old~\cite{Walser_2001}, metamaterials have a long history, which may be roughly divided in the four consecutive generations represented in \figref{fig:generations}. This history goes back to ancient times, with various composite materials, such as Lycurgus cup's dichroic glass -- made of gold and silver nanoparticles~\cite{Leonhardt_2007}, Damascus steel -- recently discovered to include nanowires and carbon nanotubes~\cite{Reibold_2006}, and medieval stained glass -- made of soda-lime-silica or potash-lime-silica compounds~\cite{Shvets_2008}. However, these early metamaterials followed empirical fabrication rules and involved little engineering principles. We may call these composites \emph{$1^{\rm st}$-generation} metamaterials (\figref{fig:generations}). Metamaterial engineering, based on Maxwell's electromagnetics theory and subsequent early microwave-optical technologies, really emerged only near the end of the XIX$^\tx{th}$ century and during the following decades, with developments such as Bose's tinfoil-paper polarizers~\cite{Bose_1898}, Lindman's chiral structures~\cite{Lindman_1920}, Kock and Cohn's  light lenses~\cite{Kock_1948,Cohn_1951}, and Rotman's wire plasmas~\cite{Rotman_1962}. These ``artificial dielectrics'' were of great practical interest, but possessed well-known physical properties. We may refer to them as \emph{$2^{\rm nd}$-generation} metamaterials (\figref{fig:generations}). At the turn of the XX$^\tx{th}$ century, metamaterials brought up new physics into the picture, initially with negative refraction~\cite{Shelby_2001,Pendry_2000}, and later with other effects such as cloaking~\cite{Schurig_2006,Leonhardt_2006}, magnetless magnetism~\cite{Kodera_APL_2011,Kodera_AWPL_2018,Caloz_2018}, and early metasurfaces (two-dimensional metamaterials with new phase, magnitude and polarization transformation capabilities)~\cite{Holloway_2012,Glybovski_2016,Achouri_NP_2018}. We shall refer here to these metamaterials - that have already generated over 300 books to date! (e.g.~\cite{Caloz_2005,Engheta_2006,Capolino_2009,Cai_2009,Marques_2013,Werner_2015}) -- as \emph{$3^{\rm rd}$-generation} metamaterials (\figref{fig:generations}).

Recent advances in metasurfaces and bianisotropic metastructures have dramatically extended the range of modern-metamaterial properties. These advances have most notably ushered the field of \emph{dynamic metamaterials}, which may be regarded as \emph{$4^\tx{th}$-generation} metamaterials (\figref{fig:generations}), and which are currently in high-speed expansion. Dynamic metamaterials are materials whose properties are partly due to the \emph{time-variation} of some of their physical parameters, induced by an external source of energy. They naturally include reconfigurable metamaterials, whose response may be changed over time, but they are much more general and more fundamental insofar as their properties at any time directly depend on the time variation, which allows a universal manipulation of the temporal and spatial spectra of electromagnetic waves, beyond the mere variation of static properties over time. We shall therefore refer to these metamaterials as \emph{spacetime metamaterials}. While such metamaterials were still unthinkable a few decades ago, the recent spectacular developments in micro/nano/quanto/bio/chemico-technologies, and even artificial intelligence, presage a brilliant future to this emerging generation of metamaterials. This paper presents the authors' vision of this area in a cohesive and pedagogical perspective.

The rest of the paper is organized as follows. Section~\ref{sec:medium_gen} introduces spacetime metamaterials as a generalization of existing time-invariant metamaterials. Section~\ref{sec:fund_phen} describes the fundamental physical phenomena occurring in general (not only metamaterial) spacetime systems. Section~\ref{sec:SR_perps} presents a relativistic perspective of spacetime systems and metamaterials, including the Minkowski diagram representation, the Fourier-direct and Fourier-inverse Lorentz transformations, and the overall approach that we next use to solve spacetime metamaterial problems. Upon this basis, Sec.~\ref{sec:scat_BCs} recalls the phenomenology of pure-space and pure-time interfaces and derives the electromagnetic spacetime boundary conditions for the subluminal and superluminal regimes from their respective purely spatial and purely temporal limit cases. Using these boundary conditions, Secs.~\ref{sec:scat_coef} and~\ref{sec:freq_trans} derive the fundamental relations for the scattering coefficients and frequency transitions, respectively, at a spacetime interface, and unravel the unusual related physics. Section~\ref{sec:ST_reversal} shows that spacetime reversal naturally emerges from our spacetime framework as a generalization of the concept of time reversal. Spacetime metamaterials may support a myriad of new physical effects, most of which are still to be discovered. Section~\ref{sec:new_physics} illustrates this new physics with a selection of three examples: the spacetime mirror and cavity, the inverse prism and chromatic birefringence, and spacetime crystals. A number of potential applications -- classified in the categories A)~frequency multiplication and mixing, B)~matching and filtering, C)~nonreciprocity and absorption, D)~cloaking, E)~electromagnetic processing, and F)~radiation --, are then discussed in Sec.~\ref{sec:new_physics}. Finally, conclusions are given in Sec.~\ref{sec:concl}.

\section{Medium Generalization}\label{sec:medium_gen}
\subsection{Non-metamaterial Spacetime Systems}\label{sec:nonmeta_ST}
There is an infinity of spacetime systems that would not qualify as spacetime metamaterials, but that are definitely relevant to consider in the context of spacetime metamaterials. This is of course the case for the entire world as experienced by us, humans, where everything, like the fall of an apple from a tree or the emission light by a star, varies in both space and time, according to Newton laws, relativity, and other laws of physics. A particularly insightful familiar spacetime phenomenon is the \emph{Doppler effect}~\cite{Eden_Doppler_1992}, produced by moving objects, such as cars or planes, emitting sounds or reflecting radar waves~\cite{Chen_MDER_2019}. At the micro/nano-scale, the emerging field of \emph{optomechanics}, which explores the classical and quantum interactions between electromagnetic radiation and moving mirrors mediated by radiation-pressure forces, represents another example of moving-media spacetime physics~\cite{Aspelmeyer_RMP_2014}.

In the above examples, the spacetime variation occurs via the \emph{motion of matter} (moving molecule blocks), as apples, stars, cars and mirrors. However, there are other spacetime systems whose variation is produced by a \emph{wave modulation} (dynamic `biasing' wave), without any transfer of matter. This is for instance the case of acousto-optic or electro-optic structures, where an external acoustic wave or light wave modulates the refractive index of the medium~\cite{Saleh_Teich_FP_2007}. A related spacetime device is the WWII traveling-wave tube, where a beam of electron amplifies a radio signal slowed down for synchronization by helical wires or coupled cavities~\cite{Hahn_1939}; in this case, the structure may in fact be seen either as a wave-modulated or moving-matter spacetime system depending on whether one considers the wave nature or particle nature of the electrons.

\subsection{Spacetime Metamaterial Illustration and Implementation}
\begin{figure}[h]
\centering
\includegraphics[width=\linewidth]{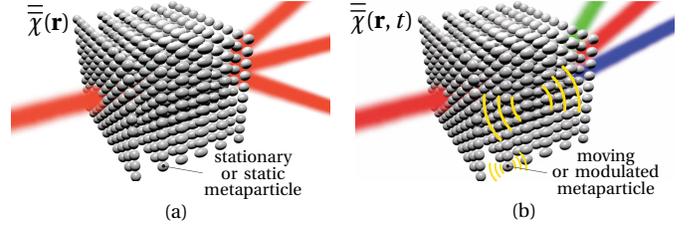}{
\psfrag{a}[c][c][0.8]{(a)}
\psfrag{b}[c][c][0.8]{(b)}
\psfrag{c}[c][c]{$\te{\chi}(\ve{r})$}
\psfrag{d}[c][c]{$\te{\chi}(\ve{r},t)$}
\psfrag{e}[l][l][0.7]{\begin{minipage}{2.5cm}\centering stationary \\ \vspace{-1mm} or static \\ \vspace{-1mm} metaparticle \end{minipage}}
\psfrag{f}[l][l][0.7]{\begin{minipage}{2.5cm}\centering moving \\ \vspace{-1mm} or modulated \\ \vspace{-1mm} metaparticle \end{minipage}}
}
\vspace{-6mm}
\caption{Generic representation of a spacetime metamaterial, using the example of a 3D periodic array of spheroidal metallic, dielectric or plasmonic metaparticles. (a)~Stationary or static (only space-varying) structure, here excited by the double-monochromatic wave $(\omega_\tx{i},\ve{k}_\tx{i})$ and producing the three diffraction orders $[(\omega_\tx{i},\ve{k}_0),(\omega_\tx{i},\ve{k}_{-1}),(\omega_\tx{i},\ve{k}_{+1})]$. (b)~Moving or modulated structure, here with the same excitation as in (a) and producing the scattered waves $[(\omega_a,\ve{k}_a),(\omega_b,\ve{k}_b),(\omega_c,\ve{k}_c)]$.}
\label{fig:generic_STM}
\end{figure}

We are here specifically interested in spacetime \emph{metamaterials}. How would such a metamaterial look like and what would it offer? A preliminary answer to this question may be formulated with the help of the illustration in Fig.~\ref{fig:generic_STM}. Figure~\ref{fig:generic_STM}(a) represents a stationary or static metamaterial, which scatters, from its periodic lattice, a few diffraction orders\footnote{Typically, a phase-gradient metamaterial or metasurface ``refracts'' an incident wave in a specified direction together with spurious diffraction orders associated to the supercell corresponding to one phase cycle of the structure. Indeed, even if the unit cell feature is much smaller than the wavelength of the incoming wave, the actual supercell period is comparable to this wavelength. This effect, which becomes severe at large angles from the normal, may be remedied by introducing proper bianisotropy~\cite{Lavigne_2018}.}. Adding a temporal variation to the spatial variation, so as to obtain a combined space and time -- or spacetime -- variation, as suggested in Fig.~\ref{fig:generic_STM}(b), produces some unique effects. For instance, the diffraction orders may be suppressed and new exotic temporal and spatial frequencies may be created.

As was the case for photonic crystals, negative materials and cloaking structures, the initial experimental spacetime metamaterials will be demonstrated at microwaves, using distributed varactors or transistors modulated by a processor, and ultimately involving artificial intelligence for adapting to the environment and realizing complex tasks. However, optical implementations, using for instance acousto-optic and electro-optic modulation, may also be envisioned in a near-future horizon.

\subsection{Spacetime Invariant -- Nondispersive Bianisotropic Media}
The behavior of an electromagnetic medium can generally be expressed by the relations
\begin{subequations}\label{eq:gen_med_resp}
\begin{equation}
\ve{D}=\epsilon_0\ve{E}+\ve{P}_\tx{e},
\end{equation}
\begin{equation}
\ve{B}=\mu_0\ve{H}+\ve{P}_\tx{m},
\end{equation}
\end{subequations}
where $\epsilon_0=8.854\cdot 10^{-12}$~As/Vm and $\mu_0=4\pi\cdot 10^{-7}$~Vs/Am are the free-space permittivity and permeability, $\ve{E}$ (V/m) and $\ve{H}$ (A/m) are the electric field and magnetic field -- considered here as the medium excitations --, $\ve{D}$ (As/m$^2$) and $\ve{B}$ (Vs/m$^2$) are the electric flux density (or displacement) field and the magnetic flux density (or induction) field -- considered here as the medium responses --, and $\ve{P}_\tx{e}$ (As/m$^2$) and $\ve{P}_\tx{m}=\mu_0\ve{M}$ (Vs/m$^2$) ($\ve{M}$: magnetization) are the electric and magnetic polarization densities corresponding to the response of the particles forming the medium while $\epsilon_0\ve{E}$ and $\mu_0\ve{H}$ represent the electric and magnetic responses of the free space between them.

$\ve{P}_\tx{e}$ may be induced not only by $\ve{E}$ but also by $\ve{H}$, while $\ve{P}_\tx{m}$ may be induced not only by $\ve{H}$ but also by $\ve{E}$. Therefore, $\ve{P}_\tx{e}$ and $\ve{P}_\tx{m}$ generally decompose as
\begin{subequations}
\begin{equation}
\ve{P}_\tx{e}=\ve{P}_\tx{ee}+\ve{P}_\tx{em},
\end{equation}
\begin{equation}
\quad\ve{P}_\tx{m}=\ve{P}_\tx{me}+\ve{P}_\tx{mm},
\end{equation}
\end{subequations}
where $\ve{P}_\tx{ee}$, $\ve{P}_\tx{em}$, $\ve{P}_\tx{me}$ and $\ve{P}_\tx{mm}$ are the electric-to-electric, magnetic-to-electric, electric-to-magnetic and magnetic-to-magnetic polarization densities induced in the medium, which we may generically refer to as $\ve{P}_{ab}$, with $ab=\tx{ee}$, $\tx{em}$, $\tx{me}$ and $\tx{mm}$. Such a medium is called \emph{biisotropic} if the responses are parallel to the excitations, i.e., $(\ve{P}_\tx{ee},\ve{P}_\tx{me})\|\ve{E}$ and $(\ve{P}_\tx{em},\ve{P}_\tx{mm})\|\ve{H}$~\cite{Lindell_EWCBM_1994,Caloz_PRAp_2019}, and \emph{bianisotropic} otherwise~\cite{Kong_EWT_2008,Caloz_PRAp_2019}.

A linear, spacetime invariant and spacetime nondispersive bianisotropic medium, where the exact definition of spacetime in/variance and non/dispersion will be given shortly, can be modeled in terms of the susceptibility dyadic $3\times 3$ tensor $\te{\chi}_{ab}$ corresponding to $\ve{P}_{ab}$ as~\cite{Caloz_PRAp_2019}
\begin{subequations}\label{eq:LSTiSTnd}
\begin{equation}\label{eq:LSTiSTnd_P}
\begin{pmatrix} \ve{P}_\tx{e} \\ \ve{P}_\tx{m} \end{pmatrix}
=\begin{pmatrix} \epsilon_0\te{\chi}_\tx{ee} & \hspace{-2mm}\sqrt{\epsilon_0\mu_0}\,\te{\chi}_\tx{em}\vspace{1mm}  \\
\sqrt{\epsilon_0\mu_0}\,\te{\chi}_\tx{me} & \hspace{-2mm}\mu_0\te{\chi}_\tx{mm} \end{pmatrix}\cdot
\begin{pmatrix} \ve{E} \\ \ve{H} \end{pmatrix}
=\te{\chi}\cdot\begin{pmatrix} \ve{E} \\ \ve{H} \end{pmatrix},
\end{equation}
where $\te{\chi}$ represents the dyadic $6\times 6$ tensor combining $\te{\chi}_\tx{ee}$, $\te{\chi}_\tx{em}$, $\te{\chi}_\tx{me}$ and $\te{\chi}_\tx{mm}$, and where we have used matrix notation for compactness and better visualization. Alternatively, inserting~\eqref{eq:LSTiSTnd} into~\eqref{eq:gen_med_resp} yields
\begin{equation}\label{eq:bianis_params}
\begin{pmatrix} \ve{D} \\ \ve{B} \end{pmatrix}
=\begin{pmatrix} \te{\epsilon} & \hspace{-1.5mm}\te{\xi} \\ \te{\zeta} & \hspace{-1mm}\te{\mu} \end{pmatrix}\cdot
\begin{pmatrix} \ve{E} \\ \ve{H} \end{pmatrix},
\:\tx{with}\:
\begin{pmatrix} \te{\epsilon} & \hspace{-1.5mm}\te{\xi} \\ \te{\zeta} & \hspace{-1mm}\te{\mu} \end{pmatrix}
=\begin{pmatrix} \epsilon_0\left(\te{I}+\te{\chi}_\tx{ee}\right) & \hspace{-1.5mm}\sqrt{\epsilon_0\mu_0}\,\te{\chi}_\tx{em} \\
\sqrt{\epsilon_0\mu_0}\,\te{\chi}_\tx{me} & \hspace{-1.5mm}\mu_0\left(\te{I}+\te{\chi}_\tx{mm}\right) \end{pmatrix},
\end{equation}
\end{subequations}
where $\te{I}=\hatv{x}\hatv{x}+\hatv{y}\hatv{y}+\hatv{z}\hatv{z}$ is the unit dyadic tensor, and $\te{\epsilon}$, $\te{\xi}$, $\te{\zeta}$ and $\te{\mu}$ are respectively the permittivity, magnetic-to-electric, electric-to-magnetic and permability tensors. \emph{Spacetime invariance} means that $\te{\chi}_{ab}$ in~\eqref{eq:LSTiSTnd} depends neither on space, i.e., $\te{\chi}_{ab}\neq\te{\chi}_{ab}(\ve{r})$, nor on time, i.e., $\te{\chi}_{ab}\neq\te{\chi}_{ab}(t)$, while \emph{spacetime nondispersion} means that $\te{\chi}_{ab}$ depends neither on the wavenumber, i.e., $\te{\chi}_{ab}\neq\te{\chi}_{ab}(\ve{k})$, nor on the (usual) frequency, $\te{\chi}_{ab}\neq\te{\chi}_{ab}(\omega)$, so that a spacetime invariant and nondispersive bianisotropic medium is one where $\te{\chi}_{ab}\neq\te{\chi}_{ab}(\ve{r},t,\ve{k},\omega)$. Despite this restriction, bianisotropy brings about great medium diversity, given its up to $4\times(3\times 3)=36$ distinct parameters, often interrelated from fundamental physical properties~\cite{Kong_1972}.

\subsection{Metamaterial Unification and Extension}\label{sec:unif_ext}

\begin{figure}[h]
\centering
\includegraphics[width=\linewidth]{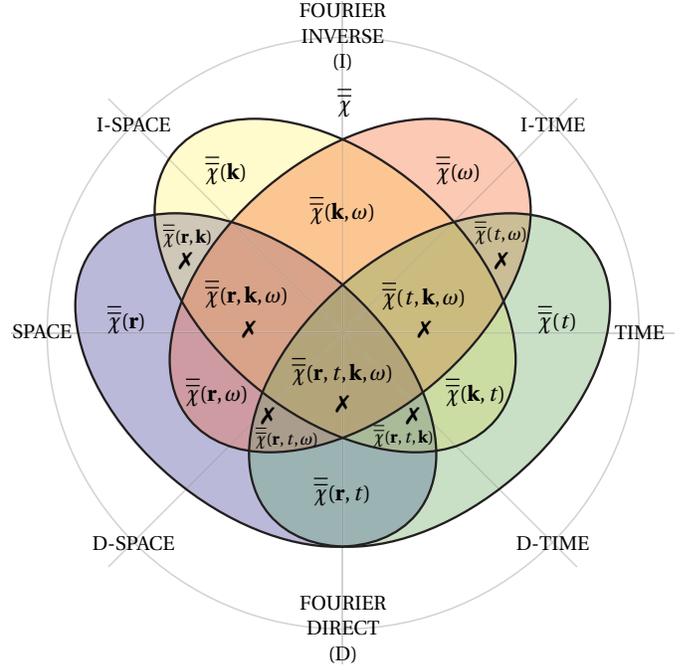}{
\psfrag{1}[c][c][0.9]{$\te{\chi}(\ve{r})$}
\psfrag{2}[c][c][0.9]{$\te{\chi}(t)$}
\psfrag{3}[c][c][0.9]{$\te{\chi}(\ve{k})$}
\psfrag{4}[c][c][0.9]{$\te{\chi}(\omega)$}
\psfrag{5}[c][c][0.9]{$\te{\chi}(\ve{r},t)$}
\psfrag{6}[c][c][0.9]{$\te{\chi}(\ve{k},t)$}
\psfrag{7}[c][c][0.9]{$\te{\chi}(\ve{r},\omega)$}
\psfrag{8}[c][c][0.9]{$\te{\chi}(\ve{k},\omega)$}
\psfrag{9}[c][c][0.75]{$\te{\chi}(\ve{r},\ve{k})$}
\psfrag{a}[c][c][0.75]{$\te{\chi}(t,\omega)$}
\psfrag{b}[c][c][0.9]{$\te{\chi}(\ve{r},\ve{k},\omega)$}
\psfrag{c}[c][c][0.9]{$\te{\chi}(t,\ve{k},\omega)$}
\psfrag{d}[c][c][0.7]{$\te{\chi}(\ve{r},t,\omega)$}
\psfrag{e}[c][c][0.7]{$\te{\chi}(\ve{r},t,\ve{k})$}
\psfrag{f}[c][c][0.9]{$\te{\chi}(\ve{r},t,\ve{k},\omega)$}
\psfrag{g}[c][c][0.9]{$\te{\chi}$}
\psfrag{x}[c][c][0.9]{\ding{55}}
\psfrag{R}[c][c][0.82]{D-SPACE}
\psfrag{T}[c][c][0.82]{D-TIME}
\psfrag{K}[c][c][0.82]{I-SPACE}
\psfrag{W}[c][c][0.82]{I-TIME}
\psfrag{D}[c][c][0.82]{\begin{minipage}{2cm}\centering FOURIER \\ DIRECT \\ (D) \end{minipage}}
\psfrag{I}[c][c][0.82]{\begin{minipage}{2cm}\centering FOURIER \\ INVERSE \\ (I) \end{minipage}}
\psfrag{A}[c][c][0.82]{SPACE}
\psfrag{B}[c][c][0.82]{TIME}
}
\vspace{-2mm}
\caption{Unified and extended representation and classification of bianisotropic spacetime metamaterials in terms of their spacetime variance and dispersion, with $\te{\chi}$ representing the global dyadic tensor defined in~\eqref{eq:LSTiSTnd}, and $\ve{r}$, $t$, $\ve{k}$ and $\omega$ being referred to as direct space or space, direct time or time, inverse space or spatial frequency and inverse time or temporal frequency, respectively. Among the $2^4=16$ spacetime variance-dispersion cases in this diagram, $8$ cases are absolutely meaningful, while the $7$ cases marked by a cross are meaningful only in situations involving very different spacetime parameter scales.}
\label{fig:classification}
\end{figure}

The spacetime invariant and nondispersive restriction can be lifted for even more diversity, as suggested by the \emph{spacetime variance and dispersion} Venn-diagram classification presented in Fig.~\ref{fig:classification}, where we holistically refer to $\ve{r}$, $t$, $\ve{k}$ and $\omega$ as \emph{direct space} or simply \emph{space}, \emph{direct time} or simply \emph{time}, \emph{inverse space} or \emph{spatial frequency} and \emph{inverse time} or \emph{temporal frequency}, respectively, with the terms `direct' and `inverse' referring to the usual Fourier transform-pair independent variables. A general and global metamaterial control of the spacetime variant-dispersive properties would unify and extend current medium science and technology, and bring about a cornucopia of novel physical effects and industrial devices. The recent spectacular developments in micro/nano/quanto/bio/chemico-technologies suggest that such a grand perspective will become a pervasive reality in the forthcoming decades.

The classification of Fig.~\ref{fig:classification} must be considered with some precaution: the relations~\eqref{eq:LSTiSTnd}, upon which it is based, do not strictly apply to all spacetime variance-dispersion cases. This may be best understood with the help of an example. Consider the case $\te{\chi}(t,\omega)$, simplified to the purely electric scalar response $\chi_\tx{ee}(t,\omega)$. This may represent for instance the Drude time-dispersive~\cite{Jackson_1998,Rothwell_2008} and time-variant~\cite{Kalluri_ETVCM_2010,Deck_arXiv_2018} dielectric medium \mbox{$\tilde{\chi}_\tx{ee}(t,\omega)=-\omega_\tx{p,ee}^2(t)/(\omega^2-j\nu_\tx{p,ee}\omega)$}, whose plasma frequency, $\omega_\tx{p,ee}$, varies in time, and where $\nu_\tx{p,ee}$ denotes the damping factor.

If the time variation of $\omega_\tx{p,ee}(t)$ is a drift that occurs on a much greater time scale, $\Delta\tau$, than the (largest Fourier) time period, $T=2\pi/\omega$, of the signal exciting the medium, i.e., $\Delta\tau\gg T$, the time dependence may be safely considered as decoupled from the frequency dependence, and the expression $\tilde{\chi}_\tx{ee}(t,\omega)$ makes sense. It may then be considered as the ``transfer function'' of the  linear quasi time-invariant system~\cite{Lathi_2017} characterized by the spectral relation \mbox{$\tilde{\ve{D}}(\omega)=\epsilon_0\tilde{\chi}_\tx{ee}(t,\omega)\cdot\tilde{\ve{E}}(\omega)$} with the ``input function'' $\tilde{\ve{E}}(\omega)$ and ``output function'' $\tilde{\ve{D}}(\omega)$ or, in the temporal domain, as the ``impulse response''\footnote{It is important to distinguish the regular time variable $t$ (here drift time variance) from the impulse time $t'$, which represents a physically meaningful time quantity only in the theoretical case where $\ve{E}(t')=\ve{E}_0\delta(t')$, for which $\ve{D}(t')=\epsilon_0\ve{E}_0\chi_\tx{ee}(t')$ (impulse response).} $\chi_\tx{ee}(t,t')$ associated with the relation \mbox{$\ve{D}(t')=\epsilon_0\chi_\tx{ee}(t,t')\ast\ve{E}(t')$}, which reads\footnote{This function is in fact acausal, and hence unphysical, since it has an infinite time support. The Drude model is indeed only an approximation of the Lorentz model, which is, in contrast, perfectly causal~\cite{Kong_EWT_2008}, but whose (time-limited) impulse response is substantially more complex.} \mbox{$\chi_\tx{ee}(t,t')=\sqrt{\pi/2}\omega_\tx{p,ee}^2(t)t'\,\tx{sgn}(t')$} in the case of negligible loss ($\nu_\tx{p,ee}\omega\ll\omega^2$).

In contrast, if the variation of $\omega_\tx{p,ee}(t)$ occurs on a time scale that is comparable to -- or smaller than -- that of the exciting wave, or $\Delta\tau\lesssim T$, the expression $\tilde{\chi}_\tx{ee}(t,\omega)$ loses its meaning. Indeed, in this case the variation scale of ``dispersion'' [$\tilde{\chi}_\tx{ee}(\omega)$ or $\chi_\tx{ee}(t,t')$] is comparable to -- or smaller than -- the variation scale of the exciting wave [$\ve{E}(t)$], and the actual dispersion is then \emph{undetermined} because the wave, whose meaningfulness requires at least one cycle ($T$), ``does not have time to probe the memory of the medium'' via $\ve{D}(t')=\epsilon_0\chi_\tx{ee}(t,t')\ast\ve{E}(t')=\epsilon_0\int_{t_0}^{t'=t'_0+\Delta\tau}\chi_\tx{ee}(t,\tau-t')\ve{E}(\tau)d\tau$, where $t_0$ is the start time of the impulse response, ``to experience a specific dispersion from it.''

A similar argument would naturally apply to a fast-varying medium with moderate to strong dispersion, $\chi(t,\omega)$, to the spatial counterpart of $\te{\chi}(t,\omega)$, $\te{\chi}(\ve{r},\ve{k})$, and, consequently, to susceptibilities with more complex dependencies. Overall, among the $2^4=16$ spacetime variance-dispersion cases in Fig.~\ref{fig:classification}, the $8$ cases in the periphery of the diagram are absolutely meaningful, whereas the $7$ cases towards the center indicated by crosses are meaningful only in the case of very different spacetime parameter scales.

These limitations are simply expressions of the \emph{uncertainty principle}~\cite{Heisenberg_1927} applied to Fourier theory, which states that one can determine the spectrogram ($t,\omega$ response) of a system only within the precision constraint~$\Delta t\cdot\Delta\omega>\pi$~\cite{Cohen_1989}. In such a case, the susceptibility concept must be  to the upstream differential equation resulting from the application of the Newton equation of motion to the relevant particle of the medium~\cite{Jackson_1998}, i.e.,
\begin{equation}\label{eq:Pol_diff_eq}
\frac{\partial^2\ve{P}_{ab}}{\partial t^2}+\gamma_{ab}\frac{\partial\ve{P}_{ab}}{\partial t}+\omega_{0,ab}^2\ve{P}_{ab}
=\omega_{\tx{p},ab}^2\ves{\Psi}_b,
\end{equation}
where $\ves{\Psi}_\tx{e}=\ve{E}$ and $\ves{\Psi}_\tx{m}=\ve{H}$, which may be solved consistently with Maxwell equations and Eqs.~\eqref{eq:gen_med_resp}~\cite{Chamanara_PRA_2018}.

\section{Fundamental Phenomena in Spacetime Systems}\label{sec:fund_phen}
\subsection{Spacetime Frequency Transitions}\label{sec:freq_trans_phen}
Spacetime systems generally alter both the temporal and spatial spectra of the electromagnetic waves with which they interact.

The simplest form of temporal-spectrum ($\omega$) transformation induced by spacetime variation is the already mentioned (Sec.~\ref{sec:nonmeta_ST}) \emph{Doppler effect}, where the wave emitted or reflected by a moving object undergoes an upward or downward temporal-frequency shift ($\Delta\omega$) in the direction of or in the direction opposite to the motion~\cite{Eden_Doppler_1992}. This effect, whose study has been extended since Doppler to arbitrary source-observer angles~\cite{Gill_1965} and to the relativistic regime~\cite{Einstein_1905}, takes a more general form and represents more diverse
embodiments in penetrable spacetime entities, and even more in spactime metamaterials, as will be shown in the sequel of the paper.

The first observed form of spatial-spectrum ($\ve{k}$) transformation induced by spacetime variation might be the \emph{aberration of light}, which causes stars to appear displaced ($\Delta\ve{k}$) towards the direction of motion of the Earth. This phenomenon was explained by Bradely in 1727~\cite{Bradley_1729}, and Einstein extended its description to the relativistic regime in his 1905 foundational paper on the special theory of relativity~\cite{Einstein_1905,Norton_2004}. As temporal-spectrum transformations, spatial-spectrum transformations become much richer in penetrable spacetime entities and spactime metamaterials, as will also be shown in the sequel of the paper.

Recently, the term \emph{frequency transition} was introduced in an analogy between photon modes in a photonic crystal with a temporally-spatially varying permittivity and optical transitions between electronic states in metals and semiconductors~\cite{Winn_1999}, where the transitions in the photonic/energy bandgap structure could be either vertical, as direct energy transitions (e.g. in GaAs and InAs) or oblique, as indirect energy transitions (e.g. in Si, Ge and AlSb). We adopt here this terminology, and we will distinguish space (horizontal), time (vertical) and spacetime (oblique) frequency transitions (see Sec.~\ref{sec:freq_trans}).
\subsection{Nonreciprocity}\label{sec:NR_phys}
Another fundamental phenomenon that is common to most spacetime varying media is \emph{nonreciprocity}~\cite{Caloz_PRAp_10_2018}. Indeed, spacetime motion or modulation generally breaks time reversal symmetry, due to the directional bias of the perturbation\footnote{In the particular case where the propagation direction of the perturbation is purely orthogonal to the propagation direction of the wave, the system remains reciprocal, since the same time-varying medium is then seen from both ends of the system by the wave. However, this corresponds to a \emph{pure-time} system, and not to a spacetime system, as will become clear in Sec.~\ref{sec:mod_med_superl}.}. Such spacetime nonreciprocity recently generated intense interest towards the realization of magnetless nonreciprocal devices~\cite{Caloz_2018}.

\subsection{Moving and Modulated System Peculiarities}\label{sec:mov_mod_specs}
As pointed out in Sec.~\ref{sec:nonmeta_ST}, the spacetime variation in a spacetime system may occur either via the motion of matter, such as for instance a moving dielectric object, or via the modulation by a wave, such as for instance an acoustic wave in a piezoelectric crystal. The latter (wave modulation) is certainly easier than the former (wave modulation) to realize and implement in devices, as it is easier to generate a propagating modulation than an actual motion of matter, particularly if very high velocities are required! Beyond this practical difference, there are also fundamental differences between these two types of spacetime systems. In particular, there are two phenomena that can occur only in the moving-medium case: the Fizeau drag and bianisotropy transformation.

The Fizeau drag is the effect according to which the velocity of a wave propagating in a moving medium -- specifically, light in flowing water in the 1851 experiment of Fizeau -- is increased or decreased from its velocity in the stationary medium~\cite{Fizeau_1851}. Specifically, Fizeau found that the velocity of a light wave in a medium of refractive index $n$ moving at the velocity $v_\tx{m}$ is $v_\pm=v_0\pm\Delta v$, where the signs $+$ and $-$ correspond to wave-medium co-directional propagation and contra-directional propagation, respectively, \mbox{$v_0=c/n$} is the velocity of the wave in the stationary medium ($v_\tx{m}=0$), and $\Delta v=v_\tx{m}(1-1/n^2)$ is the velocity increment or decrement due to motion\footnote{The \emph{Sagnac effect} is a rotative version of the Fizeau experiment~\cite{Sagnac_1_1913,Sagnac_2_1913}, which is commonly used in modern gyroscopes.}. This effect, which, together with Bradley aberration, instrumentally led Einstein to establish his theory of special relativity~\cite{Einstein_1905}, does \emph{not exist} in the case of spacetime wave modulation, since such modulation does not involve any transfer of matter and hence does not provide ``dragging molecules'' for altering the velocity of the wave to process.

The first form of bianisotropy transformation was discovered in 1888 by R{\"o}ntgen, who found that a dielectric material moving through an electric field would become magnetized~\cite{Rontgen_1888}, which corresponds to the electric-to-magnetic bianisotropic parameter $\te{\zeta}$, or $\te{\chi}_\tx{me}$, in~\eqref{eq:bianis_params}. This effect may be easily understood by inspecting the Lorentz force exerted on the particles or metaparticles that form the medium\footnote{We assume here, for simplicity, a relatively low-density medium, where the constitutive particles or metaparticles do not interact with each other.}, $\ve{F}=q\ve{E}+q\ve{v}_\tx{m}\times(\mu_0\ve{H})$, where $q$ and $\ve{v}_\tx{m}$ are the charge and velocity of a particle, and $(\ve{E},\ve{H})$ is the (free-space) electromagnetic field [as in Eqs.~\eqref{eq:gen_med_resp} to~\eqref{eq:Pol_diff_eq}] that excites the medium.

If \mbox{$\ve{v}_\tx{m}=0$} (no matter motion), $\ve{F}$ reduces to $q\ve{E}$, which typically induces an electric dipole moment parallel to $\ve{E}$, corresponding to a scalar permittivity $\epsilon=\epsilon_0(1+\chi_\tx{ee})$. If \mbox{$\ve{v}_\tx{m}\neq 0$} and $\ve{v}_\tx{m}\|\ve{v}_\tx{wave}$, we have the additional force contribution $q\ve{v}_\tx{m}\times(\mu_0\ve{H})$ inducing an electric dipole moment again parallel to $\ve{E}$, i.e., a scalar (biisotropic) magnetodielectric coupling $\xi=\sqrt{\epsilon_0\mu_0}\,\chi_\tx{em}$ in~\eqref{eq:bianis_params}\footnote{In the case of a simple nonmagnetic dielectric medium, the second force term is significant only in the relativistic regime ($|\ve{v}_\tx{m}|/c=\beta\lesssim 1$), and negligible ($|\xi|\ll|\epsilon|$) at lower velocities ($\beta\ll 1$). Indeed, according to the Maxwell-Faraday equation, $(n\omega/c)\hatv{k}\times\ve{E}=\omega\mu_0\ve{H}$ ($n$: refractive index), we have then \mbox{$(\mu_0\ve{H})\sim n\ve{E}/c$}, so that $|\ve{v}_\tx{m}\times(\mu_0\ve{H})|<|\ve{v}_\tx{m}|\cdot|\mu_0\ve{H}|\sim n(|\ve{v}_\tx{m}|/c)\cdot|\ve{E}|=n\beta|\ve{E}|$, with practically $n<10$. The situation is naturally different if the moving medium or modulation is a metamaterial, since in this case the shape of the particle can induce strong magnetoelectric coupling, as for instance helix particles in the case of biisotropic media~\cite{Caloz_2019}. In such a case, motion is not even required for magnetoelectric coupling, since the second Lorentz force term is provided by the current, $I$, induced in the metaparticles (e.g. a helices in biisotropic media~\cite{Caloz_2019}), as better seen upon rewriting it as $\int[(Id\ves{\ell})\times(\mu_0\ve{H})]$ with $Id\ves{\ell}$ being an elementary current.}, and if $\ve{v}_\tx{m}\perp\ve{v}_\tx{wave}$, we have an electric dipole moment induced perpendicularly to $\ve{E}$, leading to the tensorial (bianisotropic) coupling $\te{\xi}=\sqrt{\epsilon_0\mu_0}\,\te{\chi}_\tx{em}$ in~\eqref{eq:bianis_params}. And so on. Complete bianisotropy transformation formulas are today available~\cite{Kong_EWT_2008}. This effect, which generally changes the bianisotropy of a medium (e.g. from isotropic to bianisotropic), also does \emph{not exist} in the case of spacetime wave modulation, since the Lorentz force is associated with moving particles. More precisely, this is true in the stationary frame with respect to which the modulation is moving, whereas the opposite is true in the moving frame since that frame would see the static-frame environment moving.

\subsection{Superluminality}\label{sec:mod_med_superl}
It has been known from over 130 years of experimentation, and it is the second postulate of the theory of special relativity~\cite{Einstein_1905}, that nothing can move faster than light in free space, $c=299,792,458$~m/s. However, this does \emph{not} mean that superluminal effects are not possible. \emph{Perturbation superluminality}, associated with transfer of energy without transfer of information, is indeed perfectly possible; it may be simply achieved by injecting the perturbation into the system at a non-zero angle with respect to the direction of wave propagation.

\begin{figure}[h]
\centering
\includegraphics[width=0.8\linewidth]{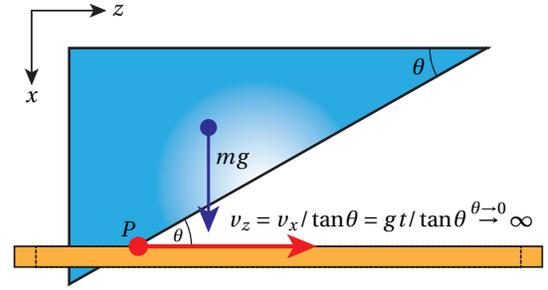}{
\psfrag{t}[c][c][0.9]{$\theta$}
\psfrag{a}[c][c][0.7]{$\theta$}
\psfrag{m}[c][c][0.9]{$mg$}
\psfrag{v}[l][l][0.9]{$v_z=v_x/\tan\theta=gt/\tan\theta\overset{\theta\rightarrow 0}{\rightarrow}\infty$}
\psfrag{x}[c][c][0.9]{$x$}
\psfrag{z}[c][c][0.9]{$z$}
\psfrag{P}[c][c][0.9]{$P$}
}
\vspace{-2mm}
\caption{Demonstration of superluminality with the example of the (simplified) guillotine, an apparatus with a blade falling under gravity through a slit operated on a base board and ipso facto cutting off in two parts the ``object'' placed on the board. The point $P$, corresponding to the intersection of the blade and the board, moves along the $z$ direction at a velocity that is unlimited. This velocity, $v_z$, is superluminal if the angle of the blade, $\theta$, is such that $\theta<\tan^{-1}(gt/c)$, where $g=9.81$~m/$\tx{s}^2$ is the gravity acceleration constant of the Earth and $t$ is the time when the blade is dropped, and it becomes infinite as $\theta\rightarrow 0$: $v_z(\theta=0)=\infty$ (pure time perturbation or variation).}
\label{fig:guillotine}
\end{figure}

To understand this, consider the macabre but insightful example of the guillotine, an apparatus used to behead royalists during the French revolution\footnote{The guillotine was in fact not \emph{invented} by Joseph-Ignace Guillotin, who was against the death penalty, but \emph{proposed} by him for capital punishment, in order to ``simply to end life rather than to inflict pain''~\cite{Dubois_1998}. This device was indeed arguably preferable to the breaking wheel, which was common before!}, which is depicted in~\figref{fig:guillotine}. A short inspection of this device reveals that the point $P$ at the intersection between the blade and the board moves along the $z$ direction at an unlimited velocity, $v_z$, which can perfectly be superluminal, i.e. $v_z>c$, and even reach infinity when the blade is perfectly flat ($\theta=0$). This effect does not violate the second postulate of relativity: the molecules of the blade exclusively move along the $x$ direction, at a velocity $v_x$ that is definitely smaller than $c$. The reason why $P$ can be superluminal is the fact that it does not move in the direction of matter motion, but perpendicularly to it, and that the blade is ``present'' everywhere over the board along the $z$ direction from the outset (no information transfer) so that it does not need to travel along $z$. This motion corresponds thus to a ``perturbation'' motion. However, such perturbation definitely carries energy, as the ghost of guillotined Queen Marie Antoinette would certainly testify!

In physics and engineering, spacetime superluminality can be realized essentially in the same manner as the mechanical system of \figref{fig:guillotine}, by applying the -- matter or modulation -- perturbation \emph{perpendicularly} to the direction of wave propagation and with an \emph{injection angle or gradient} corresponding to the desired velocity. For instance, one may modulate the refractive index of a semiconductor slab by an oblique laser pulse\footnote{The laser obliquely illuminates the board under the angle $\theta$ with respect to the $x$ axis and the width of its oblique beam front corresponds to the oblique edge of the blade.}, where the front of the pulse would play the role of the edge of the blade in the figure with the slab being at the position of the board. However, this can also be practically achieved with many other modulation techniques, involving for instance oblique ferroelectric or ferromagnetic slabs in waveguides~\cite{Holberg_1966}, arrays of varactor diodes in microwave transmission-line structures, and piezoelectric or electrooptic materials in optical structures. Note that if the modulation angle ($\theta$ in \figref{fig:guillotine}) is zero, the resulting ``infinite superluminality'' may be considered as a purely temporal or \emph{pure-time} perturbation~\cite{Kalluri_ETVCM_2010}, which represents the dual of a purely spatial perturbation, as will be seen later in the paper.

\section{Relativity Perspective}\label{sec:SR_perps}
\subsection{Spacetime Extension}\label{sec:ST_diag}

Shortly after the publication of Einstein's theory of special relativity, Minkowski formulated the concept of \emph{spacetime} as a fusion of the three dimensions of space and the one dimension of time into a single four-dimensional continuum, and developed the eponymic diagram, also simply referred to as the \emph{spacetime diagram}~\cite{Minkowski_1909}. This concept and diagram quickly became essential aspects of relativity~\cite{Einstein_1905,Einstein_1916}, and we shall see in this paper that they also represent powerful tools for handling spacetime metamaterials.

\begin{figure}[h]
\centering
\includegraphics[width=\columnwidth]{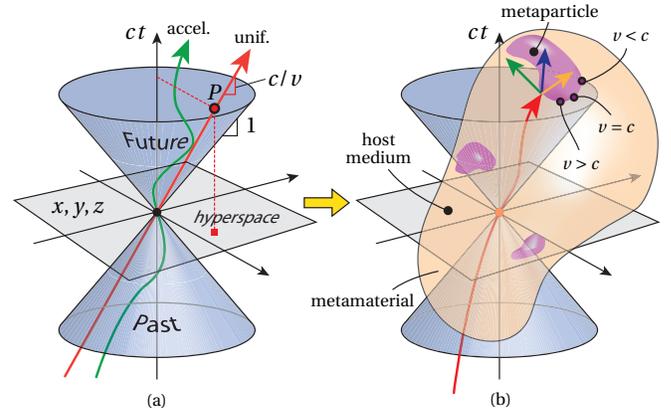}{
\psfrag{a}[c][c][0.7]{(a)}
\psfrag{b}[c][c][0.7]{(b)}
\psfrag{z}[c][c][0.9]{$ct$}
\psfrag{x}[c][c][0.9]{$x,y,z$}
\psfrag{c}[c][c][0.7]{$v>c$}
\psfrag{d}[c][c][0.7]{$v=c$}
\psfrag{e}[c][c][0.7]{$v<c$}
\psfrag{s}[c][c][0.7]{\shortstack{spacetime \\ metamaterial}}
\psfrag{h}[c][c][0.7]{\shortstack{host \\ medium}}
\psfrag{p}[c][c][0.7]{metaparticle}
\psfrag{w}[c][c][0.7]{unif.}
\psfrag{u}[c][c][0.7]{accel.}
\psfrag{P}[c][c][0.9]{$P$}
\psfrag{s}[c][c][0.9]{$c/v$}
\psfrag{1}[c][c][0.9]{$1$}
\psfrag{M}[c][c][0.7]{metamaterial}
}
\centering
\vspace{-6mm}
\caption{Spacetime diagrams. (a)~Conventional case, with a uniform trajectory (red), an accelerated trajectory (green) and the light cone (blue). (b)~Metamaterial extension, including a host medium (brown) and a few metaparticles (magenta) with subluminal ($v<c$), luminal ($v=c$) and superluminal ($v>c$) discontinuity features, here with a subluminal nonuniform (inhomogeneous host medium) wave strongly scattered by one of the spacetime metaparticles.}
\label{fig:ST_diag}
\end{figure}

Figure~\ref{fig:ST_diag}(a) shows the conventional spacetime diagram, which represents the three dimensions of space, $\ve{r}=(x,y,z)$, as a horizontal plane, called the \emph{hyperspace}, and the one dimension of (space-normalized) time, $ct$, as the direction perpendicular to that plane. In this diagram, each physical \emph{event} is localized by a unique spacetime point, $P=(\ve{r},ct)$, while the evolution of an event is represented by a continuous curve, called a \emph{trajectory} or \emph{world line}, which extends from its past to its future through its present (origin). The slope of the trajectory is $s=|\nabla_\ve{r}(ct)|=c|\nabla_\ve{r}t|=c/v$, and is hence inversely proportional to the velocity ($v$) of the event. Thus, a vertical curve ($s=\infty$) represents a stationary or static event, a straight line with $s>1$ (or $v<c$) represents a (subluminally) uniformly moving object, a curved line (with subluminal $s>1$ everywhere) represents a nonuniformly moving (accelerated/deccelerated) object, and the Dirac-like cone, with $s=1$ (or $v=c$), represents light propagation in free space\footnote{This cone explicitly represents the propagation of a circular light pulse in a 2D-spatial structure (e.g. a thin slab), with the origin of the cone corresponding to the pulse source and each horizontal circular section of the cone corresponding to the ``position'' of the pulse at a given time (see also Sec.~\ref{sec:shifted_focus}). In the case of 3D-spatial light propagation, the light pulse ``position'' is a spherical shell, and the spacetime diagram, drawn on a 2D surface, is only a \emph{symbolic} (hyperspace) representation of its propagation.}.

Relativity essentially deals with the motion of celestial bodies (e.g. stars, spacecrafts) and the propagation of electromagnetic and gravitational waves in \emph{free space}. A spacetime metamaterial may be considered as the extension of the spacetime concept, where a \emph{spacetime medium} is added to the conventional spacetime structure, as illustrated in \figref{fig:ST_diag}(b). The spacetime metamaterial is to be understood as a 3D penetrable propagating perturbation with arbitrarily complex spacetime structure\footnote{A  cosmological version of such a ``spacetime metamaterial'' would be the universe, whose spacetime varying curvature caused by gravitational waves would act as the spacetime medium across which light propagates.}. A possibly helpful illustration at this point could be that of a plasma cloud (host medium) with regions of higher density and complexity (metaparticles) passing across the observer. Such a perturbation can support continuua of subluminal and superluminal discontinuity features, and this superluminal aspect, demystified in Sec.~\ref{sec:mod_med_superl}, brings about a great diversity of unexplored physical phenomena and bears potential for many novel electromagnetic device opportunities.

\begin{figure}[h]
\centering
\includegraphics[width=\linewidth]{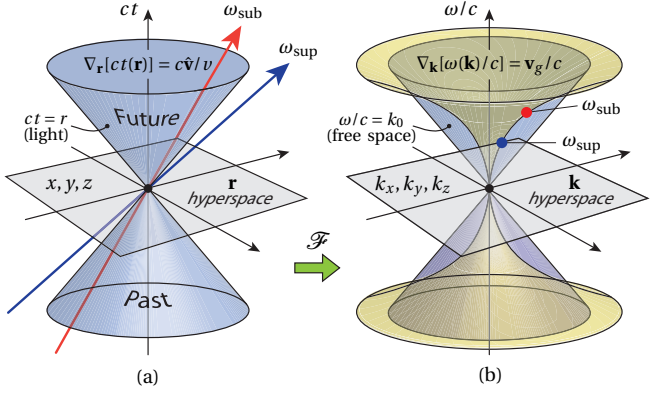}{
\psfrag{a}[c][c][0.8]{(a)}
\psfrag{b}[c][c][0.8]{(b)}
\psfrag{r}[c][c][0.8]{$\ve{r}$}
\psfrag{k}[c][c][0.8]{$\ve{k}$}
\psfrag{z}[c][c][0.8]{$ct$}
\psfrag{w}[c][c][0.8]{$\omega/c$}
\psfrag{x}[c][c][0.8]{$x,y,z$}
\psfrag{y}[c][c][0.8]{$k_x,k_y,k_z$}
\psfrag{F}[c][c][1]{${\mathcal F}$}
\psfrag{1}[c][c][0.8]{$\omega_\tx{sub}$}
\psfrag{2}[c][c][0.8]{$\omega_\tx{sup}$}
\psfrag{3}[c][c][0.8]{$\omega_\tx{sub}$}
\psfrag{4}[c][c][0.8]{$\omega_\tx{sup}$}
\psfrag{5}[c][c][0.7]{\begin{minipage}{2cm}\centering $ct=r$ \\ \vspace{-1mm} (light)\end{minipage}}
\psfrag{6}[c][c][0.7]{\begin{minipage}{2cm}\centering $\omega/c=k_0$ \\ \vspace{-1mm} (free space)\end{minipage}}
\psfrag{7}[c][c][0.75]{$\nabla_\ve{r}[ct(\ve{r})]=c\hatv{v}/v$}
\psfrag{8}[c][c][0.75]{$\nabla_\ve{k}[\omega(\ve{k})/c]=\ve{v}_g/c$}
}
\vspace{-4mm}
\caption{Fourier spacetime pair. (a)~Direct spacetime, $(\ve{r},ct)$ (host medium in brown in \figref{fig:ST_diag}(b) implicit, not shown here), with the trajectories of two harmonic plane waves propagating in the dispersive medium represented in (b) at $\omega_\tx{sub}$, where $v_\tx{sub}<c$ (subluminal), and at $\omega_\tx{sup}$, where $v_\tx{sup}>c$ (superluminal). (b)~Inverse spacetime, $(\ve{k},\omega/c)$, with the dispersion curve of the dispersive (isotropic) medium (yellow) and spectral points corresponding to the two waves in~(a).}
\label{fig:dir_inv_ST}
\end{figure}

The space $\{\ve{r},t\}$, represented in \figref{fig:ST_diag}, involves the direct-Fourier spacetime independent variables $\ve{r}$ and $t$. We therefore refer to it as the \emph{direct spacetime}, consistently with the terminology introduced in \figref{fig:classification}. This diagram is repeated in \figref{fig:dir_inv_ST}(a) with a pair of waves, a subluminal one, within the light cone, and a superluminal one, outside of the light cone. As we see here again, the direct spacetime allows to describe wave trajectories and scattering directions, and this will be used to determine the spacetime scattering coefficients in Sec.~\ref{sec:scat_coef}. However, it does not include any spacetime \emph{spectral} information, which will be needed to determine the spacetime frequency transitions in Sec.~\ref{sec:freq_trans}. That information is contained in the space of the inverse-Fourier space and time independent variables $\ve{k}$ and $\omega$, which is represented in \figref{fig:dir_inv_ST}(b) and which shall logically refer to as the \emph{inverse spacetime}. As illustrated in \figref{fig:dir_inv_ST}(b), the inverse spacetime hosts the dispersion curves of the medium, and provides a perspective that is obviously complementary to that of the direct spacetime.

\subsection{Lorentz Transformations}\label{sec:Lorentz_transf}
The Lorentz transformations are mathematical transformations that relate quantities measured in different spacetime \emph{inertial} coordinate frames~\cite{Lorentz_1904,Poincare_1906}, i.e., coordinate frames that move rectilinearly at constant velocity (or zero acceleration) and hence experience no force, according to Newton second law~\cite{Newton_1687}.

\begin{figure}[h]
\centering
\includegraphics[width=0.8\linewidth]{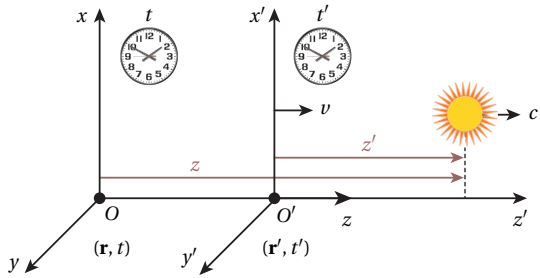}{
\psfrag{c}[c][c][0.8]{$c$}
\psfrag{x}[cb][cb][0.8]{$x$}
\psfrag{y}[cb][cb][0.8]{$y$}
\psfrag{z}[cb][cb][0.8]{$z$}
\psfrag{O}[c][c][0.8]{$O$}
\psfrag{r}[cb][cb][0.8]{$x'$}
\psfrag{s}[cb][cb][0.8]{$y'$}
\psfrag{t}[cb][cb][0.8]{$z'$}
\psfrag{Q}[c][c][0.8]{$O'$}
\psfrag{v}[c][c][0.8]{$v$}
\psfrag{1}[cb][cb][0.8]{$t$}
\psfrag{2}[cb][cb][0.8]{$t'$}
\psfrag{R}[cb][cb][0.8]{$(\ve{r},t)$}
\psfrag{S}[cb][cb][0.8]{$(\ve{r}',t')$}
\psfrag{a}[cb][cb][0.8]{\textcolor[rgb]{0.60,0.40,0.40}{$z'$}}
\psfrag{b}[cb][cb][0.8]{\textcolor[rgb]{0.60,0.40,0.40}{$z$}}
}
\caption{Pair of rectilinear inertial spacetime frames, with spacetime coordinates $(\ve{r},ct)=(x,y,z,ct)$ and $(\ve{r}',ct')=(x',y',z',ct')$, where the latter moves at the (constant) velocity $\ve{v}$ ($v=|\ve{v}|<c$) relatively to the former. Here, the two frames have been oriented in such a manner that $\hatv{z}=\hatv{z}'=\hatv{v}=v\hatv{z}$. The position of a light pulse propagating along the $\hatv{v}$ direction (with speed $c$) is measured as $z$ and $z'$ in the unprimed and primed frames, respectively.}
\label{fig:ST_frame_pair}
\end{figure}

Figure~\ref{fig:ST_frame_pair} shows a pair of such frames in direct spacetime, with spacetime coordinates $(\ve{r},ct)=(x,y,z,ct)$ and $(\ve{r}',ct')=(x',y',z',ct')$ and origins $O$ and $O'$, where the latter moves at the (constant) velocity $\ve{v}$ ($v=|\ve{v}|<c$) relatively to the former. We shall next consider, for simplicity and without loss of generality, the specific configuration shown in this figure, where the two frames have been oriented in such a manner that $\hatv{z}=\hatv{z}'=\hatv{v}=v\hatv{z}$, and where the wave of interest (light pulse in the figure) propagates in the same direction.

The Lorentz transformations of the direct spacetime variables, $(z,t)$ and $(z,t')$, follow from the \emph{second postulate} of the theory of special relativity, which states that the speed of light is the same ($c=299,792$~km/s) for all observers regardless of the source~\cite{Einstein_1905}, and read (see Appendix~\ref{app:deriv_LT_basic})
\begin{subequations}\label{eq:LT_ST_var}
\begin{equation}\label{eq:LT_ST_var_up_to_p}
\begin{Bmatrix}
z'=\gamma z-\gamma\beta(ct) \\
ct'=-\gamma\beta z+\gamma(ct)
\end{Bmatrix}
\;\;\tx{or}\;\;
\begin{pmatrix} z' \\ ct' \end{pmatrix}
=\gamma\begin{pmatrix} 1 & -v/c \\ -v/c & 1 \end{pmatrix}
\begin{pmatrix} z \\ ct \end{pmatrix},
\end{equation}
\begin{equation}\label{eq:LT_ST_var_p_to_up}
\begin{Bmatrix}
z=\gamma z'+\gamma\beta(ct') \\
ct=\gamma\beta z'+\gamma(ct')
\end{Bmatrix}
\quad\tx{or}\quad
\begin{pmatrix} z \\ ct \end{pmatrix}
=\gamma\begin{pmatrix} 1 & v/c \\ v/c & 1 \end{pmatrix}
\begin{pmatrix} z' \\ ct' \end{pmatrix},
\end{equation}
\end{subequations}
where
\begin{equation}\label{eq:gamma}
\gamma=\frac{1}{\sqrt{1-\beta^2}},\quad
\tx{with}\quad
\beta=v/c
\end{equation}
is the Lorentz factor. Note that, consistently with the symmetry of the problem, the transformation relations~\eqref{eq:LT_ST_var} are symmetric to each other, with opposite transformations differing only in the sign associated with $v$. The generalization of these relations to arbitrarily-directed motions are tensorial relations that are available for instance in~\cite{Rothwell_2008}.

The spacetime Lorentz transformations of the inverse spacetime variables, $(k_z,\omega)$ and $(k_z',\omega')$, may be found by enforcing the \emph{invariance of the phase}, $\phi$, for a time-harmonic wave, which expresses the fact that the unprimed and primed observers would agree on whether an event corresponds to a wave crest or a wave through~\cite{Pauli_1958}, i.e.,
\begin{equation}\label{eq:phase_conserv}
\phi=\phi',\quad\tx{or}\quad
k_zz-\omega t=k_zz'-\omega t'.
\end{equation}
Substituting~\eqref{eq:LT_ST_var_p_to_up} in the latter equation, grouping the coefficients of $z'$ and $t'$, and considering the fact that the resulting relation must hold for any values of the two variables leads to the sought after inverse formulas,
\begin{subequations}\label{eq:iLT_ST_var}
\begin{equation}\label{eq:iLT_ST_var_up_to_p}
\begin{Bmatrix}
k_z'=\gamma k_z-\gamma\beta\left(\dfrac{\omega}{c}\right) \\
\dfrac{\omega'}{c}=-\gamma\beta k_z+\gamma\left(\dfrac{\omega}{c}\right)
\end{Bmatrix}
\;\tx{or}\;
\begin{pmatrix} k_z' \\ \omega'/c \end{pmatrix}
=\gamma\begin{pmatrix} 1 & -v/c \\ -v/c & 1 \end{pmatrix}
\begin{pmatrix} k_z \\ \omega/c \end{pmatrix},
\end{equation}
\begin{equation}\label{eq:iLT_ST_var_p_to_up}
\begin{Bmatrix}
k_z=\gamma k_z'+\gamma\beta\left(\dfrac{\omega'}{c}\right) \\
\dfrac{\omega}{c}=\gamma\beta k_z'+\gamma\left(\dfrac{\omega'}{c}\right)
\end{Bmatrix}
\;\tx{or}\;
\begin{pmatrix} k_z \\ \omega/c \end{pmatrix}
=\gamma\begin{pmatrix} 1 & v/c \\ v/c & 1 \end{pmatrix}
\begin{pmatrix} k_z' \\ \omega'/c \end{pmatrix},
\end{equation}
\end{subequations}
which involve exactly the same transformation matrices as their direct counterparts [Eq.~\eqref{eq:LT_ST_var}].

\begin{figure}[h]
\centering
\includegraphics[width=\linewidth]{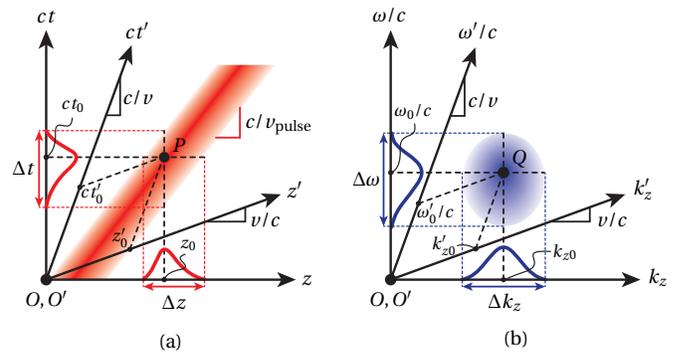}{
\psfrag{a}[c][c][0.8]{(a)}
\psfrag{b}[c][c][0.8]{(b)}
\psfrag{e}[c][c][0.8]{$c/v_\tx{pulse}$}
\psfrag{z}[c][c][0.8]{$z$}
\psfrag{t}[c][c][0.8]{$ct$}
\psfrag{u}[c][c][0.8]{$z'$}
\psfrag{v}[c][c][0.8]{$ct'$}
\psfrag{k}[c][c][0.8]{$k_z$}
\psfrag{l}[c][c][0.8]{$\omega/c$}
\psfrag{m}[c][c][0.8]{$k_z'$}
\psfrag{n}[c][c][0.8]{$\omega'/c$}
\psfrag{O}[c][c][0.8]{$O,O'$}
\psfrag{c}[c][c][0.8]{$c/v$}
\psfrag{d}[c][c][0.8]{$v/c$}
\psfrag{P}[c][c][0.8]{$P$}
\psfrag{Q}[c][c][0.8]{$Q$}
\psfrag{Z}[c][c][0.8]{$\Delta z$}
\psfrag{T}[c][c][0.8]{$\Delta t$}
\psfrag{K}[c][c][0.8]{$\Delta k_z$}
\psfrag{W}[c][c][0.8]{$\Delta\omega$}
\psfrag{1}[c][c][0.7]{$z_0$}
\psfrag{2}[c][c][0.7]{$ct_0$}
\psfrag{3}[c][c][0.7]{$z_0'$}
\psfrag{4}[c][c][0.7]{$ct_0'$}
\psfrag{5}[c][c][0.7]{$k_{z0}$}
\psfrag{6}[c][c][0.7]{$\omega_0/c$}
\psfrag{7}[c][c][0.7]{$k_{z0}'$}
\psfrag{8}[c][c][0.7]{$\omega_0'/c$}
}
\vspace{-8mm}
\caption{Lorentz unprimed and primed spacetime frames with a wave pulse. (a)~Direct spacetime [Eq.~\eqref{eq:LT_ST_var}] (Figs.~\ref{fig:dir_inv_ST}(a) and~\ref{fig:ST_frame_pair}), with $P$ representing an event of coordinates $[(z_0,ct_0),(z_0',ct_0')]$, and the graded strip representing a pulse wave of space length $\Delta z$ and time duration $\Delta t$. (b)~Inverse spacetime [Eq.~\eqref{eq:iLT_ST_var}] (\figref{fig:dir_inv_ST}(b)), with $Q$ representing a harmonic plane wave of spacetime frequencies $[(k_{z0},\omega/c),(k_{z0}',\omega'/c)]$, and the graded spot representing the spacetime spectrum with spatial bandwidth $\Delta k_z$ and temporal bandwidth $\Delta\omega$.}
\label{fig:Lorentz_dir_inv_frames}
\end{figure}

The direct and inverse spacetime unprimed and primed frames are plotted in \figref{fig:Lorentz_dir_inv_frames}, along with representative wave trajectories in~\figref{fig:Lorentz_dir_inv_frames}(a) and corresponding spectra in \figref{fig:Lorentz_dir_inv_frames}(b). The direct spacetime primed frame axes $z'$ and $ct'$ are obtained by setting $ct'=0$ and $z'=0$ in~\eqref{eq:LT_ST_var_up_to_p}, yielding $ct=(v/c)z$ and $ct=(c/v)z$, respectively, while the inverse spacetime primed frame axes $k_z'$ and $\omega'/c$ are obtained  by setting $\omega'/c=0$ and $k_z'=0$ in~\eqref{eq:iLT_ST_var_up_to_p}, yielding $\omega/c=(v/c)k_z$ and $\omega/c=(c/v)k_z$, respectively.

The graded strip in \figref{fig:Lorentz_dir_inv_frames}(a) represents the trajectory of a pulse wave of spacetime extent $(\Delta z,\Delta t)$ while the graded spot in \figref{fig:Lorentz_dir_inv_frames}(b) represents its spectrum with spacetime extent $(\Delta\ve{k},\Delta\omega)$. The trajectory of a point $(\Delta z,\Delta t)\rightarrow 0$ would correspond to a constant-phase point (e.g. a crest) of a harmonic plane wave, whose direct spacetime representation [\figref{fig:Lorentz_dir_inv_frames}(a)] would strictly involve a periodic set of parallel trajectory lines with horizontal and vertical distances corresponding to the wavelength ($\lambda=2\pi/k$) and period ($T=2\pi/\omega$), respectively. On the other hand, the inverse-spacetime point $(\Delta\ve{k},\Delta\omega)\rightarrow 0$ [\figref{fig:Lorentz_dir_inv_frames}(b)] would exactly represent a harmonic plane wave.

The Lorentz transformations of the electromagnetic fields follow from the \emph{first postulate} of the theory of special relativity, which states that the laws of physics are the same in all inertial systems~\cite{Einstein_1905}, and read (see Appendix~\ref{app:deriv_LT_fields})
\begin{subequations}\label{eq:LT_fields_up_to_p}
\begin{align}
E_x'(z')&=\gamma\left(E_x(z)-vB_y(z)\right), \\
cB_y'(z')&=\gamma\left(cB_y(z)-\frac{v}{c}E_x(z)\right), \\
H_y'(z')&=\gamma\left(H_y(z)-vD_x(z)\right), \\
cD_x'(z')&=\gamma\left(cD_x(z)-\frac{v}{c}H_y(z)\right),
\end{align}
\end{subequations}
and, inverting the primed and unprimed quantities and substituting $v\rightarrow-v$,
\begin{subequations}\label{eq:LT_fields_u_to_up}
\begin{align}
E_x(z)&=\gamma\left(E_x'(z')+vB_y'(z')\right), \\
cB_y(z)&=\gamma\left(cB_y'(z')+\frac{v}{c}E_x'(z')\right), \\
H_y(z)&=\gamma\left(H_y'(z')+vD_x'(z')\right), \\
cD_x(z)&=\gamma\left(cD_x'(z')+\frac{v}{c}H_y'(z')\right).
\end{align}
\end{subequations}
Again, the generalizations of these relations to arbitrarily-directed motions are tensorial relations that are available for instance in~\cite{Rothwell_2008}. Moreover, the spectral counterparts of these relations are straightforwardly obtainable by spacetime Fourier transformation. 
 
\subsection{Resolution Strategy}\label{sec:resol_strategy}
The conventional strategy for determining the scattering response of a medium moving with uniform velocity $\ve{v}_\tx{m}$ with respect to a given laboratory frame -- the unprimed frame in Figs.~\ref{fig:ST_frame_pair} and~\ref{fig:Lorentz_dir_inv_frames} -- consists in the following two steps~\cite{Pauli_1958,Jackson_1998,Kong_EWT_2008}. First, we solve the problem in the frame co-moving with the medium -- the primed frame in Figs.~\ref{fig:ST_frame_pair} and~\ref{fig:Lorentz_dir_inv_frames} --, where the medium is stationary or static, and hence the problem is much simpler. Second, we convert the result to the original laboratory frame, using the Lorentz transformations~\eqref{eq:LT_fields_u_to_up} for an electromagnetic problem. 

This strategy encounters two issues in the study of spacetime metamaterials: 1)~the inapplicability, in the case of superluminal systems, of the Lorentz spacetime variable transformations~\eqref{eq:LT_ST_var} and~\eqref{eq:iLT_ST_var} with the conventional definition of the Lorentz factor in~\eqref{eq:gamma}; 2)~the inappropriateness, in the case of modulated spacetime systems, to express the constitutive relations in the primed expressions of the electromagnetic fields associated with the Lorentz transformations~\eqref{eq:LT_fields_up_to_p} and~\eqref{eq:LT_fields_u_to_up}.

The first issue arises, as mentioned, when the spacetime perturbation is superluminal, i.e., \mbox{$v_\tx{m}>c$} (e.g. \mbox{$\theta<\tan(gt/c)$} in Fig.~\ref{fig:guillotine}). In this case, the Lorentz factor in~\eqref{eq:gamma}, with \mbox{$v=v_\tx{f}=v_\tx{m}$}, is imaginary, since \mbox{$\beta^2=(v_\tx{m}/c)^2>1$}, which leads to absurd imaginary direct and indirect spacetime variables according to~\eqref{eq:LT_ST_var} and~\eqref{eq:iLT_ST_var}. Can we still use Lorentz transforms while avoiding this problem? The fact that the limit case of a superluminal medium, $v_\tx{m}\rightarrow\infty$, is a pure-time medium as the limit case of a subluminal medium, $v_\tx{m}=0$, is a purely spatial, or \emph{pure space}, medium suggests that the issue may be resolved by setting the primed frame as one where the medium is purely temporal instead of purely spatial. This indeed resolves the issue, as first reported in~\cite{Deck_PRB_2018}.

\begin{figure}[h]
\centering
\includegraphics[width=\linewidth]{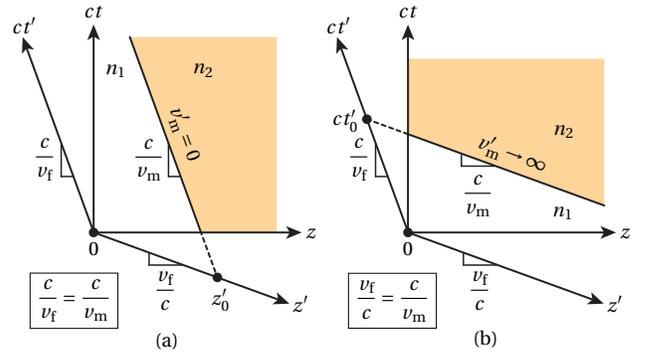}{
\psfrag{a}[c][c][0.8]{(a)}
\psfrag{b}[c][c][0.8]{(b)}
\psfrag{O}[c][c][0.8]{$0$}
\psfrag{z}[c][c][0.8]{$z$}
\psfrag{t}[c][c][0.8]{$ct$}
\psfrag{u}[c][c][0.8]{$z'$}
\psfrag{v}[c][c][0.8]{$ct'$}
\psfrag{s}[c][c][0.8]{$\boxed{\dfrac{c}{v_\tx{f}}=\dfrac{c}{v_\tx{m}}}$}	\psfrag{w}[c][c][0.8]{$\boxed{\dfrac{v_\tx{f}}{c}=\dfrac{c}{v_\tx{m}}}$}
\psfrag{p}[c][c][0.8]{$z_0'$}
\psfrag{q}[c][c][0.8]{$ct_0'$}
\psfrag{1}[c][c][0.8]{$n_1$}
\psfrag{2}[c][c][0.8]{$n_2$}
\psfrag{3}[c][c][0.8]{$\dfrac{c}{v_\tx{m}}$}
\psfrag{4}[c][c][0.8]{$\dfrac{c}{v_\tx{f}}$}
\psfrag{5}[c][c][0.8]{$\dfrac{v_\tx{f}}{c}$}
\psfrag{6}[c][c][0.8]{$v_\tx{m}'=0$}
\psfrag{7}[c][c][0.8]{$v_\tx{m}'\rightarrow\infty$}
\psfrag{8}[c][c][0.8]{$\dfrac{c}{v_\tx{m}}$}
}
\vspace{-6mm}
\caption{Choice of a proper primed Lorentz frame, moving at the velocity $v_\tx{f}$ with respect to the unprimed frame (\figref{fig:Lorentz_dir_inv_frames}) for an interface moving at the velocity $v_\tx{m}$ (here $v_\tx{m}<0$) between two semi-infinite media of respective refractive indices $n_1$ and $n_2$. (a)~Subluminal (conventional) case: $v_\tx{f}=v_\tx{m}$. (b)~Superluminal (see Sec.~\ref{sec:mod_med_superl}) case: \mbox{$v_\tx{f}=c^2/v_\tx{m}$}.}
\label{fig:sub_super_primed_frames}
\end{figure}

To show this, let us consider the problem of a spacetime-modulated interface between two semi-infinite media of respective refractive indices $n_1$ and $n_2$, which is addressed in \figref{fig:sub_super_primed_frames}\footnote{An analogy to help understanding such an interface would be that of ``a chain of standing domino tiles, sufficiently closely spaced to topple their neighbors upon falling, so that a chain reaction can be launched by knocking down the first tile. In such a reaction, one clearly sees the ``interface'' between the fallen and standing parts of the chain propagating along the structure at a specific velocity, with the domino tiles standing up or lying down on either side of the interface not moving in the direction of propagation (no transfer of energy).'' (text borrowed from~\cite{Deck_arXiv_2018}).}. This problem is fundamental because an interface between two media represents the building brick of any metamaterial, which is essentially nothing but a succession of such discontinuities. Figure~\ref{fig:sub_super_primed_frames}(a) shows the case of a \emph{subluminal interface}. In this case, following the conventional approach, we set the time axis of the primed frame parallel to the interface, so that the interface appears static ($v_\tx{m}'=0$), or pure-space, at $z'=z_0'$ in that frame. Equating the two slopes, $c/v_\tx{f}=c/v_\tx{m}$, leads then to $v_\tx{f}=v_\tx{m}$: the frame is co-moving with the interface. In contrast, in the case of a \emph{superluminal interface}, shown in~\figref{fig:sub_super_primed_frames}(b), we set the \emph{space axis} of the primed frame parallel to the interface, so that the interface appears instantaneous ($v_\tx{m}'\rightarrow\infty$), or pure-time, at $ct'=ct_0'$ in that frame. Equating the two slopes, $v_\tx{f}/c=c/v_\tx{m}$, leads this time to $v_\tx{f}=c^2/v_\tx{m}$: the frame is `co-standing' with the interface. Now $|\beta|=|v_\tx{f}|/c=c/v_\tx{m}<0$, so that $\gamma$ is real, and the Lorentz transformations in Sec.~\ref{sec:Lorentz_transf} can be safely used.

Let us now turn to the second issue, which, as mentioned, is related to the \emph{field} Lorentz transformations~\eqref{eq:LT_fields_up_to_p} and~\eqref{eq:LT_fields_u_to_up}, and which concerns \emph{modulated} (as opposed to moving) media. When the primed fields have been obtained, from applying the spacetime boundary conditions (see Sec.~\ref{sec:scat_BCs}) to~\eqref{eq:LT_fields_up_to_p} in the primed frame, as in the conventional approach, it is recommended not to decompose $\ve{D}'$ and $\ve{B}'$ in terms of the $\ve{E}'$ and $\ve{H}'$ via the primed version of the constitutive relations~\eqref{eq:bianis_params}. Indeed, since a modulation involves no transfer of matter in the laboratory frame or, equivalently, the material molecules involved in the modulation are stationary in the laboratory (unprimed) frame, they are necessarily moving, in the direction opposite to the modulation, in the primed frame, and therefore, correspond to bianisotropy-transformed parameters in that frame, according to Sec.~\ref{sec:mov_mod_specs}. For instance, if the initially modulated medium is monoisotropic in the unprimed frame, it is bianisotropic, and hence more complex, in the primed frame. To avoid this complication, it is better to apply the constitutive-relation decomposition only after expressing the primed fields in terms of their unprimed counterpart with~\eqref{eq:LT_fields_up_to_p}.

Once the fields scattered by the interface have been determined, with the precautions just described, we can solve the problem of spacetime metamaterials, as the aforementioned succession of spacetime interfaces, using in the case of a mono-directional spatial-variation a spacetime-extended version of the well-known transfer matrix method~\cite{Abeles_1950}, either isolated for computing the scattering from a finite number of spacetime layers~\cite{Deck_arXiv_2018}, or combined with the Floquet-Bloch theorem for computing the dispersion diagram~\cite{Biancalana_2007,Deck_arXiv_2018}. The problems involving spacetime corner discontinuities or higher spacetime-variation dimensions require more sophisticated analytical or computational techniques.

\subsection{Canonical Spacetime ``Metamaterials''}
\begin{figure}[h]
\centering
\includegraphics[width=\linewidth]{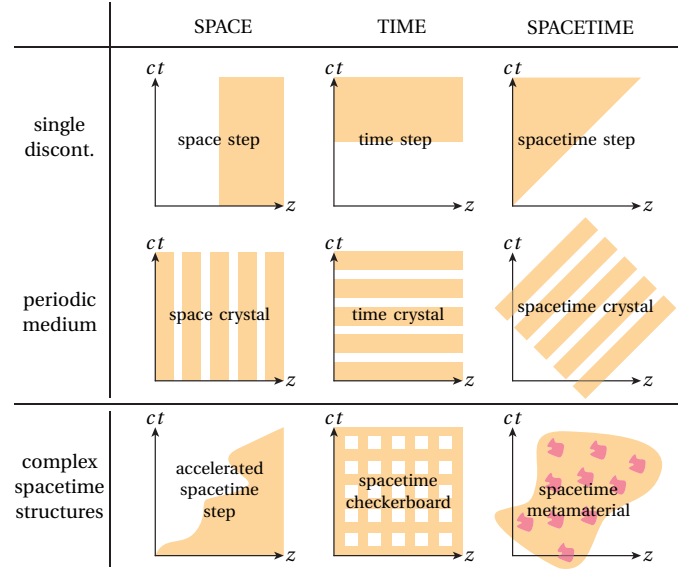}{
\psfrag{z}[c][c][0.8]{$z$}
\psfrag{t}[c][c][0.8]{$ct$}
\psfrag{S}[c][c][0.8]{SPACE}
\psfrag{T}[c][c][0.8]{TIME}
\psfrag{Q}[c][c][0.8]{SPACETIME}
\psfrag{A}[c][c][0.8]{\begin{minipage}{2cm}\centering single \\ discont. \end{minipage}}
\psfrag{B}[c][c][0.8]{\begin{minipage}{2cm}\centering periodic \\ medium \end{minipage}}
\psfrag{C}[c][c][0.8]{\begin{minipage}{2cm}\centering complex \\ spacetime \\ structures \end{minipage}}
\psfrag{1}[c][c][0.7]{space step}
\psfrag{2}[c][c][0.7]{time step}
\psfrag{3}[c][c][0.7]{spacetime step}
\psfrag{4}[c][c][0.7]{space crystal}
\psfrag{5}[c][c][0.7]{time crystal}
\psfrag{6}[c][c][0.7]{spacetime crystal}
\psfrag{7}[c][c][0.7]{\begin{minipage}{2cm}\centering accelerated \\spacetime \\ step \end{minipage}}
\psfrag{8}[c][c][0.7]{\begin{minipage}{2cm}\centering spacetime \\ checkerboard \end{minipage}}
\psfrag{9}[c][c][0.7]{\begin{minipage}{2cm}\centering spacetime \\ metamaterial \end{minipage}}
}
\vspace{-4mm}
\caption{Canonical space, time and spacetime media. The white and brown areas represent different homogenous isotropic media, for instance characterized by respective refractive indices $n_1$ and $n_2$. The axis $z$ may generally represent the hyperspace $\ve{r}=(x,y,z)$ (\figref{fig:ST_diag}), so that the spacetime dimension may be (1+1)D, (2+1)D or (3+1)D.}
\label{fig:gallery}
\end{figure}

Figure~\ref{fig:gallery} presents a number of canonical spacetime media. The first row shows, from left to right, a pure-space discontinuity, a pure-time discontinuity and a spacetime discontinuity (moving in the positive $z$ direction, contrarily to those in \figref{fig:sub_super_primed_frames}), which, as mentioned above, represents the building bricks of spacetime metamaterials, and which will therefore be analyzed in the next sections. Such discontinuities may be considered as the succession of two media, but they may alternatively be considered as a single medium involving a discontinuity; for instance, the spacetime discontinuity may be expressed as a medium with the spacetime-varying refractive index $n(z,ct)=n_1+(n_2-n_1)U(z-v_\tx{m}t)$, where $U(\cdot)$ is the Heaviside step function. Each of these media may be of dimensions (1+1)D, (2+1)D or (3+1)D, where the first and second numbers respectively represent the spatial dimension and the temporal dimension. 

The second row in~\figref{fig:gallery} shows, from left to right, a pure-space crystal, a pure-time crystal and a spacetime crystal, which are obtained by cascading the corresponding discontinuities of the first row, and which will be specifically described in Sec.~\ref{sec:crystallography}. The third row shows more complex spacetime media, namely an accelerated discontinuity\footnote{The Lorentz transforms (Sec.~\ref{sec:Lorentz_transf}) do not apply to such a medium, since they are restricted to inertial systems.}, a spacetime checkerboard~\cite{Lurie_2006,Milton_2017}, and a general spacetime metamaterial, corresponding to \figref{fig:ST_diag}.

While the media shown in \figref{fig:gallery} involve \emph{step discontinuities}, they may as well involve \emph{graded discontinuities}, characterized by a progressive variation of the medium parameter, as for instance in the core of graded-index optical fibers or in the Luneburg, Eaton and Maxwell fish-eye lenses.

\section{Pure-Space and Pure-Time Phenomenology and Electromagnetic Boundary Conditions}\label{sec:scat_BCs}
\begin{figure}[h]
	\centering
	\includegraphics[width=0.9\linewidth]{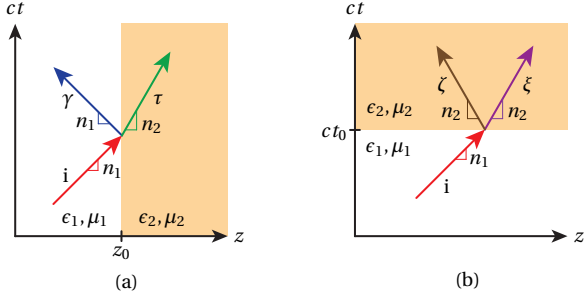}{
		\psfrag{a}[c][c][0.8]{(a)}
		\psfrag{b}[c][c][0.8]{(b)}
		\psfrag{z}[c][c][0.8]{$z$}
		\psfrag{t}[c][c][0.8]{$ct$}
		\psfrag{p}[c][c][0.8]{$z_0$}
		\psfrag{q}[c][c][0.8]{$ct_0$}
		\psfrag{1}[c][c][0.8]{$\epsilon_1,\mu_1$}
		\psfrag{2}[c][c][0.8]{$\epsilon_2,\mu_2$}
		\psfrag{3}[c][c][0.8]{i}
		\psfrag{4}[c][c][0.8]{$\gamma$}
		\psfrag{5}[c][c][0.8]{$\tau$}
		\psfrag{6}[c][c][0.8]{$\xi$}
		\psfrag{7}[c][c][0.8]{$\zeta$}
		\psfrag{8}[c][c][0.8]{$n_1$}
		\psfrag{9}[c][c][0.8]{$n_2$}
	}
	\vspace{-3mm}
	\caption{Scattering phenomenology in the most fundamental discontinuities between two electromagnetic media of permittivities and permeabilities $(\epsilon_1,\mu_1)$ and $(\epsilon_2,\mu_2)$. (a)~Pure-space step discontinuity [limit case $v_\tx{m}=0$ of the subluminal discontinuity in \figref{fig:sub_super_primed_frames})(a)], with incident (i), reflected ($\gamma$) and transmitted ($\tau$) waves. (b)~Pure-time step discontinuity [limit case $v_\tx{m}\rightarrow\infty$ of the superluminal discontinuity in \figref{fig:sub_super_primed_frames})(a)], with incident (i) or earlier, later-backward ($\zeta$) and later-forward ($\xi$) waves.}
	\label{fig:scat_phen}
\end{figure}

The problem of scattering by a spacetime discontinuity, as that shown in the top right panel of \figref{fig:gallery}, can be solved from the solutions of the pure-space discontinuity and pure-time discontinuity problems in the top left and top center panels of \figref{fig:gallery} combined with the Lorentz transforms provided in Sec.~\ref{sec:Lorentz_transf}. We shall therefore start with these two problems, whose scattering phenomenology is illustrated in \figref{fig:scat_phen}. Note that the pure-space discontinuity is the limit case \mbox{$v_\tx{m}=0$} of the subluminal discontinuity in \figref{fig:sub_super_primed_frames})(a), while the pure-space discontinuity is the limit case $v_\tx{m}\rightarrow\infty$ of the superluminal discontinuity in \figref{fig:sub_super_primed_frames})(b).

The scattering of a wave by a pure-space discontinuity, whose spacetime-diagram is shown in \figref{fig:scat_phen}(a), is a high-school textbook problem: the incident (pulse or harmonic) wave (i) splits into a wave reflected with reflection coefficient $\gamma$ and a wave transmitted (or refracted) with transmission coefficient $\tau$ at the discontinuity position, $z_0$. The scattering of a wave by a pure-time discontinuity is the dual problem, whose spacetime-diagram is shown in \figref{fig:scat_phen}(b)~\cite{Kalluri_ETVCM_2010}. Here, the incident wave is called the \emph{earlier} wave. At the discontinuity time, $ct_0$, this wave splits into a \emph{later-backward} wave with coefficient $\zeta$ and a \emph{later-forward} wave with coefficient $\xi$. It is important to realize that, in duality with the pure-space problem where a harmonic wave scatters at the position $z_0$ \emph{at all times} ($ct$ axis), such a harmonic wave in the pure-time problem scatters at the instant $ct_0$ \emph{at all positions} ($z$ axis).

The boundary conditions follow from the fundamental physical constraint that all physical quantities must remain bounded everywhere and at every time. Specifically, they are found from inspecting the space and time derivatives in Maxwell equations,
\begin{equation}\label{eq:Maxwell_eqs}
\nabla\times\ve{E}=-\frac{\partial\ve{B}}{\partial t}
\quad\tx{and}\quad
\nabla\times\ve{H}=\frac{\partial\ve{D}}{\partial t},
\end{equation}
which are here assumed to concern a sourceless region \mbox{($\ve{J}=0$)} for simplicity.

In the case of the pure-space discontinuity, represented in \figref{fig:scat_phen}(a), one must naturally focus on the spatial derivative along the same direction, i.e., $\partial/\partial z$, which corresponds to the tangential ($x,y$) part of~\eqref{eq:Maxwell_eqs},
\begin{equation}\label{eq:Maxwell_eqs_tan}
\hatv{z}\times\frac{\partial\ve{E}_\tx{t}}{\partial z}=-\frac{\partial\ve{B}_\tx{t}}{\partial t}
\quad\tx{and}\quad
\hatv{z}\times\frac{\partial\ve{H}_\tx{t}}{\partial z}=\frac{\partial\ve{D}_\tx{t}}{\partial t},
\end{equation}
where the $t$ subscript denotes tangential field components. If $\ve{E}_\tx{t}$ or $\ve{H}_\tx{t}$ were discontinuous at $z_0$, then the fields $\ve{B}_\tx{t}$ or $\ve{D}_\tx{t}$ would be singular, i.e., unbounded, at that position, which is physically not allowed. Therefore, as well-known from electromagnetics textbooks, $\ve{E}_\tx{t}$ and $\ve{H}_\tx{t}$, must be continuous at a space discontinuity. 

An analogous argument applies to pure-time discontinuity in \figref{fig:scat_phen}(b). However, the critical derivative for ensuing bounded fields is now the temporal derivative in~\eqref{eq:Maxwell_eqs}. If $\ve{B}$ or $\ve{D}$ were discontinuous at $ct_0$, then the fields $\ve{E}$ or $\ve{D}$ would be singular, i.e., unbounded, at that instant, which is physically not allowed. Therefore, $\ve{B}$ or $\ve{D}$ must be continuous at a time discontinuity.

\begin{subequations}\label{eq:BCs}
As a result, the \emph{boundary conditions} at a pure-space discontinuity, perpendicular to the direction~$z$, are
\begin{equation}\label{eq:BC_PS}
\left.\hatv{z}\times(\ve{E}_2-\ve{E}_1)\right|_{z'=z_0'}=0
\quad\tx{and}\quad
\left.\hatv{z}\times(\ve{H}_2-\ve{E}_1)\right|_{z'=z_0'}=0,
\end{equation}
while they are at a pure-time discontinuity 
\begin{equation}\label{eq:BC_PT}
\left.(\ve{B}_2-\ve{B}_1)\right|_{ct'=ct_0'}=0
\quad\tx{and}\quad
\left.(\ve{D}_2-\ve{D}_1)\right|_{ct'=ct_0'}=0,
\end{equation}
\end{subequations}
where the subscripts $1$ and $2$ correspond to the media $1$ and $2$, respectively (\figref{fig:scat_phen}). 

These boundary conditions will next allow us to determine the scattering coefficients and transition frequencies for spacetime interfaces. For simplicity, we will restrict here our attention to the case of (1+1)D problems, whose \emph{wave equation} may be written as
\begin{subequations}\label{eq:wave_eq}
\begin{equation}
\left\{\frac{\partial^2}{\partial z^2}-\frac{\partial^2}{\partial^2[v(z,ct)t]}\right\}\psi(z,ct)=0,
\end{equation}
with
\begin{equation}
v(z,ct)=\frac{c}{n(z,ct)},
\end{equation}
\end{subequations}
where $\psi$ represents the continuous field component, and $n(z,ct)$ is the spacetime-varying refractive index of the `medium' formed by the discontinuity.

\section{Scattering Coefficients}\label{sec:scat_coef}
\subsection{Pure-Space Medium}
The determination of the electromagnetic reflection and transmission coefficients -- or Fresnel coefficients -- for the pure-space discontinuity depicted in Fig.~\ref{fig:scat_phen}(a) is a basic textbook problem, but we shall still address it here, both to establish a foundation for the forthcoming spacetime interface problem and to provide deep insight into the duality of this problem with the pure-time interface problem. The spatial-discontinuity continuous fields are here, according to~\eqref{eq:BC_PS}, $\ve{E}_\tx{t}$ and $\ve{H}_\tx{t}$. We therefore set in~\eqref{eq:wave_eq} $\psi=E,H$ (e.g. with $\ve{E}\|\hatv{x}\rightarrow E=E_x$ and $\ve{H}\|\hatv{y}\rightarrow H=H_y$). The wave equation may then be rewritten in each of the two media as
\begin{equation}\label{eq:we_PS}
\left[\frac{1}{n_{1,2}^2}\frac{\partial^2}{\partial z^2}-\frac{\partial^2}{\partial(ct)^2}\right]
\begin{Bmatrix}E_{1,2}(z,ct) \\ H_{1,2}(z,ct)\end{Bmatrix}=0.
\end{equation}
The solutions of these equations may be found using the method of separation of variables and proper wave propagation directions\footnote{We use here the time-harmonic complex dependence $e^{+j\omega t}$, corresponding to the spatial expressions $e^{-jk z}$ and $e^{+jk z}$ for the forward and backward waves, respectively.} as
\begin{subequations}\label{eq:fields_PS}
\begin{align}
E_1&=e^{-jn_1\frac{\omega_\tx{i}}{c}z}e^{j\omega_\tx{i}t}+\gamma e^{jn_1\frac{\omega_\gamma}{c}z}e^{j\omega_\gamma t}, \\
E_2&=\tau e^{-jn_2\frac{\omega_\tau}{c}z}e^{j\omega_\tau t}, \\
H_1&=\left(e^{-jn_1\frac{\omega_\tx{i}}{c}z}e^{j\omega_\tx{i} t}-\gamma e^{jn_1\frac{\omega_\gamma}{c}z}e^{j\omega_\gamma t}\right)/\eta_1, \\
H_2&=\tau e^{-jn_2\frac{\omega_\tau}{c}z}e^{j\omega_\tau t}/\eta_2,
\end{align}
\end{subequations}
where the subscripts i, $\gamma$ and $\tau$ correspond to the incident, reflected and transmitted waves, respectively [\figref{fig:scat_phen}(a)], and $\eta_{1,2}=\sqrt{\mu_{1,2}/\epsilon_{1,2}}$ is the intrinsic impedance of the medium. The unknown reflection and transmission coefficients, $\gamma$ and $\tau$, are then found by inserting these relations into the boundary conditions [Eqs.~\eqref{eq:BC_PS}]
\begin{equation}\label{eq:BCs_PS_E}
\left.E_1\right|_{z=z_0}=\left.E_2\right|_{z=z_0}
\quad\tx{and}\quad
\left.H_1\right|_{z=z_0}=\left.H_2\right|_{z=z_0},
\end{equation}
which yields the well-known pure-space scattering coefficients
\begin{equation}\label{eq:scat_coeff_PS}
\gamma=\frac{\eta_2-\eta_1}{\eta_1+\eta_2}
\quad\tx{and}\quad
\tau=\frac{2\eta_2}{\eta_1+\eta_2}.
\end{equation}

\begin{figure}[h]
\centering
\includegraphics[width=0.9\linewidth]{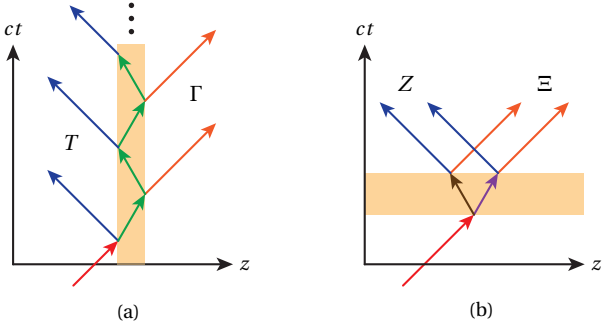}{
\psfrag{a}[c][c][0.8]{(a)}
\psfrag{b}[c][c][0.8]{(b)}
\psfrag{z}[c][c][0.8]{$z$}
\psfrag{t}[c][c][0.8]{$ct$}
\psfrag{A}[c][c][0.9]{$T$}
\psfrag{B}[c][c][0.9]{$\Gamma$}
\psfrag{C}[c][c][0.9]{$Z$}
\psfrag{D}[c][c][0.9]{$\Xi$}
}
\vspace{-3mm}
\caption{Direct spacetime representation scattering in (step) slabs. (a)~Pure-space slab, involving an infinity of multiple-scattering trajectories, contributing to the total reflection coefficient $\Gamma$ and transmission coefficient~$T$. (b)~Pure-time slab, involving only four final scattering trajectories, contributing to the total later-backward scattering coefficient $Z$ and later-forward scattering coefficient $\Xi$.}
\label{fig:slab_scat}
\end{figure}

The problem of scattering by a \emph{pure-space slab} is then conventionally solved by applying boundary conditions of the type~\eqref{eq:BCs_PS_E} at each of the two interfaces of the slab with the proper field expressions involving the scattering coefficients~\eqref{eq:scat_coeff_PS} (e.g.~\cite{Rothwell_2008}). The corresponding direct spacetime representation is shown in \figref{fig:slab_scat}(a). It involves an infinity of scattering events, and therefore an infinite number of trajectory contributions to the global reflection and transmission coefficients, $\Gamma$ and $T$.

\subsection{Pure-Time Medium}\label{sec:PT_refl_coeff}
The determination of the electromagnetic later-forward and later-backward coefficients for the pure-time discontinuity depicted in Fig.~\ref{fig:scat_phen}(b) is less known, although it is now also a textbook topic~\cite{Kalluri_ETVCM_2010}. The temporal-discontinuity continuous fields are, according to~\eqref{eq:BC_PT}, $\ve{D}$ and $\ve{B}$. We therefore set in~\eqref{eq:wave_eq} $\psi=D,B$ (e.g. with $\ve{D}\|\hatv{x}\rightarrow D=D_x$ and $\ve{B}\|\hatv{y}\rightarrow B=B_y$). The wave equation may then be rewritten in each of the two media as
\begin{equation}\label{eq:we_PT}
\left[\frac{\partial^2}{\partial z^2}-n_{1,2}^2\frac{\partial^2}{\partial(ct)^2}\right]
\begin{Bmatrix}D_{1,2}(z,ct) \\ B_{1,2}(z,ct)\end{Bmatrix}=0.
\end{equation}
The solutions of these equations may be found using again the method of separation of variables and proper wave propagation directions\footnote{We use here the space-harmonic, or plane wave, complex dependence $e^{-jk_z z}$, corresponding to the temporal expressions $e^{j\omega t}$ and $e^{-j\omega t}$ for the forward and backward waves, respectively.}, as
\begin{subequations}\label{eq:fields_PT}
\begin{align}
D_1&=e^{-jk_{z,\tx{i}}z}e^{j\frac{k_{z,\tx{i}}}{n_1}ct}, \\
D_2&=\zeta_D e^{-jk_{z,\zeta}z}e^{-j\frac{k_{z,\zeta}}{n_1}ct}+\xi_D e^{-jk_{z,\xi}z}e^{j\frac{k_{z,\xi}}{n_1}ct}, \\
B_1&=e^{-jk_{z,\tx{i}}z}e^{j\frac{k_{z,\tx{i}}}{n_1}ct}/\eta_1, \\
B_2&=\left(-\zeta_D e^{-jk_{z,\zeta}z}e^{-j\frac{k_{z,\zeta}}{n_1}ct}+\xi_D e^{-jk_{z,\xi}z}e^{j\frac{k_{z,\xi}}{n_1}ct}\right)/\eta_2,
\end{align}
\end{subequations}
where the subscripts i, $\zeta$ and $\xi$ correspond to the earlier (incident), later-backward and later-forward waves, respectively [\figref{fig:scat_phen}(b)]. The unknown later-backward and later-forward coefficients, $\zeta_D$ and $\xi_D$, are then found by inserting these relations into the boundary conditions [Eqs.~\eqref{eq:BC_PT}]
\begin{equation}\label{eq:BCs_PT_D}
\left.D_1\right|_{ct=ct_0}=\left.D_2\right|_{ct=ct_0}
\;\tx{and}\;
\left.B_1\right|_{ct=ct_0}=\left.B_2\right|_{ct=ct_0},
\end{equation}
which yields the less known pure-time $D$-related scattering coefficients
\begin{equation}\label{eq:scat_coeff_PT}
\zeta_D=\frac{\eta_2-\eta_1}{2\eta_1}
\quad\tx{and}\quad
\xi_D=\frac{\eta_1+\eta_2}{2\eta_1}.
\end{equation}

The asymmetry between these relations and the pure-space scattering coefficients~\eqref{eq:scat_coeff_PS} is due to the asymmetry between the space and time scattering phenomenology, which is apparent in \figref{fig:scat_phen} and which is itself due to \emph{causality}: since scattering back to the past is impossible, the mathematically correct trajectory pointing towards the bottom-right direction in both \figref{fig:scat_phen}(a) and~\figref{fig:scat_phen}(b) is physically unacceptable (which is why it is not drawn). As result, given the $\pi/2$ angle difference between pure-space and pure-time discontinuities, the former case involves one scattered wave in medium~1 (the reflected wave) and one scattered waves in medium~2 (the transmitted wave), whereas the latter case has the two scattered waves (later-forward and later-backward) in medium~2.

The $E$-related form of the pure-time scattering parameters~\eqref{eq:scat_coeff_PT}, more appropriate for combination with the pure-space scattering parameters~\eqref{eq:scat_coeff_PS}, are simply found by substituting $D_{1,2}=\epsilon_{1,2}E_{1,2}$ and $B_{1,2}=\mu_{1,2}B_{1,2}$ in~\eqref{eq:we_PT} to~\eqref{eq:BCs_PT_D}, which yields~\cite{Morgenthaler_1958}
\begin{equation}\label{eq:scat_coeff_PT_E}
\zeta=\frac{n_1}{n_2}\frac{\eta_2-\eta_1}{2\eta_1}
\quad\tx{and}\quad
\xi=\frac{n_1}{n_2}\frac{\eta_1+\eta_2}{2\eta_1},
\end{equation}
where $n_{1,2}=\sqrt{\epsilon_{1,2}\mu_{1,2}}$.

The problem of scattering by a \emph{pure-time slab} is then solved by applying boundary conditions of the type~\eqref{eq:BCs_PT_D} at each of the two interfaces of the slab with the proper field expressions involving the scattering coefficients~\eqref{eq:scat_coeff_PT} or~\eqref{eq:scat_coeff_PT_E}~\cite{Deck_PRB_2018}. The corresponding direct spacetime representation is shown in \figref{fig:slab_scat}(b). This representation is fundamentally different from its pure-space counterpart in \figref{fig:slab_scat}(a):
it involves a finite number of scattering events, and only $4$ trajectory contributions in the global later-backward and later-forward coefficients, $Z$ and $\Xi$.

\subsection{Spacetime Medium}\label{sec:ST_scat_coeff}
\begin{figure}[h]
\centering
\includegraphics[width=0.9\linewidth]{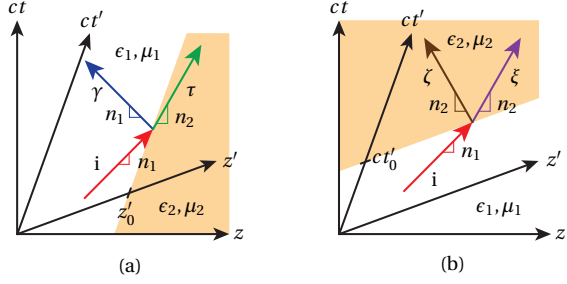}{
\psfrag{a}[c][c][0.8]{(a)}
\psfrag{b}[c][c][0.8]{(b)}
\psfrag{z}[c][c][0.8]{$z$}
\psfrag{t}[c][c][0.8]{$ct$}
\psfrag{u}[c][c][0.8]{$z'$}
\psfrag{v}[c][c][0.8]{$ct'$}
\psfrag{p}[c][c][0.8]{$z_0'$}
\psfrag{q}[c][c][0.8]{$ct_0'$}
\psfrag{1}[c][c][0.8]{$\epsilon_1,\mu_1$}
\psfrag{2}[c][c][0.8]{$\epsilon_2,\mu_2$}
\psfrag{3}[c][c][0.8]{i}
\psfrag{4}[c][c][0.8]{$\gamma$}
\psfrag{5}[c][c][0.8]{$\tau$}
\psfrag{6}[c][c][0.8]{$\xi$}
\psfrag{7}[c][c][0.8]{$\zeta$}
\psfrag{8}[c][c][0.8]{$n_1$}
\psfrag{9}[c][c][0.8]{$n_2$}
}
\vspace{-3mm}
\caption{Scattering in spacetime discontinuities between two electromagnetic media of permittivities and permeabilities $(\epsilon_1,\mu_1)$ and $(\epsilon_2,\mu_2)$. Here $v_\tx{m}>0$ (in contrast to the case of \figref{fig:sub_super_primed_frames}); the wave and the interface are co-directional. (a)~Subluminal (space-like) spacetime step discontinuity, with incident~(iW), reflected ($\gamma$) and transmitted ($\tau$) waves. (b)~Superluminal (time-like) spacetime step discontinuity, with incident (i) or earlier, later-backward ($\zeta$) and later-forward ($\xi$) waves.}
\label{fig:ST_scat_coeff}
\end{figure}

We have now the required tools to solve the most fundamental spacetime problem associated with spacetime metamaterials: the scattering of waves by a spacetime discontinuity, represented in \figref{fig:ST_scat_coeff}. The fact that, as pointed out in Sec.~\ref{sec:scat_BCs}, the pure-space interface is the $v_\tx{m}=0$ limit of a subluminal interface while the pure-time interface is the $v_\tx{m}\rightarrow\infty$ limit of a superluminal interface suggests that there are fundamental similarities between a subluminal spacetime interface and a pure-space interface and between a superluminal spacetime interface and a pure-time interface. Indeed, comparing Figs.~\ref{fig:sub_super_primed_frames} and~\ref{fig:scat_phen} reveals that the former both support reflected and transmitted (or refracted) waves, as shown in \figref{fig:ST_scat_coeff}(a), while the latter both support later-backward and later-forward waves, as shown in \figref{fig:ST_scat_coeff}(b)\footnote{In fact, there is an intermediate regime between the subluminal and superluminal regimes, specifically in the velocity interval $c/n_2<v_\tx{m}<c/n_1$, assuming $n_2>n_1$, called the \emph{interluminal} regime. Here, we do not consider this regime, whose phenomenology is described in~\cite{Ostrovskii_1975,Deck_arXiv_2019}.}. We therefore sometimes refer to a subluminal discontinuity as \emph{space-like} and to a superluminal discontinuity as \emph{time-like}, and we will be able to solve each problem from its limit counterpart.

Let us start with the subluminal interface [\figref{fig:ST_scat_coeff}(a)]. Here, the interface is static, or pure-space, at $z'=z_0'$, in the primed frame [see \figref{fig:sub_super_primed_frames}(a)], and we can therefore apply the pure-space boundary conditions~\eqref{eq:BC_PS} in that frame, i.e.,
\begin{equation}\label{eq:BCs_sub_E}
\left.E_1'\right|_{z'=z_0'}=\left.E_2'\right|_{z'=z_0'}
\quad\tx{and}\quad
\left.H_1'\right|_{z'=z_0'}=\left.H_2'\right|_{z'=z_0'}.
\end{equation}
Following the prescriptions in Sec.~\ref{sec:resol_strategy}, we next substitute the primed fields in these relations by their unprimed counterparts, provided by~\eqref{eq:LT_fields_up_to_p}. This leads, upon setting $v=v_\tx{f}=v_\tx{m}$, canceling the Lorentz factor appearing on both sides, and setting $D_{1,2}=\epsilon_{1,2}E_{1,2}$ and $B_{1,2}=\mu_{1,2}H_{1,2}$ for a modulated\footnote{In the case of a \emph{moving} interface, the constitutive relations would typically be expressed in the primed frame, where matter is not moving, and where the medium is hence most often monoisotropic.} interface, to
\begin{align}
\left.E_1-v_\tx{m}(\mu_1H_1)\right|_{z'=z_0'}&=\left.E_2-v_\tx{m}(\mu_2H_2)\right|_{z'=z_0'}, \\
\left.H_1-v_\tx{m}(\epsilon_1E_1)\right|_{z'=z_0'}&=\left.H_2-v_\tx{m}(\epsilon_2E_2)\right|_{z'=z_0'}.
\end{align}
Substituting in these relations the fields $E_{1,2}$ and $H_{1,2}$ by their pure-space expressions~\eqref{eq:fields_PS} finally provides the subluminal interface reflection and transmission coefficients
\begin{equation}\label{eq:scat_coeff_sub}
\gamma=\frac{\eta_2-\eta_1}{\eta_1+\eta_2}\left(\frac{1-n_1v_\tx{m}/c}{1+n_1v_\tx{m}/c}\right)
\;\;\tx{and}\;\;
\tau=\frac{2\eta_2}{\eta_1+\eta_2}\left(\frac{1-n_1v_\tx{m}/c}{1-n_2v_\tx{m}/c}\right),
\end{equation}
with $v_{1,2}=c/n_{1,2}$, which properly reduce to~\eqref{eq:scat_coeff_PS} at $v_\tx{m}=0$. The first of these equations indicates that increasing $v_\tx{m}$ decreases $\gamma$, which makes intuitive sense since as the interface moves faster co-directionally with the wave, the scattering gets weaker, as in the collision of two co-directionaly versus contra-directionally driving cars. The dependence of $\tau$ on $v_\tx{m}$ is less pronounced and more subtle.

The subluminal scattering coefficients~\eqref{eq:scat_coeff_sub} correspond to the \emph{co-directional} problem where the incident wave ($v_{1,\tx{i}}>0$) and the medium ($v_\tx{m}>0$) propagate in the \emph{same} direction. The \emph{contra-directional} problem, corresponding to the situation in \figref{fig:sub_super_primed_frames}(a) with the medium moving in the direction opposite to the wave ($v_\tx{m}<0$), is quite different, due to the breaking of spacetime symmetry. Noting the co-directional coefficients~\eqref{eq:scat_coeff_sub} $\gamma^+$ and $\tau^+$, the contra-directional coefficients $\gamma^-$ and $\tau^-$ are obtained from them by simply exchanging $n_1$ and $n_2$ and reversing the sign of $v_\tx{m}$~\cite{Deck_arXiv_2018}.

Let us now consider the superluminal interface [\figref{fig:ST_scat_coeff}(b)]. Here, the interface is instantaneous, or pure-time, at $ct'=ct_0'$, in the primed frame [see \figref{fig:sub_super_primed_frames}(b)], and we therefore apply the pure-time boundary conditions~\ref{eq:BC_PT} in that frame, i.e.,
\begin{equation}\label{eq:BCs_sub_E}
\left.D_1'\right|_{ct'=ct_0'}=\left.D_2'\right|_{ct'=ct_0'}
\quad\tx{and}\quad
\left.B_1'\right|_{ct'=ct_0'}=\left.B_2'\right|_{ct'=ctz_0'}.
\end{equation}
Following the prescriptions in Sec.~\ref{sec:resol_strategy}, we next substitute the primed fields in these relations by their unprimed counterparts, provided by~\eqref{eq:LT_fields_up_to_p}. This leads, upon setting $v=v_\tx{f}=c^2/v_\tx{m}$, canceling the Lorentz factor appearing on both sides, and setting $E_{1,2}=D_{1,2}/\epsilon_{1,2}$ and $H_{1,2}=B_{1,2}/\mu_{1,2}$ for (again) a modulated interface, to
\begin{align}
\left.D_1-(v_\tx{m}/c^2)B_1/\mu_1\right|_{ct'=ct_0'}&=\left.D_2-(v_\tx{m}/c^2)B_2/H_2\right|_{ct'=ct_0'}, \\
\left.B_1-(v_\tx{m}/c^2)D_1/\epsilon_1\right|_{ct'=ct_0'}&=\left.B_2-(v_\tx{m}/c^2)D_2/\epsilon_2\right|_{ct'=ct_0'}.
\end{align}
Substituting in these relations the fields $D_{1,2}$ and $B_{1,2}$ by their pure-time expressions~\eqref{eq:fields_PT} finally provides the ($E$-related) superluminal interface later-backward and later-forward scattering coefficients
\begin{equation}\label{eq:scat_coeff_super}
\zeta=\frac{\eta_1-\eta_2}{2\eta_1}\left(\frac{1-n_1v_\tx{m}/c}{1+n_2v_\tx{m}/c}\right)
\;\tx{and}\;
\xi=\frac{\eta_1+\eta_2}{2\eta_1}\left(\frac{1-n_1v_\tx{m}/c}{1-n_2v_\tx{m}/c}\right),
\end{equation}
which properly reduce to~\eqref{eq:scat_coeff_PT} at $v_\tx{m}\rightarrow\infty$. The first of these equations indicates that increasing $v_\tx{m}$ decreases $\zeta$, which makes intuitive sense for the same reason as given above for $\gamma$ decreasing with increasing $v_\tx{m}$, since $\zeta$ also corresponds to some kind of reflection, with the wave being sent back towards the source due to the time discontinuity.

The contra-directional superluminal scattering coefficients, $\zeta^-$ and $\xi^-$, are obtained from their co-directional counterparts, $\zeta^+$ and $\xi^+$, by only reversing the sign of $v_\tx{m}$~\cite{Deck_arXiv_2018}, assuming that the early medium is still medium~1.

In this section, we have considered only straight, or linear, spacetime interface trajectories, $(ct)=(c/v_\tx{m})z$ with $v_\tx{m}=\tx{constant}$  [Figs.~\ref{fig:sub_super_primed_frames} and \ref{fig:ST_scat_coeff}], i.e., interfaces with constant (or uniform) velocity. However, spacetime interfaces may be curved, as shown in the left-most bottom panel of \figref{fig:gallery}, which corresponds to a variable velocity, $v_\tx{m}=v_\tx{m}(z,t)$, and hence to an accelerated interface, $\partial v_\tx{m}/\partial t=a(z,t)\neq 0$, where $a(z,t)$ is the acceleration. For instance, a uniformly accelerated interface, i.e., and interface with acceleration $a(t)=a_0$ (constant) or velocity $v_\tx{m}(t)=a_0t$ (assuming $v_\tx{m}(0)=0$), would correspond to the curved trajectory $(ct)=\pm(c/\sqrt{a_0})\sqrt{z}$~\cite{Tanaka_1982,Meter_2001}. In such a case, the scattering coefficients~\eqref{eq:scat_coeff_sub} and \eqref{eq:scat_coeff_super}, become naturally spacetime-dependent, i.e., $(\gamma,\tau,\zeta,\xi)=(\gamma,\tau,\zeta,\xi)(z,t)$, and the Lorentz transformations (Sec.~\ref{sec:Lorentz_transf}), which are restricted to uniform-velocity systems, i.e. special relativity~\cite{Einstein_1905}, are not applicable here anymore. One may then either solve the problem classically using Maxwell equation or, more efficiently and ore elegantly, by using the tools of general relativity~\cite{Einstein_1916} (differential geometry).

\section{Frequency Transitions}\label{sec:freq_trans}
\subsection{Pure-Space Medium}\label{sec:spat_freq_trans}
The determination of the electromagnetic frequency transitions for a pure-space discontinuity are usually not explicitly mentioned in textbooks, where they are generally taken for granted, and not even being called by a specific name. Since the wavelength gets compressed in the denser medium, the scattered wavenumber, $k_z$, obviously increases in that medium, while the scattered frequency, $\omega$, can only be conserved, given the absence of nonlinearity. Let us nevertheless address here this problem from a rigorous point of view, again for completeness and insight.

Such a treatment was already initiated in the pure-space fields given by~\eqref{eq:fields_PS}, where, avoiding any a priori assumption, we have admitted the possibility of having different temporal frequencies, $\omega_\gamma$ and $\omega_\tau$, in media~1 and~2. It is the fact that the boundary conditions~\eqref{eq:BCs_PS_E} must hold \emph{for all times} that implies $\omega_\tau=\omega_\tx{i}$, or $\Delta\omega=\omega_\tau-\omega_\tx{i}=0$. Indeed, a purely spatial discontinuity cannot induce any temporal frequency transformation. At the same time, these boundary conditions lead to spatial frequency transformations, corresponding to the traditional phase matching condition. Specifically, we find from solving~\eqref{eq:BCs_PS_E} 
\begin{subequations}\label{eq:PS_freqs_conds}
\begin{equation}\label{eq:PS_omega_conds}
\omega_\gamma=\omega_\tau=\omega_\tx{i}\quad(\Delta\omega=0),
\end{equation}
\begin{equation}
k_{z,\gamma}=-k_{z,\tx{i}}
\quad\tx{and}\quad
k_{z,\tau}=\frac{n_2}{n_1}k_{z\tx{i}},
\end{equation}
\end{subequations}
where the reflected spatial frequency, $k_{z,\gamma}$, is equal to the incident spatial frequency, $k_{z\tx{i}}$, since reflection occurs in the same medium, while the transmitted spatial frequency, $k_{z,\tau}$, gets compressed or expanded depending on whether the second medium is denser or rarer.

\begin{figure}[h]
\centering
\includegraphics[width=\linewidth]{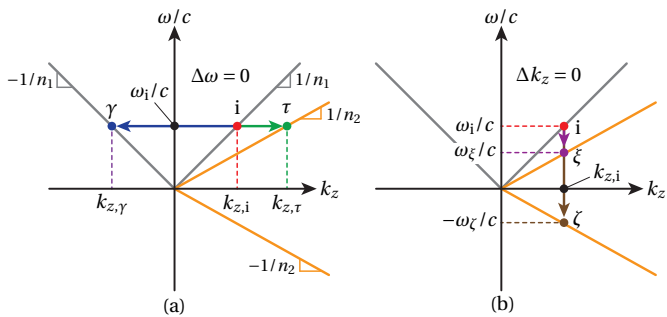}{
\psfrag{a}[c][c][0.8]{(a)}
\psfrag{b}[c][c][0.8]{(b)}
\psfrag{k}[c][c][0.8]{$k_z$}
\psfrag{w}[c][c][0.8]{$\omega/c$}
\psfrag{i}[c][c][0.8]{i}
\psfrag{g}[c][c][0.8]{$\gamma$}
\psfrag{t}[c][c][0.8]{$\tau$}
\psfrag{1}[c][c][0.8]{$k_{z,\tx{i}}$}
\psfrag{2}[c][c][0.8]{$k_{z,\tau}$}
\psfrag{3}[c][c][0.8]{$k_{z,\gamma}$}
\psfrag{4}[c][c][0.8]{$\omega_\tx{i}/c$}
\psfrag{5}[c][c][0.8]{$\omega_\tx{i}/c$}
\psfrag{6}[c][c][0.8]{$\omega_\xi/c$}
\psfrag{7}[c][c][0.8]{$-\omega_\zeta/c$}
\psfrag{8}[c][c][0.8]{$k_{z,\tx{i}}$}
\psfrag{z}[c][c][0.8]{$\zeta$}
\psfrag{x}[c][c][0.8]{$\xi$}
\psfrag{W}[c][c][0.8]{$\Delta\omega=0$}
\psfrag{K}[c][c][0.8]{$\Delta k_z=0$}
\psfrag{c}[c][c][0.7]{$1/n_1$}
\psfrag{d}[c][c][0.7]{$1/n_2$}
\psfrag{e}[c][c][0.7]{$-1/n_1$}
\psfrag{f}[c][c][0.7]{$-1/n_2$}
}
\vspace{-4mm}
\caption{Pure spacetime frequency transitions between two electromagnetic media, corresponding to \figref{fig:scat_phen}. The gray and brown lines correspond to the dispersion relations of media~1 and~2, respectively. (a)~Horizontal transition in the pure-space step discontinuity, corresponding to \figref{fig:scat_phen}(a). (b)~Vertical transition in the pure-time step discontinuity, corresponding to \figref{fig:scat_phen}(b).}
\label{fig:PST_transitions}
\end{figure}

In terms of the frequency transitions discussed in Sec.~\ref{sec:freq_trans_phen}, the pure-space discontinuity relations~\eqref{eq:PS_freqs_conds} translate into the dispersion-diagram \emph{horizontal transitions} represented in \figref{fig:PST_transitions}(a). Such transitions correspond to conserved temporal frequency ($\Delta\omega=0$), and involve the transformation from the incident wave at $k_{z,\tx{i}}$ to the reflected waves at $k_{z,\gamma}$ and to the transmitted wave at $k_{z,\tau}$.

\subsection{Pure-Time Medium}\label{sec:temp_freq_trans}
The determination of the electromagnetic frequency transitions for a pure-time discontinuity may be rigorously computed from the pure-time fields given by~\eqref{eq:fields_PT}, where, avoiding again any a priori assumption, we have admitted the possibility of having different spatial frequencies, $k_{z,\tx{i}}$ and $k_{z,\zeta}$, $k_{z,\xi}$, in media~1 and~2. This time, the boundary conditions, given by~\eqref{eq:BCs_PT_D}, must hold \emph{for all positions}, which implies that $k_{z,\zeta}=k_{z,\xi}=k_{z,\tx{i}}$, or $\Delta k_z=k_{z,\zeta}-k_{z,\tx{i}}=k_{z,\xi}-k_{z,\tx{i}}=0$. Indeed, a purely temporal discontinuity cannot induce any spatial frequency transformation. At the same time, these boundary conditions lead to temporal frequency transformations. Specifically, we find form solving~\eqref{eq:BCs_PT_D}~\cite{Morgenthaler_1958}
\begin{subequations}\label{eq:PT_freqs_conds}
\begin{equation}\label{eq:PT_kz_conds}
k_{z,\zeta}=k_{z,\xi}=k_{z,\tx{i}}\quad (\Delta k_z=0)
\end{equation}
\begin{equation}\label{eq:PT_omega_conds}
\omega_\zeta=-\frac{n_1}{n_2}\omega_\tx{i}
\quad\tx{and}\quad
\omega_\xi=\frac{n_1}{n_2}\omega_\tx{i},
\end{equation}
\end{subequations}
where both the later-backward and the later-forward temporal frequencies are different from the incident one, since the both propagate in the second medium, and both undergo the same frequency expansion or compression depending on whether the second medium is denser or rarer.

In terms of the frequency transitions discussed in Sec.~\ref{sec:freq_trans_phen}, the pure-time discontinuity relations~\eqref{eq:PT_freqs_conds} translate into the dispersion-diagram \emph{vertical transitions} represented in \figref{fig:PST_transitions}(b). Such transitions correspond to conserved spatial frequency ($\Delta k_z=0$), and involve the transformation from the incident frequency at $\omega_\tx{i}/c$ to the later-forward positive frequency at $\omega_\xi/c$ and later-backward frequency at $\omega_\zeta/c$.

Note that the generation of the new temporal frequencies in a time discontinuity occurs without resorting to nonlinearity. The new frequencies are indeed produced by the product of the time-varying constitutive parameters and the time-varying excitation fields in the time-varying version of~\eqref{eq:bianis_params}, rather than from field intensity-dependent constitutive parameters. The physical process that induces the modulation is typically nonlinear, but the assumed \emph{small-signal regime}, where the wave fields are much weaker than the modulation pump, ensures linearity, with the fundamental benefit of the response being independent from the incident wave intensity, as long as it is sufficiently smaller than that of the pump. Note also that, in contrast to horizontal transitions, vertical transitions are \emph{energy non-conservative} operations\footnote{This is fundamentally incarnated in the de Broglie relation ($\Delta E=\hbar\Delta\omega$, $E$: energy), which states that a difference in frequency corresponds to a transfer of energy.}, where energy is injected into or taken away from the wave by the external source of the modulation. A time-varying system can therefore involve instability and produce gain or attenuation, as will be seen in Sec.~\ref{sec:applications}. 

\subsection{Spacetime Medium}\label{sec:ST_freq_trans}
As the spacetime scattering coefficients, the spacetime frequency transitions can be obtained from their pure-space counterparts in the subluminal case and from their pure-time counterparts in the superluminal case.

In the subluminal case, the interface is static, or pure-space, in the primed frame, where it can therefore be treated exactly as in Sec.~\ref{sec:spat_freq_trans}, i.e., from~\eqref{eq:PS_omega_conds},
\begin{equation}\label{eq:PS_omegap_conds}
\omega_\gamma'=\omega_\tau'=\omega_\tx{i}'\quad(\Delta\omega'=0).
\end{equation}
We then simply substitute the second relation of~\eqref{eq:iLT_ST_var_up_to_p} in this equation to find the temporal-frequency result 
\begin{equation}\label{eq:sub_omega}
\omega_\gamma=\frac{1-n_1v_\tx{m}/c}{1+n_1v_\tx{m}/c}\omega_\tx{i}
\quad\tx{and}\quad
\omega_\tau=\frac{1-n_1v_\tx{m}/c}{1-n_2v_\tx{m}/c}\omega_\tx{i},
\end{equation}
from which the spatial-frequency results is obtained, by substituting $\omega=ck_z/n$ for each wave, as
\begin{equation}\label{eq:sub_kz}
k_{z,\gamma}=\frac{1-n_1v_\tx{m}/c}{1+n_1v_\tx{m}/c}k_{z,\tx{i}}
\quad\tx{and}\quad
k_{z,\tau}=\frac{n_2}{n_1}\frac{1-n_1v_\tx{m}/c}{1-n_2v_\tx{m}/c}\omega_{z,\tx{i}},
\end{equation}
and the contra-directional result is obtained again by exchanging $n_1$ and $n_2$ and by reversing the sign of $v_\tx{m}$.

\begin{figure}[h]
\centering\textsl{}
\includegraphics[width=\linewidth]{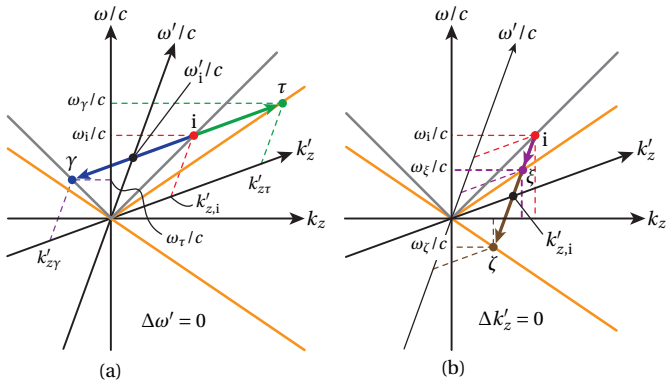}{
\psfrag{a}[c][c][0.8]{(a)}
\psfrag{b}[c][c][0.8]{(b)}
\psfrag{k}[c][c][0.8]{$k_z$}
\psfrag{w}[c][c][0.8]{$\omega/c$}
\psfrag{u}[c][c][0.8]{$k_z'$}
\psfrag{v}[c][c][0.8]{$\omega'/c$}
\psfrag{i}[c][c][0.8]{i}
\psfrag{g}[c][c][0.8]{$\gamma$}
\psfrag{t}[c][c][0.8]{$\tau$}
\psfrag{1}[c][c][0.7]{$k_{z,\tx{i}}'$}
\psfrag{2}[c][c][0.7]{$k_{z\gamma}'$}
\psfrag{3}[c][c][0.7]{$k_{z\tau}'$}
\psfrag{4}[c][c][0.8]{$\omega_\tx{i}'/c$}
\psfrag{A}[r][r][0.7]{$\omega_\tx{i}/c$}
\psfrag{B}[r][r][0.7]{$\omega_\gamma/c$}
\psfrag{C}[r][r][0.7]{$\omega_\tau/c$}
\psfrag{D}[r][r][0.7]{$\omega_\xi/c$}
\psfrag{E}[r][r][0.7]{$\omega_\zeta/c$}
\psfrag{5}[c][c][0.8]{$k_{z,\tx{i}}'$}
\psfrag{8}[c][c][0.8]{$k_{z,\tx{i}}$}
\psfrag{z}[c][c][0.8]{$\zeta$}
\psfrag{x}[c][c][0.8]{$\xi$}
\psfrag{W}[c][c][0.8]{$\Delta\omega'=0$}
\psfrag{K}[c][c][0.8]{$\Delta k_z'=0$}
}
\vspace{-3mm}
\caption{Spacetime frequency transitions between two electromagnetic media, corresponding to the co-directional situation of \figref{fig:ST_scat_coeff}. (a)~Subluminal step discontinuity, corresponding to \figref{fig:ST_scat_coeff}(a). (b)~Superluminal step discontinuity, corresponding to \figref{fig:ST_scat_coeff}(b).}
\label{fig:ST_transitions}
\end{figure}

This result may also be found geometrically and interpreted graphically with the inverse spacetime scattering of \figref{fig:ST_transitions}(a), which now displays \emph{oblique space-like transitions}. In the co-directional situation corresponding to this figure, the reflected temporal frequency ($\omega_\gamma$) is lower than the incident temporal frequency ($\omega_\tx{i}$), in agreement with \eqref{eq:sub_omega} and with the Doppler effect for a receding moving reflector, while the figure reveals that the transmitted temporal frequency  ($\omega_\tau$) is higher than the incident temporal frequency. It may be easily understood from the graph, that the transmitted frequency is up-shifted if the second medium is denser (as here) and down-shifted if it is rarer (medium dispersion curve between the vertical and medium~1 dispersion curve).

In the superluminal case, the interface is dually instantaneous, or pure time, in the primed frame, where it can therefore be treated exactly as in Sec.~\ref{sec:temp_freq_trans}, i.e., from~\eqref{eq:PT_kz_conds},
\begin{equation}\label{eq:PT_kzp_conds}
k_{z,\zeta}'=k_{z,\tau}'=k_{z,\tx{i}}'\quad (\Delta k_z'=0).
\end{equation}
We then simply substitute the first relation of~\eqref{eq:iLT_ST_var_up_to_p} in this equation to find the spatial-frequency result
\begin{equation}\label{eq:super_kz}
k_{z,\zeta}=\frac{1-c/(n_1v_\tx{m})}{1+c/(n_2v_\tx{m})}k_{z,\tx{i}}
\quad\tx{and}\quad               
k_{z,\xi}=\frac{1-c/(n_1v_\tx{m})}{1-c/(n_2v_\tx{m})}k_{z,\tx{i}},
\end{equation}
from which the temporal-frequency result is obtained, by substituting $k_z=n\omega/c$ for each wave, as
\begin{equation}\label{eq:super_omega}
\omega_\zeta=-\frac{1-(n_1v_\tx{m})/c}{1+(n_2v_\tx{m})/c}\omega_\tx{i}
\quad\tx{and}\quad               
\omega_\xi=\frac{1-(n_1v_\tx{m})/c}{1-(n_2v_\tx{m})/c}\omega_\tx{i},
\end{equation}
where the contra-directional results are obtained again by simply reversing the sign of $v_\tx{m}$.

As for the subluminal case, this result may also be found geometrically and interpreted graphically, this time with the inverse spacetime scattering of \figref{fig:ST_transitions}(b), which displays \emph{oblique time-like transitions}. In the co-directional situation, corresponding to this figure, the lower later-backward temporal frequency ($\omega_\zeta$) and a lower later-forward temporal frequency ($\omega_\xi$) are both lower than the incident or earlier frequency ($\omega_\tx{i}$) due to the higher density of the later medium, whereas a lower-density later medium would produce larger scattered frequencies.

In the case of an accelerated interface, mentioned in the last paragraph of Sec.~\ref{sec:ST_scat_coeff} in connection with the scattering coefficients, the frequency transitions can generally not be deduced from the uniform-interface frequency transitions [Eqs.~\eqref{eq:sub_omega}, \eqref{eq:sub_kz}, \eqref{eq:super_kz} and~\eqref{eq:super_omega}], due to the fundamental indeterminacy associated with the uncertainty principle, $\Delta t\cdot\Delta\omega>\pi$ and $\Delta k_z\cdot\Delta z>\pi$ (Sec.~\ref{sec:unif_ext}). However, this can still be approximately done in the particular case where the variation spacetime scale of the medium variation is much smaller than the smallest spacetime frequency of the wave, in which case the variation of the scattered frequency is called a \emph{chirp}, whose \emph{instantaneous} frequency can be simply obtained by substituting $v_\tx{m}=\tx{constant}$ by  $v_\tx{m}(t)$ in~\eqref{eq:sub_omega}, \eqref{eq:sub_kz}, \eqref{eq:super_kz} and~\eqref{eq:super_omega}, 

\section{Spacetime Reversal and Compansion}\label{sec:ST_reversal}
The oldest reference to \emph{spacetime reversal} might be the Loschmidt paradox~\cite{Weinert_2016}. In 1876, Loschmidt argued that if the velocities of all the particles of a system evolving in time were reversed, all of these particles would naturally return to their previous positions, as a movie played back returns to previous scenes~\cite{Loschmidt_1876}. Such reversibility is generally forbidden by the second law of thermodynamics, where inelastic collisions, due to random thermal fluctuations, induce irreversible heat dissipation\footnote{Dissipation loss could only be compensated by gain, but this would require an external source of energy, which, by definition, is not present in a closed system.}~\cite{Casimir_1963}. However, the basic laws of physics are time-reversal symmetric [$\ves{\Psi}(-t)=\ves{\Psi}(t)$, with $\ves{\Psi}(t)$ being the operator describing the time evolution of the physical system], in the absence of external forces; for instance, Maxwell equations [Eqs.~\eqref{eq:Maxwell_eqs}] are invariant under time-reversal~\cite{Jackson_1998}, and it is only the constitutive-relation [Eq.~\eqref{eq:bianis_params}] decomposition of $\ve{D}$ and $\ve{B}$ in terms of $\ve{E}$ and $\ve{H}$ that possibly induces irreversibility, via loss or bias present in the constitutive parameters~\cite{Caloz_PRAp_10_2018}. If dissipation is moderate, time reversal is practically realizable, and it has been abundantly applied for wave focusing in ultrasonics and electromagnetics, using arrays of transducers (engineers' ``Loschmidt demons'') properly distributed on a Hugyens surface around the source~\cite{Fink_1989,Bacot_2016}.

Time reversal is a particular incarnation of the more general concept of \emph{phase conjugation}~\cite{Ishimaru_2017}. What produces focusing is the inversion of the \emph{phase}, $\phi=\ve{k}\cdot\ve{r}-\omega t$, which does not only include a time part [$-\omega t=(\partial\phi/\partial t)t$], but also a space part [$\ve{k}\cdot\ve{r}=(\nabla_\ve{r}\phi)\cdot\ve{r}$], so reversal effects may occur not only via time but also via space and, more generally, via both space and time, or spacetime, and spacetime systems involve a diversity of extra phenomena, that may generally be referred to as \emph{compansion}, a portemanteau representing -- positive or negative\footnote{Negative compansion is equivalent to  compansion and (space, time or spacetime) reversal.} -- compression or expansion~\cite{Chamanara_arXiv_2018}, particularly useful in real-time wave processing~\cite{Caloz_MM_2013,Saleh_Teich_FP_2007}.

\begin{figure}[h]
\centering\textsl{}
\includegraphics[width=\linewidth]{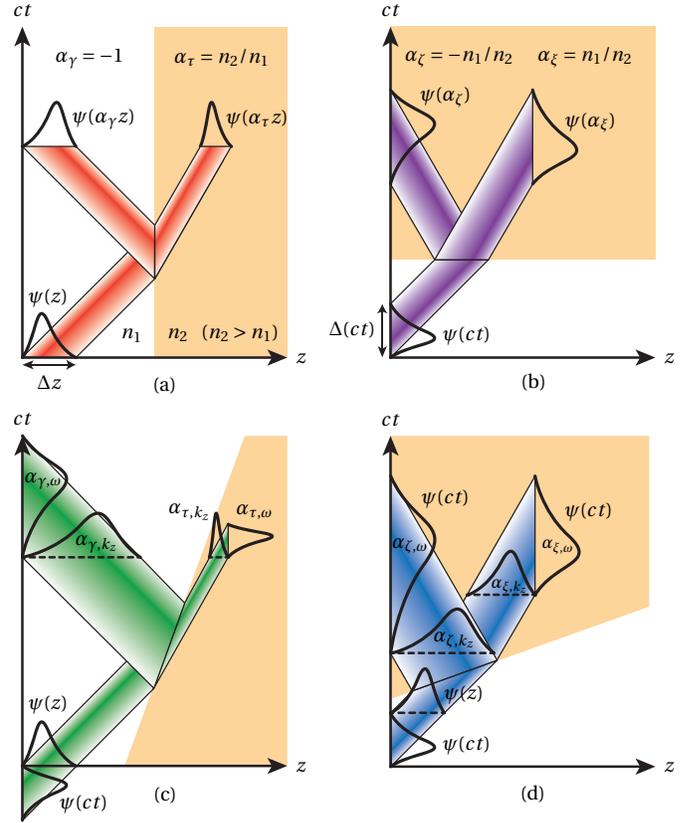}{
\psfrag{a}[c][c][0.8]{(a)}
\psfrag{b}[c][c][0.8]{(b)}
\psfrag{c}[c][c][0.8]{(c)}
\psfrag{d}[c][c][0.8]{(d)}
\psfrag{z}[c][c][0.8]{$z$}
\psfrag{t}[c][c][0.8]{$ct$}
\psfrag{1}[c][c][0.8]{$n_1$}
\psfrag{2}[c][c][0.8]{$n_2$}
\psfrag{3}[c][c][0.8]{$\psi(z)$}
\psfrag{4}[c][c][0.8]{$\psi(\alpha_\gamma z)$}
\psfrag{5}[c][c][0.8]{$\psi(\alpha_\tau z)$}
\psfrag{E}[c][c][0.8]{$\alpha_\gamma=-1$}
\psfrag{F}[c][c][0.8]{$\alpha_\tau=n_2/n_1$}
\psfrag{6}[c][c][0.8]{$\psi(ct)$}
\psfrag{7}[c][c][0.8]{$\psi(\alpha_\zeta)$}
\psfrag{8}[c][c][0.8]{$\psi(\alpha_\xi)$}
\psfrag{G}[c][c][0.8]{$\alpha_\zeta=-n_1/n_2$}
\psfrag{H}[c][c][0.8]{$\alpha_\xi=n_1/n_2$}
\psfrag{9}[c][c][0.8]{$\alpha_{\gamma,k_z}$}
\psfrag{0}[c][c][0.8]{$\alpha_{\gamma,\omega}$}
\psfrag{I}[c][c][0.8]{$\alpha_{\tau,k_z}$}
\psfrag{J}[c][c][0.8]{$\alpha_{\tau,\omega}$}
\psfrag{x}[c][c][0.8]{$\psi(ct)$}
\psfrag{y}[c][c][0.8]{$\psi(ct)$}
\psfrag{p}[c][c][0.8]{$z_0$}
\psfrag{q}[c][c][0.8]{$ct_0$}
\psfrag{B}[c][c][0.8]{$(n_2>n_1)$}
\psfrag{C}[c][c][0.8]{$\Delta z$}
\psfrag{D}[c][c][0.8]{$\Delta(ct)$}
\psfrag{K}[c][c][0.8]{$\alpha_{\zeta,\omega}$}
\psfrag{L}[c][c][0.7]{$\alpha_{\xi,k_z}$}
\psfrag{M}[c][c][0.8]{$\alpha_{\zeta,k_z}$}
\psfrag{N}[c][c][0.7]{$\alpha_{\xi,\omega}$}
}
\vspace{-6mm}
\caption{Spacetime reversal and pulse compansion (reflection or expansion) at an interface between two electromagnetic media of refractive indices $n_1$ and $n_2$, here with $n_2>n_1$ and co-directionality [$\text{sgn}(v_\tx{m})=\text{sgn}(v_\tx{wave})$] in the spacetime cases. The variable $\alpha_\chi$ ($\chi=\gamma,\tau,\zeta,\xi$) is the compansion factor, whose negativeness corresponds to reversal. The spacetime evolution of the pulse is easily understood by following the trajectory of specific points of the pulse. (a)~Pure-space interface. (b)~Pure-time interface. (c)~Subluminal spacetime interface. (d)~Superluminal spacetime interface.}
\label{fig:reversal}
\end{figure}

Figure~\ref{fig:reversal} presents a general perspective of spacetime reversal and compansion of a pulse wave at a spacetime (step) interface in the direct spacetime\footnote{The inverse spacetime representation counterpart of \figref{fig:reversal}, corresponding to \figref{fig:Lorentz_dir_inv_frames}(b) versus \figref{fig:Lorentz_dir_inv_frames}(a), can be obtained by the Fourier-transformed version of~\eqref{eq:LT_fields_up_to_p} and~\eqref{eq:LT_fields_u_to_up}, and would naturally provide a complementary spacetime \emph{frequency} perspective.}. In such a problem, the spacetime-diagram graphical tool is particularly efficient, as it provides a visual perspective of the overall physics. This section restricts to (1+1)D pulse compansion, while spacetime focusing for a (2+1)D system will be addressed in Sec.~\ref{sec:shifted_focus}. We shall also quantify compansion, beyond the qualitative geometrical results of the figure, in terms of the variation of the spacetime spectra induced by the discontinuity, in reference to the results of Sec.~\ref{sec:freq_trans}. Fundamentally, increasing (resp. decreasing) the inverse spacetime variables ($k_z$ and $\omega$) by a given factor increases (resp. decreases) the spacetime spectra ($\Delta k_z$ and $\Delta\omega$) and hence \emph{decreases} (resp. \emph{increases}) the corresponding spacetime support ($\Delta z$ and $\Delta t$) by the same factor, that we shall refer to here as the \emph{compansion factor}, $\alpha_\chi$, with $\chi=\gamma,\tau,\zeta,\xi$, with values respectively given by Eqs.~\eqref{eq:PS_freqs_conds}, \eqref{eq:PT_freqs_conds}, \eqref{eq:sub_omega}-\eqref{eq:sub_kz} and \eqref{eq:super_kz}-\eqref{eq:super_omega} for the pure-space, pure-time, subluminal spacetime and superluminal spacetime cases. To avoid overcrowding, \figref{fig:reversal} explicitly gives the expressions of these factors only for the pure-space and pure-time problems, those for the subluminal and superluminal spacetime problems being accessible in \eqref{eq:sub_omega}-\eqref{eq:sub_kz} and \eqref{eq:super_kz}-\eqref{eq:super_omega}.

Let us now describe the different cases of \figref{fig:reversal}. In the pure-space interface problem, shown in \figref{fig:reversal}(a), the incident pulse is space-reversed upon reflection and space-compressed (resp. space-expanded) upon transmission for incidence to a denser (resp. rarer) medium; in contrast, the temporal waveform undergoes neither reversal nor compansion, due to the absence of time discontinuity. In the pure-time interface problem, shown in \figref{fig:reversal}(b), the incident pulse is time-expanded (resp. time-compressed) by the same factor for the later-backward and later-forward waves for incidence to a denser (resp. rarer) medium, with the later-backward wave being also time-reversed, while the spatial waveform undergoes neither reversal nor compansion due to the absence of space discontinuity.

Spacetime compansion is more complex than pure-space and pure-time compansion because it involves both spatial and temporal reversal and compansion. In the subluminal spacetime interface problem, shown in \figref{fig:reversal}(c), the reflected wave is space-reversed and space-expanded (resp. space-compressed) and time-expanded (resp. time-compressed) for incidence to a co-directional (resp. contra-directional) medium; on the other hand, the transmitted wave is space-compressed (resp. space-expanded) and time-compressed (resp. time-expanded) for incidence to a denser (resp. rarer) medium. The problem of the superluminal spacetime interface, shown in \figref{fig:reversal}(d), is left as an exercise to the reader.

\section{New Physics}\label{sec:new_physics}
\subsection{Mirror and Cavity}\label{sec:shifted_focus}
In the previous three sections (Secs.~\ref{sec:scat_coef}, \ref{sec:freq_trans} and~\ref{sec:ST_reversal}), we have restricted our attention to the problem of scattering by spacetime interfaces between two semi-infinite media, as the building bricks of any spacetime metamaterial. We shall now address the problem of scattering, and more specifically, of \emph{reflection} -- meant here to encompass both subluminal reflected waves (space reflection) and superluminal later-backward waves (``time reflection'') -- from a spacetime \emph{slab} in free space (or in a simple host medium), and explore related exotic physical effects. We will mostly consider the transformation of the angular spectrum, with applications such as real-time imaging or radiotherapy in mind.

Specifically, we shall consider a slab that is \emph{thin} in terms of its spacetime length-duration dimensions, $(\ell,d)$, i.e., $(\ell,d)\ll(\lambda,T)$, where $\lambda$ and $T$ are respectively the smallest wavelength and the smallest period of the spatial and temporal spectra of the incident-early wave. Given its deeply subwavelength and subperiod feature, the slab does not support any Fabry-Perot resonance, and the transmitted or later-forward -- or, globally, `transmitted' -- fraction of the energy therefore propagates as if no slab existed. The whole physical interest resides thus in the reflection phenomenology. Consequently, the reflection coefficients, $\gamma$ [first relations in~\eqref{eq:scat_coeff_PS} or~\eqref{eq:scat_coeff_sub}] and $\zeta$ [first relations in~\eqref{eq:scat_coeff_PT_E} or~\eqref{eq:scat_coeff_super}], should be maximized to avoid functional loss; this is achieved by using the most abrupt possible slab interfaces, namely the \emph{step} interfaces considered in the previous sections. In the following, the slab is assumed to move at the velocity $\ve{v}_\tx{m}$ in the negative $z$ direction ($\ve{v}_\tx{m}=-|v_\tx{m}|\hatv{z}$).

Let us first consider an incident-early \emph{plane wave} propagating in the direction $\ve{k}_0$, i.e. a wave with the trivial spatial spectrum $\delta(\ve{k}-\ve{k_0})$, corresponding to the angle $\theta=\sin^{-1}(k_x/k_0)$ with respect to the normal of the slab. The limit case of a stationary or static slab ($v_\tx{m}=0$) reduces to the problem of conventional specular reflection, already described by Euclides 300~BC. The problem of a subluminal slab ($v_\tx{m}<c$) essentially corresponds to the Bradely-Einstein aberration phenomenon (Sec.~\ref{sec:freq_trans_phen}), where the specularly reflected wave deflects towards the normal of a contra-directional slab as its velocity increases towards $c$ ($|v_\tx{m}|\rightarrow c$). We recently extended this problem to a superluminal slab in~\cite{Deck_PRB_2018}. In this case, the ``reflected'' wave is slower than the contra-directional slab; therefore, it passes across it, and the later-backward wave propagates backward, towards the source, \emph{at the other side of the slab}. We found that as the slab velocity increases beyond the speed of light towards infinity ($c<|v_\tx{m}|\rightarrow\infty$), the later-backward wave (at the other side of the slab) eventually returns exactly backward to the source, as expected, in the limit ($v_\tx{m}$) of a pure-time slab.

\begin{figure}[h]
\centering
\includegraphics[width=\linewidth]{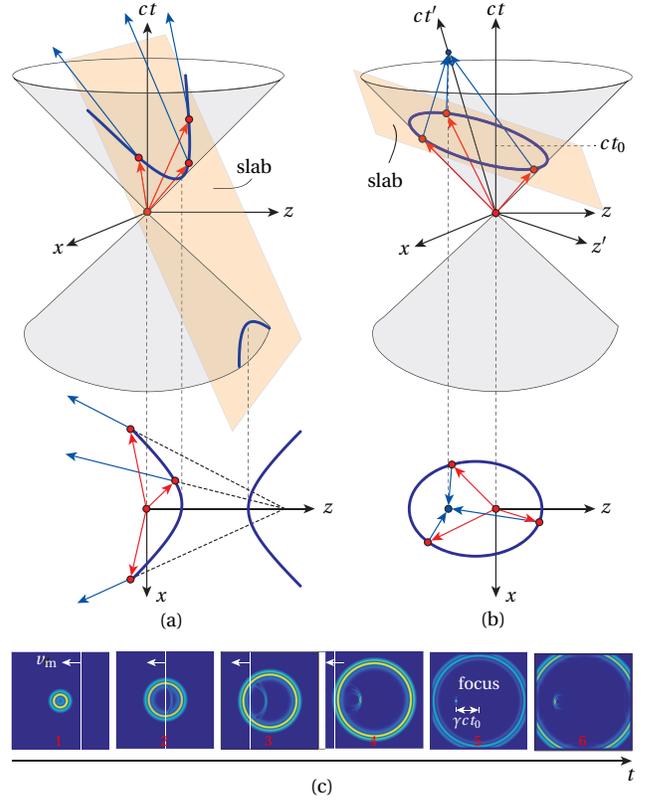}{
\psfrag{a}[c][c][0.8]{(a)}
\psfrag{b}[c][c][0.8]{(b)}
\psfrag{c}[c][c][0.8]{(c)}
\psfrag{x}[c][c][0.8]{$x$}
\psfrag{z}[c][c][0.8]{$z$}
\psfrag{t}[c][c][0.8]{$ct$}
\psfrag{Z}[c][c][0.8]{$z'$}
\psfrag{T}[c][c][0.8]{$ct'$}
\psfrag{s}[c][c][0.8]{slab}
\psfrag{v}[c][c][0.7]{\textcolor{white}{$v_\tx{m}$}}
\psfrag{7}[c][c][0.8]{$t$}
\psfrag{o}[c][c][0.8]{$ct_0$}
\psfrag{d}[c][c][0.6]{\textcolor{white}{$\gamma ct_0$}}
\psfrag{F}[c][c][0.7]{\textcolor{white}{focus}}
\psfrag{1}[c][c][0.6]{\textcolor{red}{$1$}}
\psfrag{2}[c][c][0.6]{\textcolor{red}{$2$}}
\psfrag{3}[c][c][0.6]{\textcolor{red}{$3$}}
\psfrag{4}[c][c][0.6]{\textcolor{red}{$4$}}
\psfrag{5}[c][c][0.6]{\textcolor{red}{$5$}}
\psfrag{6}[c][c][0.6]{\textcolor{red}{$6$}}
}
\vspace{-8mm}
\caption{Reflection of a circular (or spherical) wave by a thin (deeply subwavelength and subperiod) penetrable spacetime slab~\cite{Deck_PRB_2018}. (a)~Subluminal slab: hyperbolic-mirror diffraction. (b)~Superluminal slab: elliptic-cavity shifted focusing. (c)~Full-wave simulated shifted focusing by the superluminal slab of (b), represented by the vertical white line, at the following instants, from left to right: 1)~before scattering, 2)~when the slab reaches the source origin ($t_0$), 3)~when the slab has almost crossed the entire pulse, 4)~after crossing, 5)~at the focus time, 6)~after focusing.}
\label{fig:shifted_focusing}
\end{figure}

We are now prepared to address the more involved problem of scattering of a \emph{circular (or spherical) wave} from a spacetime slab, which is depicted in \figref{fig:shifted_focusing}. Let us start, as usual, with the subluminal case, represented here in \figref{fig:shifted_focusing}(a). The trajectory of a circular wave corresponds to the free-space cone of the spacetime diagram [see \figref{fig:ST_diag})(a)]. On the other hand, the thin slab, essentially corresponds plane rotated about the $x$ axis and with a slope of $c/|v_\tx{m}|$, with $c/|v_\tx{m}|>1$, in the $(ct,z)$ plane [see \figref{fig:sub_super_primed_frames}(a)]. Therefore, the scattering events coincide with the intersection of the incident-wave cone and of the slab plane, which, according to the analytic-geometry theory of conic section~\cite{Pedoe_GCC_2013}, forms a hyperbola, with the origin of the cone, and hence the wave source, coinciding with one of the foci of the hyperbola. The subluminal slab is therefore a \emph{hyperbolic mirror}, which diffracts the incident wave, as shown in the figure, with the virtual source of the reflected wave placed at the other focus of the hyperbola.

Figure~\ref{fig:shifted_focusing} represents the case of a superluminal slab. The intersection with the circular-wave cone and the plane of the slab, with slope $c/|v_\tx{m}|$, is now, according to the theory of conical sections~\cite{Pedoe_GCC_2013}, an ellipse, with the origin of the cone or wave source coinciding with one of the foci of the ellipse. The superluminal slab is therefore an \emph{elliptic cavity}, which reflects, or \emph{focuses}, the incident wave, as shown in the figure, at the other focus of the ellipse. Such \emph{shifted focusing} may be considered as a spacetime generalization of the pure-time time-reversal focusing~\cite{Bacot_2016}. Snapshots of the shifted focusing are shown in~\figref{fig:shifted_focusing}(c), where, as in the case of the plane wave, the ``reflected'' wave (or the later-backward wave) has passed to the other side of the slab. Note that the focal point is shifted with respect to the source origin by the amount $\gamma c t_0$ in the direction of the slab propagation, where $t_0$ is the time when the slab passes across the source origin. Formulas for the temporal frequency transformations of the reflected wave are available in~\cite{Deck_PRB_2018}.

\subsection{Inverse Prim and Chromatic Birefringence}\label{sec:inv_prism}
\begin{figure}[h]
\centering
\includegraphics[width=0.95\linewidth]{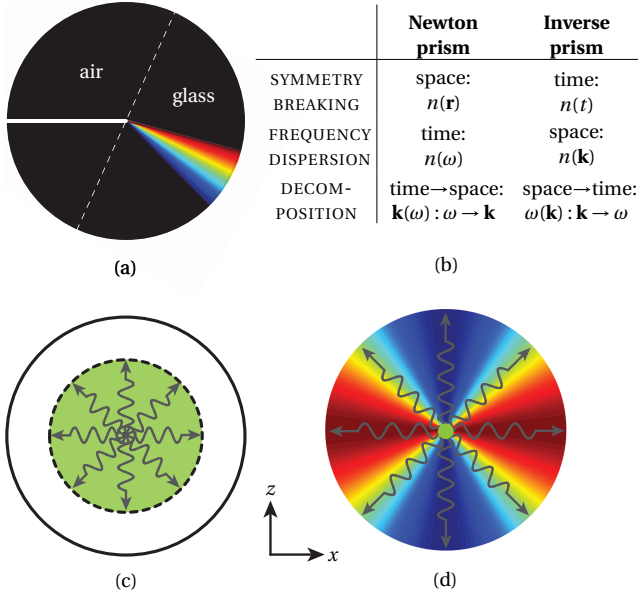}{
\psfrag{a}[c][c][0.8]{(a)}
\psfrag{b}[c][c][0.8]{(b)}
\psfrag{c}[c][c][0.8]{(c)}
\psfrag{d}[c][c][0.8]{(d)}	
\psfrag{z}[c][c][0.8]{$z$}
\psfrag{x}[c][c][0.8]{$x$}	
\psfrag{A}[c][c][0.8]{\textcolor{white}{air}}
\psfrag{B}[c][c][0.8]{\textcolor{white}{glass}}
\psfrag{X}[c][c][0.8]{
\begin{minipage}{7cm}
\centering
\begin{center}
{\renewcommand{\arraystretch}{2.2}
\begin{tabular}{c|cc}
& \begin{minipage}{1.4cm}\centering \textbf{Newton} \\ \textbf{prism} \end{minipage} 
& \begin{minipage}{1.4cm}\centering \textbf{Inverse} \\ \textbf{prism} \end{minipage} 
\\
\hline
\begin{minipage}{1.6cm}\centering {\sc symmetry} \\ {\sc breaking} \end{minipage} 
& \begin{minipage}{1.8cm}\centering space: \\ $n(\ve{r})$ \end{minipage}
& \begin{minipage}{1.8cm}\centering time: \\ $n(t)$ \end{minipage} \\  
\begin{minipage}{1.6cm}\centering {\sc frequency} \\ {\sc dispersion} \end{minipage}
& \begin{minipage}{1.8cm}\centering time: \\ $n(\omega)$ \end{minipage}
& \begin{minipage}{1.8cm}\centering space: \\ $n(\ve{k})$ \end{minipage} \\  
\begin{minipage}{1.6cm}\centering {\sc decom-} \\ {\sc position} \end{minipage}
& \begin{minipage}{1.8cm}\centering time$\rightarrow$space: \\ $\ve{k}(\omega):\omega\rightarrow\ve{k}$ \end{minipage}
& \begin{minipage}{1.8cm}\centering space$\rightarrow$time: \\ $\omega(\ve{k}):\ve{k}\rightarrow\omega$ \end{minipage}
\end{tabular}}
\end{center}
\end{minipage}}
}
\vspace{-3mm}
\caption{Concept of the inverse prism~\cite{Akbarzadeh_2018}. (a)~Conventional Newton prism. (b)~Duality between the Newton prism and the inverse prism. (c)~Circular wave propagation, with single temporal frequency, $\omega=\omega_0$, represented here by the green color, before the time discontinuity. (d)~Transformation of the wave in (c) after temporal transition to a spatially dispersive medium, with decomposition of the spatial frequencies into temporal frequencies, $\omega=\omega(\ve{k})$, distinguished here by the different colors.}
\label{fig:inv_prism}
\end{figure}

The spacetime mirror and cavity presented in Sec.~\ref{sec:shifted_focus} are based on spacetime reflection and later-backward propagation, or global reflection. We shall now present another new physical phenomenon, that is based on spacetime transmission and later-forward propagation, or global transmission: the inverse prism. The principle of this device, which was introduced in~\cite{Akbarzadeh_2018}, is described in \figref{fig:inv_prism}, where `inverse' is meant here in reference -- or duality -- with the Newton prism effect. Given its transmissive nature, this phenomenon would require, for minimal functional loss, to maximize the transmission coefficients $\tau$ [second relations in~\eqref{eq:scat_coeff_PS} or~\eqref{eq:scat_coeff_sub}] and $\xi$ [second relations in~\eqref{eq:scat_coeff_PT_E} or~\eqref{eq:scat_coeff_super}], which can be perfectly achieved (up to zero reflection) by using a sufficiently \emph{tapered} spacetime transition\footnote{The concept of \emph{spacetime} tapering is a fundamental generalization of the well-known concept of space tapering, which is abundantly used in electromagnetics engineering for matching~\cite{Pozar_ME_2011}.}.

The conventional prism, represented in \figref{fig:inv_prism}(a), is a transparent device that decomposes incident white light into its constituent colors by refracting them into different directions. This effect was explained by Newton in his book Opticks \cite{Newton_1721} to be caused by the frequency dispersive nature of the glass medium forming the prism, whereby the refractive index is a function of frequency. From a general spacetime viewpoint, the Newton prism is, as mentioned in the first column of the table in \figref{fig:inv_prism}(b), a device whose operation results from a combination of spatial-symmetry breaking ($n(\ve{r})$, corresponding to the air-glass interface) and temporal-frequency dispersion ($n(\omega)$, frequency dispersion of the glass pointed out by Newton), which decomposes the temporal frequencies of the incident temporal-spectrum broadband wave into spatial frequencies ($\ve{k}(\omega):\omega\rightarrow\ve{k}$).

In~\cite{Akbarzadeh_2018}, we addressed the following fundamental question: can spacetime be manipulated so as to provide the inverse properties of the Newton prism? The device realizing such a feat would exhibit the dual properties listed in the second column of the table in \figref{fig:inv_prism}(b): its operation would result from a combination of temporal-symmetry breaking [$n(t)$] and spatial-frequency dispersion [$n(\ve{k})$], and would decompose the spatial frequencies of the earlier spatial-spectrum broadband wave into temporal frequencies ($\omega(\ve{k}):\ve{k}\rightarrow\omega$). We showed that such a device is indeed possible~\cite{Akbarzadeh_2018}, as we shall next explain how.

Temporal-symmetry breaking is precisely what happens at a pure-time interface (Secs.~\ref{sec:PT_refl_coeff} and~\ref{sec:temp_freq_trans}), which is realized by switching between an earlier medium and a later different medium, for instance by ionizing a plasma, exciting a nonlinearity, producing a shock wave or modulating an array of vacactors. Since the sought after effect is here transmissive, this switching should ideally be a smooth tapered time transition, as mentioned above, rather than an abrupt time step. This is an advantageous feature of the inverse prism since slow time variations are practically easier to produce than fast ones. 

On the other hand, spatial-frequency dispersion may be achieved by a uniaxial medium (e.g. a 2D-lattice wire-like medium), of tensorial refractive index $\te{n}=\text{diag}\left\{n_t,n_t,n_z\right\}$, where $n_t$ and $n_z$ are the refractive indices of the (assumed nonmagnetic) medium along the transverse ($x,y$) and longitudinal ($z$) directions of the medium, respectively. In such a medium (see coordinate system in \figref{fig:inv_prism}), an $s$-wave ($\ve{E}\|\hatv{z}$) sees only the refractive index $n_t$, and hence undergoes the simple isotropic Doppler shift $\omega_2^s=c\sqrt{(k_x/n_t)^2+(k_z/n_t)^2}=(n_1/n_t)\omega_1$, where $\omega_1$ is the frequency of the earlier wave. In contrast, a $p$-wave ($\ve{E}\|(x,y)$-plane) sees a combination of the refractive indices $n_t$ and $n_z$, leading to the anisotropic frequency shift $\omega_2^p=c\sqrt{(k_x/n_t)^2+(k_z/n_z)^2}=\sqrt{(\omega_x/n_t)^2+(\omega_z/n_z)^2}$. The overall phenomenon is illustrated in Figs.~\ref{fig:inv_prism}(c) and~\ref{fig:inv_prism}(d), which represent circular propagation of a monochromatic wave before and after the time transition from an earlier isotropic medium to a later spatially dispersive medium, for the $p$-polarization. This effect might find a diversity of applications in optics. Given the fact that only one of the polarizations undergoes frequency decomposition into temporal frequencies, it may be referred to as \emph{chromatic birefringence}\footnote{An interesting effect associated to this chromatic birefringence is the fact the polarization of the a mixed $s,p$-wave follows a Lissajous polarization trajectory~\cite{Akbarzadeh_2018}.}. This effect also bears some spacetime duality with the Snell law, whereby spatial transformations are replaced by temporal transformations. Although we have discussed here a pure-time transition, a spacetime transition would be interesting to study, and would surely lead to even more exotic effects.

\subsection{Crystals}\label{sec:crystallography}
Section~\ref{sec:shifted_focus}, on the spacetime mirror and cavity, dealt with a spacetime reflective slab, while Sec.~\ref{sec:inv_prism}, on the inverse prism, dealt with a spacetime transmissive interface. Arguably, neither of these two systems corresponds to a full-fledged metamaterial. In contrast, this section describes spacetime \emph{periodic structures} or \emph{crystals}~\cite{Cassedy_1963,Biancalana_2007,Li_2012,Deck_arXiv_2018}, which may be considered as real spacetime metamaterial systems, with much greater complexity and richer physics. Such crystals, which are generically represented in the right-most middle panel of \figref{fig:gallery}, may be considered as generalizations of the pure-space periodic structures and photonic crystals~\cite{Collin_1990,Joannopoulos_2008} and of the  pure-time periodic structures and systems~\cite{Holberg_1966,Zurita_2009}. 

\begin{figure}[h]
\centering
\includegraphics[width=0.95\linewidth]{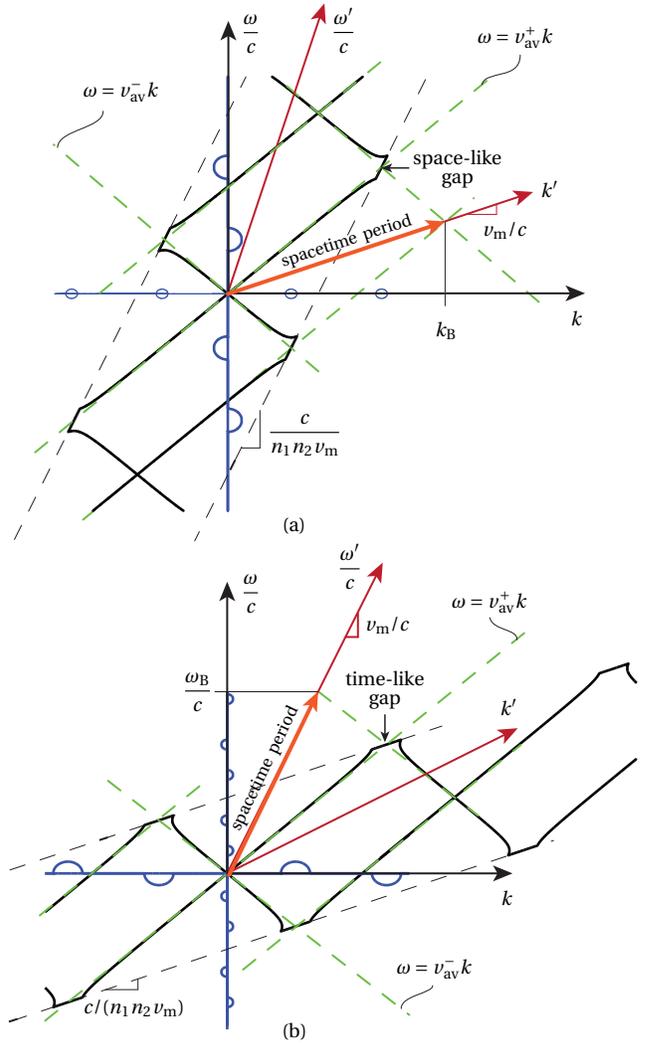}{
\psfrag{a}[c][c][0.8]{(a)}
\psfrag{b}[c][c][0.8]{(b)}
\psfrag{w}[c][c][0.8]{$\dfrac{\omega }{c}$}
\psfrag{k}[c][c][0.8]{$k$}
\psfrag{p}[c][c][0.8]{$\dfrac{\omega'}{c}$}
\psfrag{P}[c][c][0.8]{$k'$}
\psfrag{v}[c][c][0.8]{$v_\text{m}/c$}
\psfrag{V}[c][c][0.8]{$c/(n_1n_2v_\text{m})$}
\psfrag{B}[c][c][0.8]{$\dfrac{c}{n_1n_2v_\text{m}}$}
\psfrag{F}[c][c][0.8]{$\omega=v_\text{av}^+ k$}
\psfrag{N}[c][c][0.8]{$\omega=v_\text{av}^- k$}
\psfrag{f}[c][c][0.8]{$\omega=v_\text{av} k$}
\psfrag{n}[c][c][0.8]{$\omega=v_\text{av} k$}
\psfrag{u}[c][c][0.8]{$k_\text{B}$}
\psfrag{U}[c][c][0.8]{$\dfrac{\omega_\text{B}}{c}$}
\psfrag{m}[c][c][0.8]{$\vec{p}$}
\psfrag{G}[c][c][0.8]{\begin{minipage}{2cm}\centering space-like \\ \vspace{-1mm} gap\end{minipage}}
\psfrag{H}[c][c][0.8]{\begin{minipage}{2cm}\centering time-like \\ \vspace{-1mm} gap\end{minipage}}
\psfrag{Z}[c][c][0.7]{spacetime period}
}
\vspace{-2mm}
\caption{Dispersion diagrams of a (1+1)-D bilayer spacetime-modulated crystals of equal-electrical-length layers (to avoid extra complexity) with refractive index ratio $n_2/n_1=1.5$. The green dashed lines represent the average dispersion structure~\cite{Deck_arXiv_2018}. The black skewed diamond-like structure represents the real spacetime frequencies and the blue half ellipses represent the complex spacetime frequencies. (a)~Subluminal case, with $v_\tx{m}=c/3$ ($k_\tx{B}=2\pi/\ell_B$, where $\ell_\tx{B}$ is the spatial period of the crystal). (b)~Superluminal case, with $v_\tx{m}=3c$  ($\omega_\tx{B}=2\pi/T_B$, where $T_\tx{B}$ is the temporal period of the crystal).}
\label{fig:crystals}
\end{figure}

Spacetime crystals may be (1+1)D, (2+1)D or (3+1)D, where the dimension refers here to the periodicity rather than the structure itself; for instance, (1+1)D refers to a crystal with monodimensional spatial and temporal periodicity whose structure can extend in the three (volume) dimensions of space and support tridimensional wave propagation. We shall restrict here our attention to \emph{unbounded} crystals, the truncation of spacetime crystals being a relatively complex matter, which is treated elsewhere~\cite{Deck_arXiv_2018}. Our study of these unbounded spacetime crystals will be largely based on the dispersion diagrams that are plotted in \figref{fig:crystals} using a spacetime extension of the transmission matrix technique combined with the Floquet-Bloch theorem~\cite{Biancalana_2007,Deck_arXiv_2018}. We also restrict here our attention to spacetime-\emph{modulated} crystals, without covering their spacetime-moving (Secs.~\ref{sec:nonmeta_ST}) counterparts~\cite{Skorobogatiy_2008}, which would be probably less amenable to electromagnetic technology (Sec.~\ref{sec:mov_mod_specs}).

Conventional photonic crystals are \emph{pure-space} periodic structures, whose periodicity may be 1D, 2D or 3D~\cite{Joannopoulos_2008}, corresponding to 1, 5 and 14 possible distinct (Bravais) lattice structures, respectively~\cite{Kittel_1998}. They have been used in a great diversity of applications, including low-loss waveguides and cavities~\cite{Joannopoulos_2008}, diffraction gratings~\cite{Saleh_Teich_FP_2007}, spatial filters~\cite{Munk_FSS_2005}, and multifunctional metasurfaces~\cite{Holloway_2012,Glybovski_2016,Achouri_NP_2018}. Since they do not involve any time variation, they are energy conservative ($\Delta E\propto\Delta\omega=0$) and only alter the spatial spectrum, or momentum, of waves ($\Delta\ve{k}\neq 0$). They exhibit a horizontal $\omega-\ve{k}$ bandgap structure, whose stopband complex-number temporal frequencies corresponds to reflection-type attenuation. 

A \emph{subluminal spacetime crystal}, whose (1+1)D diagram is shown in \figref{fig:crystals}(a), is a generalization of a pure-space crystal~\cite{Cassedy_1963,Yu_2009,Deck_arXiv_2018}. Such a crystal exhibits the same spatial-lattice diversity as a pure-space crystal along with the additional time periodicity. In contrast to its pure-space counterpart, it exhibits an \emph{oblique bandgap structure} (dispersion curves, gaps and Brillouin zone), and is energy non-conservative, due temporal variation. Given the breaking of spacetime symmetry, it is fundamentally nonreciprocal~\cite{Yu_2009,Caloz_PRAp_10_2018}, and may therefore be used to realize magnetless nonreciprocal devices~\cite{Caloz_2018}. The slopes of the forward and backward spacetime harmonics in a spacetime crystal are different from each other because of the different average times spent by the wave in the different media composing the crystal\footnote{For instance, in a bilayer (1+1)D crystal (case of \figref{fig:crystals}), in the extreme case where the wave velocity, $v_\tx{wave}$, is equal to the crystal velocity, $v_\tx{m}$, and the velocity of one of the media, e.g. $v_1=c/n_1$, i.e., $v_\tx{wave}=v_1=v_\tx{m}$, a pulse wave from medium~1 would always remain in medium~1 in the co-directional regime, leading an an average index of $n_1$, whereas it would periodically cross the two media in the contra-directional regime, and hence experience in this regime an average refractive index depending on the two media~\cite{Deck_arXiv_2018}.}. Some periodic acousto-optic and electro-optic systems are in fact (1+1)D spacetime crystals, although they have possibly never been described from such a general perspective to date. It should be noted that a subluminal crystal can amplify reflected waves but only attenuate transmitted waves. This may be shown analytically~\cite{Deck_arXiv_2018}, but also intuitively understood from the analogy of a surfer who can be carried by a wave only if it not slower than him.

A \emph{superluminal spacetime crystal}, whose (1+1)D diagram is shown in \figref{fig:crystals}(b), shares most of the features of its subluminal counterpart. The most striking difference is that its gaps are more vertical, which results in a stronger ratio of complex $\omega$ versus complex $\ve{k}$, as seen by comparing the half-ellipses in Figs.~\ref{fig:crystals}(b) and~\ref{fig:crystals}(a). Moreover, superluminal crystals can amplify not only the reflected (contra-directional) wave, but also the transmitted (co-directional) wave~\cite{Deck_arXiv_2018}, which is the effect used in some distributed parametric amplifiers, the circuital counterparts of actual (medium) superluminal spacetime crystals. Finally, whereas subluminal crystals are essentially diffractive, as subluminal slabs (see mirroring in Sec.~\ref{sec:shifted_focus}), superluminal crystal might support interesting focusing effects, as superluminal slabs (see focusing in Sec.~\ref{sec:shifted_focus}), akin to some spacetime generalization of Anderson-like localization in photonic crystals~\cite{John_1987}. The pure-time limit case of a superluminal crystal, which recently even attracted attention as a new potential state of matter in theoretical physics~\cite{Wilczek_2012,Shapere_2012}, also prosaically corresponds to the circuital form to a lumped parametric amplifier, using a synchronized modulation (pump) with frequency at twice the signal frequency.

\section{Applications}\label{sec:applications}
Given their fundamental nature and immense diversity, spacetime metamaterials have a potential for virtually unlimited applications across the entire electromagnetic spectrum, from radio to optics through terahertz, and beyond electromagnetic waves, with mechanical (acoustic, fluid and seismic) waves, quantum mechanical waves, and gravitational waves. This section lists a number of these applications, some of which have been recently reported and others of which are still only conceptual at this point\footnote{We do not  specifically cover the application of amplification because 1)~it has already been abundantly demonstrated in distributed parametric amplifiers~\cite{Collin_2007} and 2)~because it is implicitly inherent to all spacetime systems.}. 
\subsection{Frequency Multiplication and Mixing}
Most modern electromagnetic devices are time-invariant, and their time-invariant nature imposes fundamental bounds on their properties. This is for instance the case of \emph{frequency multipliers}, which are used in terahertz or optical wave sources~\cite{Pozar_ME_2011,Saleh_Teich_FP_2007}. Frequency multipliers may be categorized as reactive multipliers (based on varactor or step-recovery diodes), resistive multipliers (based on Schottky diodes and other nonlinear I-V elements), or transistor (FET or BJT) multipliers~\cite{Pozar_ME_2011}. 

Reactive multipliers feature small loss and can theoretically reach an efficiency of 100$\%$. However, their power is distributed among the different harmonics according to the time-invariance-based Manley-Row relation~\cite{Manley_1956} $\sum_{m=0}^{\infty}\sum_{n=-\infty}^{\infty}
\frac{mP_{mn}}{m\omega_1+n\omega_2}=0$, which shows that transferring 100$\%$ of the fundamental ($m=0$) frequency energy onto the $m^\tx{th}$ harmonic frequency, i.e., reaching
\begin{equation}
\eta_\tx{reactive}=\frac{P_{m0}}{P_{10}}\rightarrow 1,
\end{equation}
implies terminating (shorting or opening, typically with stubs) all the other harmonics, which is a formidable task when $m\gg 1$. Moreover, such multipliers suffer from poor bandwidth. Resistive multipliers offer larger bandwidth (with lower efficiency), but their  harmonic generation is bound by the time-invariance-based Pantell relation~\cite{Pantell_1958} $\sum_{m=0}^{\infty}\sum_{n=-\infty}^{\infty}m^2P_{mn}\ge 0$, which shows that the conversion efficiency of the $m^\tx{th}$ harmonic frequency is restricted as
\begin{equation}\label{eq:eta_res_mult}
\eta_\tx{resistive}=\frac{P_{m0}}{P_{10}}\le\frac{1}{m^2},
\end{equation}
which is an obvious issue for large values of $m$. Transistor multipliers can reach efficiencies surpassing 100$\%$ while consuming smaller DC power, but they have their own time-invariance-related bounds. For instance, for a FET with drain current waveform $i_\tx{D}(t)=\cos(\pi t/\tau)\Pi(t/\tau)$, where $\Pi(\cdot)$ is the rectangular pulse function and $\tau$ the duration of the drain current pulse, the conversion efficiency is found from Fourier-series decomposition to scale with $m$ as
\begin{equation}
\frac{P_{m0}}{P_{10}}
=\left[\frac{4\tau}{\pi T}\frac{\cos(m\pi\tau/T)}{1-(2m\tau/T)^2}\right]^2
\propto\frac{1}{m^4},
\end{equation}
where $T=2\pi/\omega_1$ where $\omega_1$ is the fundamental, which is even worse than the efficiency of resistive multipliers [Eq.~\eqref{eq:eta_res_mult}] while requiring the same harmonic termination effort as reactive multipliers.

Time-variant systems, and hence spacetime metamaterial-based devices, are not subjected to any of the aforementioned bounds. Therefore, they intrinsically bear potential for more efficient and agile multipliers. In the (modulated) pulse regime, frequency multiplication can be achieved by a simple time discontinuity (see Secs.~\ref{sec:PT_refl_coeff} and~\ref{sec:temp_freq_trans}), and not only without any spurious harmonics and unrestricted efficiency, but also with arbitrary anharmonic conversion ratios, $\alpha=n_1/n_2$ [Eq.~\eqref{eq:PT_omega_conds}], depending only on the electromagnetic densities of the two involved media~\cite{Kalluri_2012}. Alternatively, multiplication can be accomplished by a spacetime medium (e.g. a spacetime-modulated waveguide structure) with conversion ratio $\alpha=\frac{1-(n_1v_\tx{m})/c}{1\pm(n_2v_\tx{m})/c}$ [Eq.~\eqref{eq:super_omega}] that can be tuned via $v_\tx{m}$. In the harmonic regime, frequency multiplication would require a continuous spacetime perturbation, and could be realized with a spacetime crystal structure (Sec.~\ref{sec:crystallography})~\cite{Chamanara_PRA_2018}.

Conventional, time-invariant frequency \emph{mixers} are restricted by similar bounds as frequency multipliers in both the microwave and optical frequency ranges~\cite{Pozar_ME_2011,Saleh_Teich_FP_2007}. As a result, they produce intermodulation frequencies that are difficult to suppress when a simple mixing product is required, from two (or three) input frequencies to a third (or fourth) one, as for instance in up/down-converters in radio systems and four-wave mixers in nonlinear optics.

Spacetime metamaterial structures can overcome these limitations in various ways. A first possibility is to use a spacetime periodic waveguide structure with a dispersive background medium, i.e., with a refractive index of the form
\begin{equation}
n(z,t)=n_0(\omega)\left[1+M\cos(\omega_\tx{m}t-\beta_\tx{m}z)\right],
\end{equation}
where $\beta_\tx{m}$ and $\omega_\tx{m}$ are the spatial and temporal modulation frequencies, respectively, $M$ is the modulation index, and $n_0(\omega)$ is the background-medium dispersion. In this system, the dispersion curves are engineered in such a way as to support an oblique transition (see Sec.~\ref{sec:ST_freq_trans}) to the desired mixing frequency while avoiding a cascade effect that would induce spurious intermodulation products~\cite{Chamanara_PRA_2018}. Another possibility, in the same vein, is the avoid the harmonic cascade effect with a nonuniform spatial waveguide structure~\cite{Taravati_2018}, corresponding to a refractive index of the form
\begin{equation}
n(z,t)=n_0(z)\left[1+M\cos(\omega_\tx{m}t-\beta_\tx{m}z)\right].
\end{equation}

\subsection{Matching and Filtering}
Another fundamental bound in electromagnetic devices is the \emph{matching-bandwidth} limit, which was introduced by Bode~\cite{Bode_1945} and generalized by Fano~\cite{Fano_1950} for passive, linear and time-invariant networks. The Bode-Fano limit may be summarized by the sum-rule~\cite{Schwinger_1998}
\begin{equation}\label{eq:BodeFano}
\int_0^\infty f(\omega)\ln\left[\frac{1}{\left|\Gamma(\omega)\right|}\right]\,d\omega<\Omega_\tx{lim},
\end{equation}
where $\left|\Gamma(\omega)\right|$ is the frequency-dependent magnitude of the reflection coefficient at any port of the network, $f(\omega)$ is a frequency function characterizing the network and $\Omega_\tx{lim}$ is the corresponding limit. For instance, in the case of a simple $RC$ parallel load, we have $f(\omega)=1$ and $\Omega_\tx{lim}=\pi/(RC)$, so that Eq.~\eqref{eq:BodeFano} reduces to $\int_0^\infty\ln[1/|\Gamma(\omega)|]\,d\omega<\pi/(RC)$. Such a relation essentially indicates that increasing the bandwidth ($\int_0^\infty\,d\omega$) of a network implies decreasing the inverse of its reflection magnitude (specifically, $\ln[1/|\Gamma(\omega)]$), and hence increasing the reflection level ($|\Gamma(\omega)|$), i.e., sacrificing matching. This is the so-called matching-bandwidth limit. This limit explains why a quarter-wave length transformer achieves perfect matching only at one frequency and why arbitrarily sophisticated matching networks are subjected to a trade-off between the matching flatness and the matching level (and loss, of course), which mathematically correspond to tuning $f(\omega)$ in~\eqref{eq:BodeFano}.

Revoking any of the conditions at the origin of the Bode-Fano limit, and particularly, in our current interest, the condition of time invariance, naturally removes the limit, and potentially -- although without guarantee! -- allows to achieve broader bandwidth matching. A heuristic time-variant approach to beat the Bode-Fano limit in the pulse regime would be, using for instance an array of varactors, to switch the characteristic impedance of the transmission line to the impedance of the mismatched load before the pulse reaches it, since this would suppress the reflection from the spatial discontinuity on the load. However, the switching induces then a temporal discontinuity (see Sec.~\ref{sec:PT_refl_coeff}), whose `reflected' wave [the later-backward wave, corresponding to $\zeta$ in~\eqref{eq:scat_coeff_PT_E}] might cause even more mismatch than the initial spatial discontinuity. Fortunately, as was shown in~\cite{Shlivinski_2018}, the matching efficiency can be optimized in terms of source, load and modulation powers, to indeed beat the Bode-Fano limit, even when conservatively accounting for the modulation power in the definition of the efficiency. The result in~\cite{Shlivinski_2018} could be most likely further improved by replacing the step impedance jump by an optimized tapered time discontinuity. Note that this approach works only in the pulse regime, and that, again, a periodic (crystal-type) approach (Sec.~\ref{sec:crystallography}) would be required for the continuous wave regime. In this regime, a transmission line with inductance varying as the tangent function so as to emulate a receding load for matching was theoretically presented in~\cite{Mirmoosa_2019}.

A matching-network related device that is conventionally limited by time-invariance is the \emph{filter}, and particularly microwave filters, which have a history of over 100 years of development~\cite{Matthaei_1980,Cameron_2018}. The quasi-totality of microwave filters are reflective, i.e., the energy that is prohibited from propagating in the stopbands is reflected towards the source. In some applications this may cause detuning, spurious modulation and even component destruction, hence necessitating the addition of an isolator in front of the filter, which naturally increases the insertion loss of the overall circuit. 

An alternative approach would be to elaborate \emph{absorptive filters}. Such filters arise as a natural application of subluminal codirectional spacetime crystals (see Sec.~\ref{sec:crystallography}), where the stopband energy is taken away from the wave by the modulator\footnote{An analogy to understand this intuitively would be the tennis racket, representing the (periodic) medium. If the racket moves towards the incoming ball, representing the wave, the ball gets bounced back (`reflected') with a higher speed and energy than it had before hitting the racket; in such a contra-directional scenario, the reflected wave is thus amplified. In contrast, if the racked is properly moved backwards, the incoming ball is cushioned, and hence not bounced back; in this co-directional scenario, the energy of the wave is thus absorbed by the system (here the tennis player).}, and hence dissipated from the viewpoint of the filter, instead of being reflected to the source. The periodic implementation would only be a starting point. As in pure-space filters, where spatial periodicity is a very non-optimal way to design filters and aperiodic configurations, such as those obtained with insertion loss method, are practically largely preferable~\cite{Pozar_ME_2011}, spacetime periodicity will not be the optimal way to design such absorptive spacetime filters. This is an open field of research.

\subsection{Nonreciprocity and Absorption}
As pointed out in Sec.~\ref{sec:NR_phys}, spacetime systems are inherently nonreciprocal, and space time nonreciprocity represents thus an obvious route to magnetless nonreciprocity~\cite{Caloz_2018}. Spacetime-based nonreciprocity was first suggested over 50 years ago, in the context of microwave parametric amplifiers (e.g. \cite{Kamal_1960,Baldwin_1961}). It was recently revived in optics, along with the elegant analogy with semiconductor transitions mentioned in Sec.~\ref{sec:freq_trans_phen} in~\cite{Winn_1999}, and led to novel types of magnetless isolators and circulators (e.g.~\cite{Yu_2009_2,Yu_2011,Estep_2014,Taravati_PRB_10_2017,Chamanara_PRB_10_2017,Sounas_2018}). Most interestingly, recently papers reported multidimensional nonreciprocal operations, such as nonreciprocal specular reflection or/and transmission~\cite{Shaltout_OME_11_2015,Hadad_2015} and spatial nonreciprocal circulation~\cite{Shi_2017}.

The spacetime approach to nonreciprocity is typically less practical than that provided by transistor-loaded metamaterials at microwaves~\cite{Kodera_APL_2011,Sounas_2013,Taravati_TAP_07_2017,Kodera_AWPL_2018} because it requires a modulation source that is dynamic and difficult to inject into the structure, while transistor-based magnetless nonreciprocal metamaterials use a much simpler and more economical DC source, and have already been experimentally largely demonstrated in real metamaterial structures~\cite{Kodera_AWPL_2018}. However, the implementation of transistor-based magnetless nonreciprocal devices still awaits the availability of optical transistors~\cite{Neumeier_2013,Chen_AOST_2013}.

Nonreciprocity is one of the most fundamental ways to realize absorption, since a wave that is not returned to the source is seen as absorbed from its viewpoint. We recently discovered that it can offer angle-independent absorption, amplification and phase-shifting (gyrotor-type operation) in properly designed crossanisotropic [$\chi_\tx{ee}=\chi_\tx{mm}=0$ and $\chi_\tx{em},\chi_\tx{me}\neq 0$ in~\eqref{eq:bianis_params}] metasurface configurations~\cite{Lavigne_2019}, whether using spacetime or other types of nonreciprocity. The implementation of such functionalities, still to be experimentally demonstrated, would be quite revolutionary in electromagnetics engineering.

\subsection{Electromagnetic Cloaking}
Pure-space electromagnetic cloaking, and related manipulations based on coordinate transformation designs, represent particularly spectacular electromagnetic operations~\cite{Schurig_2006,Leonhardt_2009}. This concept has been recently theoretically extended to spacetime~\cite{McCall_2010}. Unfortunately, such cloaking structures are overly complex to realize, due to their three-dimensional anisotropic and nonuniform characteristics. Poor-man versions of them, such as for instance metasurface penetrable cloaking~\cite{Dehmollaian_2018}, seem practically more realistic and still extremely useful. We recently demonstrated a space-time camouflaging version of it, where the wave incident from the interrogator undergoes temporal frequency spectrum spreading, from a phase-modulated mirror constituted by grounded patches with diode loading, so as to render the scattered wave undetectable by an interrogator not possessing the spreading key~\cite{Wang_2019}.

\subsection{Electromagnetic Processing}
Spacetime metamaterials represent a most natural opportunity for processing signal waves in an ultrafast and flexible manner. They could, for instance, dramatically extend the range of pure-space real-time processing electromagnetic and metamaterial systems~\cite{Caloz_IEEEProc_2011,Caloz_MM_2013,Silva_2014,Estakhri_2019}. This will be particularly true when they will, soon, be driven by processors and involve artificial intelligence. Until this becomes a reality, simpler but again still very useful versions of them can be realized. We shall here only describe two examples of spacetime processing systems that we recently developed.

The first example is the spacetime compander (Sec.~\ref{sec:ST_reversal}) reported in~\cite{Chamanara_arXiv_2018}. Compansion, i.e., pulse compression or expansion, is widely used in photonic technologies, for instance in high-resolution spectroscopy, high-power pulse generation, and ultrafast signal processing~\cite{Saleh_Teich_FP_2007}. Conventional compansion is based on nonlinear phase chirping followed by group velocity dispersion with the opposite or identical chirp slope for compression or expansion, respectively. However, this technique suffers from signal distorsion and superposition incapability, due to its nonlinear nature. The system in~\cite{Chamanara_arXiv_2018} is based on a new, spacetime, mechanism that is immune to these issues. It uses an antisymmetric (monocycle-shaped for a Gaussian-like pulse) spacetime-modulated medium that co-propagates with the pulse, typically in a modulated waveguide structure, and compands it by continuously modifying the velocity of the rising and trailing sections of the pulse. 

Such a compander is a nonuniform-velocity medium, whose acceleration profile is a monocycle with an acceleration part and a deceleration part alternating at the centroid of the pulse. More elaborated forms of such a medium could lead to other interesting applications, such as for instance quantum noise squeezing~\cite{Teich_1989}, which increases the precision of interferometers by reshaping the waveform so as to trade magnitude accuracy for phase accuracy, at the limit of the uncertainty principle, for the detection of signals with extremely low signal-to-noise ratios, such as gravitational waves~\cite{Castelvecchi_2016}.

The second example is the \emph{spacetime-spectra transforming metasurface}, reported in~\cite{Chamanara_TAP_2019}. This device results from the recent development of metasurface synthesis techniques, and particularly the General Sheet Transition Conditions (GSTCs) technique coupled with bianisotropic susceptibilities~\cite{Achouri_TAP_2085,Achouri_NP_2018}. The GSTCs are a generalization of the conventional boundary conditions~\cite{Jackson_1998} for discontinuities supporting \emph{surface} polarization currents. They take, in the time domain, the general form~\cite{Idemen_1987}
\begin{subequations}\label{eq:GSTC}
\begin{equation}
\hatv{n}\times\Delta\ve{H}
=\ve{J}+\dfrac{\partial\ve{P}_{\tx{s,e},t}}{\partial t}
-\hatv{n}\times\nabla P_{\tx{s,m},n},
\end{equation}
\begin{equation}
\hatv{n}\times\Delta\ve{E}
=-\ve{K}-\dfrac{\partial\ve{P}_{\tx{s,m},t}}{\partial t}
-\hatv{n}\times\nabla P_{\tx{s,m},n},
\end{equation}
\end{subequations}
where $\hatv{n}$ is a unit vector perpendicular to the metasurface, the subscripts $n$ and $t$ respectively denote the components that are normal and tangential to the metasurface, the superscripts $\pm$ denote the positions infinitesimally just before and beyond the metasurface (e.g. $z=0^\pm$ if $\hatv{n}=\hatv{z}$ and the metasurface is positioned at $z=0$), $P_{\tx{s,e}}$~(As/m) and $P_{\tx{s,m}}$~(Vs/m) are the surface polarization densities related to the volume polarization densities in~\eqref{eq:gen_med_resp} via 
$P_{\tx{s,e}}=P_\tx{e}\delta(z)$ and $P_{\tx{s,m}}=P_\tx{m}\delta(z)$ ($\delta(z)$ (1/m): Dirac delta distribtion), having the surface version (surface instead of volume susceptibility) constitutive relations of the volume constitutive relations~\eqref{eq:LSTiSTnd}, and where
\begin{subequations}\label{eq:Diff}
\begin{equation}
\Delta\ve{H}=\left[\ve{H}^+(\rho,t)-\ve{H}^-(\rho,t)\right],
\end{equation}
\begin{equation}
\Delta\ve{E}=\left[\ve{E}^+(\rho,t)-\ve{E}^-(\rho,t)\right].
\end{equation}
\end{subequations}
Time-integrating~\eqref{eq:GSTC} with~\eqref{eq:Diff} and the surface version of~\eqref{eq:LSTiSTnd}, and solving the result for the surface susceptibilities yields the spacetime surface susceptibility functions, $\te{\chi}_{\tx{s},ab}(\ve{\rho},t)$ ($ab=$ ee, mm, em, me), required to theoretically transform an arbitary incident wave into an arbitrary scattered wave in terms of both temporal spectrum ($\omega$) and spatial spectrum ($\ve{k}$)~\cite{Chamanara_TAP_2019}. The paper~\cite{Chamanara_TAP_2019} shows the example of a normally incident modulated pulse wave being refracted to two different directions of space, with the time-reversed version of the incident waveform in one direction and the time-derivative version of the incident waveform in the other direction. This method is universal. Such spacetime metasurfaces are still challenging to fabricate at the present time, but some promising simpler versions of them have already been experimentally demonstrated~\cite{Zhang_Nat_2018,Wang_2019}.

\subsection{Radiation}
The forthcoming decade will witness the development of a flurry of spacetime antennas, reflectors, polarizers and metasurface radomes with very diverse functionalities. At the time of this writing, this is still an emerging area with very little reports. We shall only mention here, again, two examples of related systems that we recently developed.

The first example is that of a \emph{spacetime leaky-wave structure} that accomplishes the function of a full radio front-end system~\cite{Taravati_TAP_02_2017}. The system is a two-port periodically spacetime modulated structure, with one port used for transmitting and the other for receiving. In the uplink, the base-band signal is injected into the transmit port, converted to the RF band by oblique transition into the radiation cone, and hence radiated to the desired direction of space, without leaking any significant energy towards the receive port due to exponential leakage decay along the structure. In the downlink, the incoming signal is picked up by the leaky-wave antenna structure, unidirectionally routed to the receive port -- as if passing through a circulator -- due to the inherent nonreciprocity of the spacetime modulation, and downconverted to base-band when reaching the receive port.

The second example is in fact accomplishing two or three very distinct and powerful operations with the combination of a single antenna (rather than an array) and a metasurface~\cite{Wang_2_2019,Wang_3_2019}. The metasurface in these applications is a structure whose elements (or rows for 2D-plane instead of 3D processing) are individually addressed by spacetime modulating waves. The first metasurface-antenna system performs Direction-of-Arrival (DoA) detection: the elements of the metasurface are time-modulated by different orthogonal codes, and DoA is accomplished by measuring the delay between the different waveforms obtained by multiplication of the received mixed signal
with the different codes~\cite{Wang_2_2019}. The second metasurface-antenna system performs spatial multiplexing for wireless communication using a same modulated mirror as in~\cite{Wang_2019}: different wireless signal waves, carrying different orthogonal codes, are received by the metasurface that is modulated by the same sequences so as to radiate the different data stream towards the intended receivers placed at different positions of space. This concept may in fact also be applied to dynamically deflect incoming waves to different directions of space as a general approach of tunable generalized reflection-refraction with gradient bianisotropic metasurfaces~\cite{Lavigne_2018}.

\section{Conclusions}\label{sec:concl}
We have presented our vision of the emerging field of spacetime metamaterials in a cohesive and -- hopefully! -- pedagogical perspective, systematically building up the physics, modeling and applications of these media from their pure-space and pure-time counterparts. The main conclusions and results may be summarized as follows:

\begin{enumerate}
\item \label{it:STM_def} A spacetime metamaterial is a metamaterial (see definition in Sec.~\ref{sec:intro}) whose structure vary in both space and time.
\item According to~\ref{it:STM_def}), spacetime metamaterials represent a generalization of conventional metamaterials, which most often vary only in space (1D, 2D or 3D array of metaparticles) or, sometimes, only in time, within the spatial extent of interest.
\item There is an infinity of natural spacetime systems in our universe that may be considered as trivial particular cases of spacetime metamaterials, since everything (known to human science) evolves in space and time. 
\item There is also a number of engineering systems, such as opto-mechanical and acousto/electro-optic devices, that may be considered as particular cases of spacetime metamaterials.
\item \label{it:matter_wave} The spacetime variation in spacetime metamaterials may occur via two fundamentally distinct mechanisms: matter motion (moving blocks of molecules, as in optomechanics) and wave modulation (dynamic `biasing' waves, as in acouto/electro-optics).
\item \label{it:ST_spectra} While spatial variation transforms only the spatial spectrum ($\ve{k}$) of waves, temporal variation transforms only their temporal spectrum ($\omega$). The combination of the two variations opens up new opportunities to simultaneously manipulate the spatial and temporal spectra, or the spacetime spectrum ($\ve{k},\omega$), of waves (see example in \figref{fig:generic_STM}). 
\item \label{it:ST_beyond_bianis} Conventional pure-space electromagnetic metamaterials already offer unprecedented opportunities for controlling waves from their general 36 bianisotropic parameters. The addition of time variation in spacetime metamaterials brings about an extra level of diversity and a fundamental generalization of pure-space bianisotropic metamaterials through the transformation of the complete spacetime spectrum of waves mentioned in~\ref{it:ST_spectra}).
\item \label{it:dirinv_ST} The generalization mentioned in~\ref{it:ST_beyond_bianis}) prompts for a unified representation of metamaterials, whose bianisotropic parameters generally depend on direct space~($\ve{r}$), direct time~($t$), inverse space~($\ve{k}$) and inverse time~($\omega$) (see \figref{fig:classification}).
\item \label{it:var_disp} The spacetime dependences in~\ref{it:dirinv_ST}) may be logically referred to as spacetime variance for the direct space~($\ve{r},t$) and spacetime dispersion for the inverse space ($\ve{k},\omega$).
\item The principle of uncertainty imposes some fundamental restrictions on the possible types of spacetime variance and dispersion in~\ref{it:var_disp}); among the 16 existing combinations (see \figref{fig:classification}), 7 are determined only in the case where the spacetime scales involved are very different from each other.
\item \label{it:nr} Spacetime metamaterials are inherently nonreciprocal, except in the direction perpendicular to their motion or modulation.
\item The phenomena of Fizeau drag and bianisotropy transformation occur in moving-matter spacetime media, but not in modulated-wave media, due to the absence of transfer of matter in them (see \ref{it:matter_wave}) above).
\item Spacetime systems can be superluminal, in the sense of energy (but not information) transfer, without violating any law of physics. This can occur in directions that are not parallel to the direction of motion of the system (e.g. guillotine). Such superluminality opens up new horizons for wave manipulations with spacetime metamaterials.
\item Spacetime diagrams constitute a powerful tool for the description of spacetime metamaterials, where the metamaterial, of arbitrary spacetime complexity, is added onto the conventional `empty' spacetime of cosmology, which restrictively deals with the motion of celestial bodies and the propagation (electromagnetic and gravitational) waves in \emph{free space}. The direct and inverse spacetime diagrams offer a complementary perspective of spacetime medium phenomenology.
\item A subluminal discontinuity involves the same scattered waves as a pure-space discontinuity (space-like), namely incident, reflected and transmitted waves; a superluminal discontiniuty involves the same scattered waves as a pure-time discontinuity (time-like), namely earlier, later-backward and later-forward waves. In both the subluminal and superluminal cases, the scattering coefficients and frequency transitions (or transformations) are drastically different in the co-directional and contra-directional cases (wave and medium moving in same or opposite directions).
\item The computation of electromagnetic scattering in spacetime media involve two difficulties that do not exist in conventional relativity: a)~in the superluminal regime, the moving frame must be taken as instantaneous instead of stationary ; b)~in a wave-modulated spacetime medium, the constitutive relations should not be applied in the moving frame, where they involve bianisotorpy complication, but in the laboratory frame.
\item As pure-space transitions between different media, spacetime transitions between different media may be abrupt (step-type) or progressive (graded-type). Some applications, such as spacetime reversal focusing, are based on reflection, and require therefore abrupt spacetime transitions, whereas others, such as the inverse prism, are based on transmission, and require therefore tapered (matching) spacetime transitions.
\item In electromagnetics, while the continuous fields at a spatial discontinuities are the tangential $\ve{E}$ and $\ve{H}$ fields, the continuous field at a temporal discontinuity are the total $\ve{D}$ and $\ve{B}$ fields. The boundary conditions for subluminal and superluminal media respectively derive from the pure-space and pure-time boundary conditions as a distinct combination of all the fields.
\item While scattering from a pure-space slab involves an infinite number of spacetime wave contributions, a pure-time slab involves exactly four spacetime wave contributions. The space and time phenomenologies are thus fundamentally different, which is due to the unidirectionality of the time arrow (or causality). Subluminal and superluminal media follow their pure-space and pure-time limit cases, with extra complexity and diversity, as for instance the nonreciprocity mentioned in Sec.~\ref{it:nr}.
\item Time reversal naturally generalizes to spacetime reversal and, further, to spacetime compansion (compression or expansion), with negative space and time compansion  respectively corresponding to space reversal and time reversal, and the type and amount of spacetime compansion depending on the electromagnetic density of the media involved in the discontinuity.
\item \label{it:physics}Spacetime metamaterials host a cornucopia of unexplored physical phenomena. Examples of recently discovered physics include spacetime mirroring and focusing, inverse prism transformation and chromatic birefringence, and generalized spacetime crystallography.
\item As a consequence of~\ref{it:physics}), spacetime metamaterials harbor a myriad of potential applications, many of which result from the revocation of the constraint of time invariance that dominates current technology. We have briefly described a number of such applications, both recently reported and still embryonic, classified in the categories of frequency multiplication and mixing, matching and filtering, nonreciprocity and absorption, cloaking, electromagnetic processing and radiation.
\item From the research point of view, the field of space metamaterials is bursting with new theoretical, computational and technological problems and challenges, whose resolution might take decades of effort but prove most rewarding.
\end{enumerate}

It may be safely predicted that spacetime metamaterials, given their fundamental nature and virtually unlimited diversity, along with recent spectacular advances in micro/nano/quanto/bio/chemico-technologies, will represent a most vibrant field of research and development in the forthcoming decades. We hope that this paper will contribute to this exciting adventure!

\appendices

\section{Derivation of the Spacetime Variable \\ Lorentz Transforms}\label{app:deriv_LT_basic}
Given the assumed homogeneity of space and time, the space and time intervals in the unprimed and primed frames (\figref{fig:ST_frame_pair}) are proportional to each other, i.e., $(\Delta z',\Delta t')\propto(\Delta z,\Delta t)$, and the relations between the space and time variables are therefore \emph{linear}, i.e.,
\begin{subequations}\label{eq:lin_ST_rel}
\begin{equation}\label{eq:lin_ST_rel_zp}
z'=Az+Bt,
\end{equation}
\begin{equation}\label{eq:lin_ST_rel_tp}
t'=Cz+Dt,
\end{equation}
\end{subequations}
where $A$, $B$, $C$ and $D$ are unknown constants, whose determination requires four independent relations.

By definition, $O'$ corresponds to the point $z'=0$ in that primed frame, and hence to the point $z=vt$ in the unprimed frame. Substituting these values into~\eqref{eq:lin_ST_rel_zp} yields a first relation:
\begin{equation}\label{eq:BA_rel}
B=-Av.
\end{equation}

According the \emph{second postulate of relativity} (constancy of speed velocity of light for all observers regardless of the source)~\cite{Einstein_1905}, we have that $c=z/t$ and $c=z'/t'$, assuming coinciding space and time origins in each frame. Substituting~\eqref{eq:lin_ST_rel} into the latter relation, replacing in the result $z$ by $ct$ following from the former relation, and simplifying yields, as a second relation,
\begin{equation}\label{eq:BCDA_rel}
B=c^2C-c(D-A).
\end{equation}

As we shall see, enforcing the condition that a transformation from the unprimed system to the primed system followed by a transformation from the primed system back to the unprimed system must be equivalent to no transformation provides the missing two relations. The first transformation is given by~\eqref{eq:lin_ST_rel}. The second transformation may be related to the first one by substituting $(A,D)\rightarrow(A,D)$ and $(B,C)\rightarrow-(B,C)$, following from the fact that $z$ and $z'$ change sign whereas $t$ and $t'$ do not change sign between the two transformations. We may then write, using for convenience matrix notation,
\begin{equation}
\begin{pmatrix} A & -B \\ -C & D \end{pmatrix}
\begin{pmatrix} z' \\ t' \end{pmatrix}
=\begin{pmatrix} A & -B \\ -C & D \end{pmatrix}
\begin{pmatrix} A & B \\ C & D \end{pmatrix}
\begin{pmatrix} z \\ t \end{pmatrix}
=\begin{pmatrix} 1 & 0 \\ 0 & 1 \end{pmatrix}
\begin{pmatrix} z \\ t \end{pmatrix},
\end{equation}
which yields
\begin{equation}
\begin{pmatrix} A^2-BC & AB-BD \\ -CA+DC & -CB+D^2 \end{pmatrix}
=\begin{pmatrix} 1 & 0 \\ 0 & 1 \end{pmatrix},
\end{equation}
that reduce to the two announced independent relations
\begin{align}
\label{eq:ABC1_rel}
A^2-BC&=1, \\
\label{eq:AD_rel}
A&=D.
\end{align}
Solving the system of equations~\eqref{eq:BA_rel}, \eqref{eq:BCDA_rel}, \eqref{eq:ABC1_rel} and~\eqref{eq:AD_rel} provides the sought after constants as $A=D=\gamma$, $B=-\gamma v$ and $C=-\gamma v/c^2$, where $\gamma=1/\sqrt{1-(v/c)^2}$. Inserting these results into~\eqref{eq:lin_ST_rel} finally provides the Lorentz transformations~\eqref{eq:LT_ST_var}.

\section{Derivation of the Electromagnetic Field \\ Lorentz Transformations}\label{app:deriv_LT_fields}
According to the \emph{first postulate of relativity} (identicality of the laws of physics in all inertial systems)~\cite{Einstein_1905}, Maxwell equations have the same form in the primed and unprimed frames, i.e.,
\begin{subequations}\label{eq:Maxwell_upm_pm}
\begin{equation}
\nabla\times\ve{E}(\ve{r},t)=-\frac{\partial\ve{B}(\ve{r},t)}{\partial t},
\quad
\nabla\times\ve{H}(\ve{r},t)=\frac{\partial\ve{D}(\ve{r},t)}{\partial t}+\ve{J}(\ve{r},t),
\end{equation}
\begin{equation}
\nabla'\times\ve{E}'(\ve{r}',t')=-\frac{\partial\ve{B}'(\ve{r}',t')}{\partial t'},
\quad
\nabla'\times\ve{H}'(\ve{r}',t')=\frac{\partial\ve{D}'(\ve{r}',t')}{\partial t'}+\ve{J}'(\ve{r}',t'),
\end{equation}
\end{subequations}
where it is to be noted that the primed and unprimed variables generally have \emph{different} values\footnote{Consider for instance a distribution of charges that is stationary in the primed frame, and hence moving at the velocity $\ve{v}$ in the unprimed frame. Such a charge distribution, of density $\rho$, produces in the unprimed frame the current $\ve{J}=\rho\ve{v}$, but $\ve{v}'=0$ implies $\ve{J}'=\rho'\ve{v}'=0$ in the primed frame. So, $\ve{J}'(\ve{r}',t')\neq \ve{J}(\ve{r},t)$.}. Assuming an $x$-polarized plane wave propagating in the $z$ direction (as the light pulse in \figref{fig:ST_frame_pair}) and no variation in the $xy$-plane ($\partial/\partial x=\partial/\partial y=0$), Eqs.~\eqref{eq:Maxwell_upm_pm} reduce to
\begin{subequations}
\begin{equation}\label{eq:Maxwell_upm_pm_ExHy}
\frac{\partial E_x(z)}{\partial z}=-\frac{\partial B_y(z)}{\partial t},
\quad
-\frac{\partial H_y(z)}{\partial z}=\frac{\partial D_x(z)}{\partial t}+J_x(z),
\end{equation}
\begin{equation}\label{eq:Maxwell_upm_pm_ExpHyp}
\frac{\partial E_x'(z')}{\partial z'}=-\frac{\partial B_y'(z')}{\partial t'},
\quad
-\frac{\partial H_y'(z')}{\partial z'}=\frac{\partial D_x'(z')}{\partial t'}+J_x'(z').
\end{equation}=
\end{subequations}

To find the relations between the fields in the unprimed and primed frames, or the electromagnetic field transformation formulas, we have to establish some comparison between Eqs.~\eqref{eq:Maxwell_upm_pm_ExHy} and~\eqref{eq:Maxwell_upm_pm_ExpHyp}. This may be done by expressing the unprimed derivation operators in~\eqref{eq:Maxwell_upm_pm_ExHy} in terms of the primed derivation operators in~\eqref{eq:Maxwell_upm_pm_ExpHyp}. For this purpose, we successively use the chain rule and apply~\eqref{eq:LT_ST_var_up_to_p} as
\begin{subequations}
\begin{equation}
\frac{\partial}{\partial z}
=\frac{\partial}{\partial z'}\frac{\partial z'}{\partial z}+\frac{\partial}{\partial t'}\frac{\partial t'}{\partial z}
=\gamma\frac{\partial}{\partial z'}-\gamma\frac{v}{c^2}\frac{\partial}{\partial t'},
\end{equation}
\begin{equation}
\frac{\partial}{\partial t}
=\frac{\partial}{\partial z'}\frac{\partial z'}{\partial t}+\frac{\partial}{\partial t'}\frac{\partial t'}{\partial t}
=-\gamma v\frac{\partial}{\partial z'}+\gamma\frac{\partial}{\partial t'}.
\end{equation}
\end{subequations}

Substituting the last equalities of these relations into~\eqref{eq:Maxwell_upm_pm_ExHy} and grouping the terms with the same operators yields
\begin{subequations}
\begin{equation}
\frac{\partial}{\partial z'}\left[\gamma\left(E_x(z)-vB_y(z)\right)\right]
=-\frac{\partial}{\partial t'}\left[\gamma\left(B_y(z)-\frac{v}{c^2}E_x(z)\right)\right],
\end{equation}
\begin{equation}
-\frac{\partial}{\partial z'}\left[\gamma\left(H_y(z)-vD_x(z)\right)\right]
=\frac{\partial}{\partial t'}\left[\gamma\left(D_x(z)+\frac{v}{c^2}H_y(z)\right)\right].
\end{equation}
\end{subequations}
Comparing these relations with those in~\eqref{eq:Maxwell_upm_pm_ExpHyp} finally provides the electromagnetic field transformations~\eqref{eq:LT_fields_up_to_p}.

\bibliographystyle{IEEEtran}
\bibliography{Spacetime_Metamaterials_Caloz}

\end{document}